\documentclass[%
 reprint,
superscriptaddress,
 amsmath,amssymb,
 aps,
 prapplied,
 longbibliography,
 floatfix,
]{revtex4-2}

\usepackage{graphicx}% Include figure files
\usepackage{dcolumn}% Align table columns on decimal point
\usepackage{bm}% bold math
\usepackage{hyperref}% add hypertext capabilities
\usepackage{braket}
\usepackage{physics}
\usepackage{amsmath}
\usepackage{xcolor}
\usepackage{qcircuit}
\renewcommand{\vec}[1]{\boldsymbol{#1}}
\makeatletter
\newcommand*{\rom}[1]{\expandafter\@slowromancap\romannumeral #1@}
\makeatother

\begin{document}

\preprint{APS/123-QED}

\title{Quantum error correction with metastable states of trapped ions using erasure conversion}

\author{Mingyu Kang}
\email{mingyu.kang@duke.edu}
\affiliation{Duke Quantum Center, Duke University, Durham, NC 27701, USA}
\affiliation{Department of Physics, Duke University, Durham, NC 27708, USA}
\author{Wesley C. Campbell}%
\affiliation{Department of Physics and Astronomy, University of California, Los Angeles, CA 90095, USA}
\affiliation{Challenge Institute for Quantum Computation, University of California, Los Angeles, CA 90095, USA}
\affiliation{Center for Quantum Science and Engineering, University of California, Los Angeles, CA 90095, USA}
\author{Kenneth R. Brown}%
\email{ken.brown@duke.edu} 
\affiliation{Duke Quantum Center, Duke University, Durham, NC 27701, USA}
\affiliation{Department of Physics, Duke University, Durham, NC 27708, USA}
\affiliation{Department of Electrical and Computer Engineering, Duke University, Durham, NC 27708, USA}
\affiliation{Department of Chemistry, Duke University, Durham, NC 27708, USA}

\date{\today}

\begin{abstract}
Erasures, or errors with known locations, are a more favorable type of error for quantum error-correcting codes than Pauli errors. Converting physical noise into erasures can significantly improve the performance of quantum error correction. Here we apply the idea of performing erasure conversion by encoding qubits into metastable atomic states, proposed by Wu, Kolkowitz, Puri, and Thompson [Nat. Comm. \textbf{13}, 4657 (2022)], to trapped ions. We suggest an erasure-conversion scheme for metastable trapped-ion qubits and develop a detailed model of various types of errors. We then compare the logical performance of ground and metastable qubits on the surface code under various physical constraints, and conclude that metastable qubits may outperform ground qubits when the achievable laser power is higher for metastable qubits.
\end{abstract}

\maketitle

\section{Introduction}

The implementation of quantum error correction (QEC) is a necessary path toward a scalable and fault-tolerant quantum computer, as quantum states are often inherently fragile and physical operations on quantum states have limited fidelities. QEC protects quantum information from errors by encoding a logical qubit into entangled states of multiple physical qubits~\cite{Knill97}.

There have been exciting efforts on manipulating and exploiting the \textit{type} of physical error such that the performance of QEC is improved. One example is the engineering of qubits and operations that have strong bias between the $X$ and $Z$ Pauli noise~\cite{Mirrahimi14, Ofek16, Puri20, Cong22} and designing QEC codes that benefit from such bias by achieving higher thresholds~\cite{Aliferis08, Li19, Guillaud19, Huang20compass, Bonilla21, Darmawan21, Dua22, Xu23}.

Another example is converting physical noise into \textit{erasures}, i.e., errors with known locations~\cite{Grassl97, Bennett97, Lu08}. It is clear that erasures are easier to correct than Pauli errors for QEC codes, as a code of distance $d$ is guaranteed to correct at most $d-1$ erasures but only $\lfloor (d-1)/2 \rfloor$ Pauli errors. 

Erasure conversion is performed by cleverly choosing the physical states encoded as qubits, such that physical noise causes leakage outside the qubit subspace. Crucially, such leakage should be detectable using additional physical operations~\cite{Alber01, Vala05, Wu22, Kubica22}. Typically, undetected leakage errors can have even more detrimental effects on QEC than Pauli errors, as traditional methods for correcting Pauli errors do not apply~\cite{Ghosh13} and methods such as leakage-reducing operations~\cite{Wu02, Byrd04, Byrd05} and circuits~\cite{Fowler13, Ghosh15, Suchara15, Brown18, Brown19, Brown20} require significant overhead. However, when leakage errors are detectable, they can be converted to erasures by resetting the leaked qubit to a known state, e.g., the maximally mixed state, within the qubit subspace. Erasure conversion is expected to achieve significantly higher QEC thresholds for hardware platforms such as superconducting qubits~\cite{Kubica22} and Rydberg atoms~\cite{Wu22}.

Trapped ions are leading candidate for a scalable quantum computing platform~\cite{Brown16}. In particular, QEC has been demonstrated in various trapped-ion experiments~\cite{Schindler11, Nigg14, Linke17, Egan21, RyanAnderson21, RyanAnderson22,  Postler22, Erhard21}, which include fault-tolerant memory~\cite{Egan21, RyanAnderson21} and even logical two-qubit gates~\cite{Erhard21, Postler22, RyanAnderson22} on distance-3 QEC codes. Here we address the question of whether the idea of erasure conversion can be applied to trapped-ion systems. 

In fact, the erasure-conversion method in Ref.~\cite{Wu22}, designed for Rydberg atoms, can be directly applied to trapped ions. In Ref.~\cite{Wu22}, a qubit is encoded in the metastable level, such that the majority (approximately $98\%$) of the spontaneous decay of the Rydberg states during two-qubit gates does not return to the qubit subspace. Additional operations can detect such decay, thereby revealing the locations of errors. For trapped ions, the spontaneous decay of the excited states during laser-based gate operations is also the fundamental source of errors, which we aim to convert to erasures in this paper. Note that an earlier work~\cite{Campbell20} has proposed a method of detecting a different type of error for trapped ions using qubits in the metastable level.  

While the most popular choice of a trapped-ion qubit is the ground qubit encoded in the $S_{1/2}$ manifold, the metastable qubit encoded in the $D_{5/2}$ or $F_{7/2}^o$ manifold is also a promising candidate~\cite{Allcock21}. Recently, high-fidelity coherent conversion between ground and metastable qubits has been experimentally demonstrated using Yb$^+$ ions~\cite{Yang22}. Also, it has been proposed that when ground and metastable qubits are used together, intermediate state measurement and cooling can be performed within the same ion chain~\cite{Allcock21}. Therefore, it is a timely task to add erasure conversion to the list of functionalities of metastable qubits. 

A careful analysis is needed before concluding that metastable qubits will be more advantageous in QEC than ground qubits. As discussed below, the erasure conversion relies on the fact that the excited states are more strongly coupled to the ground states than the metastable states. However, this fact may also cause the Rabi frequency of metastable qubits to be significantly lower than that of ground qubits, which leads to longer gate time required for metastable qubits. Also, most metastable states decay to the ground manifold after a finite lifetime, while ground qubits have practically infinite lifetime. Whether the advantage of having a higher threshold overcomes these drawbacks needs to be verified. 

This paper is organized as follows. In Sec.~\ref{sec:sec2}, we introduce the method of laser-based gate operation and erasure conversion on metastable qubits. In Sec.~\ref{sec:sec3}, we show the model of various types of errors for ground and metastable qubits and discuss the criteria of comparison. In Sec.~\ref{sec:sec4}, we briefly introduce the surface code and the simulation method. In Sec.~\ref{sec:sec5}, we present the results of comparing the QEC performance between ground and metastable qubits. Specifically, we conclude that metastable qubits may outperform ground qubits when metastable qubits allow higher laser power than ground qubits, which is reasonable considering the material loss due to lasers. In Sec.~\ref{sec:sec6}, we compare the erasure-conversion scheme on trapped ions and Rydberg atoms, as well as discuss future directions. We conclude with a summary in Sec.~\ref{sec:sec7}.

%%%%%%%%%%%%%%%%%%%%%%%%%
%%%%%%%SECTION 2%%%%%%%%%
%%%%%%%%%%%%%%%%%%%%%%%%%

\section{Erasure-conversion scheme} \label{sec:sec2}

In this paper, we denote the hyperfine quantum state as $|L, J; F, M\rangle$, where $L$, $J$, $F$, and $M$ are the quantum numbers in the standard notation. Also, we denote a set of all states with the same $L$ and $J$ as a manifold. 

We define the ground qubit as hyperfine clock qubit encoded in the $S_{1/2}$ manifold as $\ket{0}_g := \ket{0, 1/2; I-1/2, 0}$ and $\ket{1}_g := \ket{0, 1/2; I+1/2, 0}$, where $I$ is the nuclear spin~\cite{Ozeri07}. Similarly, we define the metastable qubit as hyperfine clock qubit encoded in the $D_{5/2}$ manifold as $\ket{0}_m := \ket{2, 5/2; F_0, 0}$ and $\ket{1}_m := \ket{2, 5/2; F_0+1, 0}$, which is suggested for $\rm{Ba}^+$, $\rm{Ca}^+$, and $\rm{Sr}^+$ ions~\cite{Allcock21}. Here, $F_0$ can be chosen as any integer that satisfies $|J-I| \leq F_0 < J+I$. Both qubits are insensitive to magnetic field up to first order as $M=0$.

Unlike ground qubits, metastable qubits are susceptible to idling error due to their finite lifetime. As a $D_{5/2}$ state spontaneously decays to the $S_{1/2}$ manifold, such an error is a \textit{leakage} outside the qubit subspace. The probability that idling error occurs during time duration $t$ after state initialization is given by
\begin{equation} \label{eq:idleerror}
    p^{\rm (idle)}(t) = 1 - e^{-t/\tau_m},
\end{equation}
where $\tau_m$ is the lifetime of the metastable state. Typically, $\tau_m$ is in the order of a few to tens of seconds for $D_{5/2}$ states~\cite{Allcock21}.

%%%%%%%%%%%%%%%FIGURE 1%%%%%%%%%%%%%%
\begin{figure*}[ht]
\includegraphics[width=\linewidth]{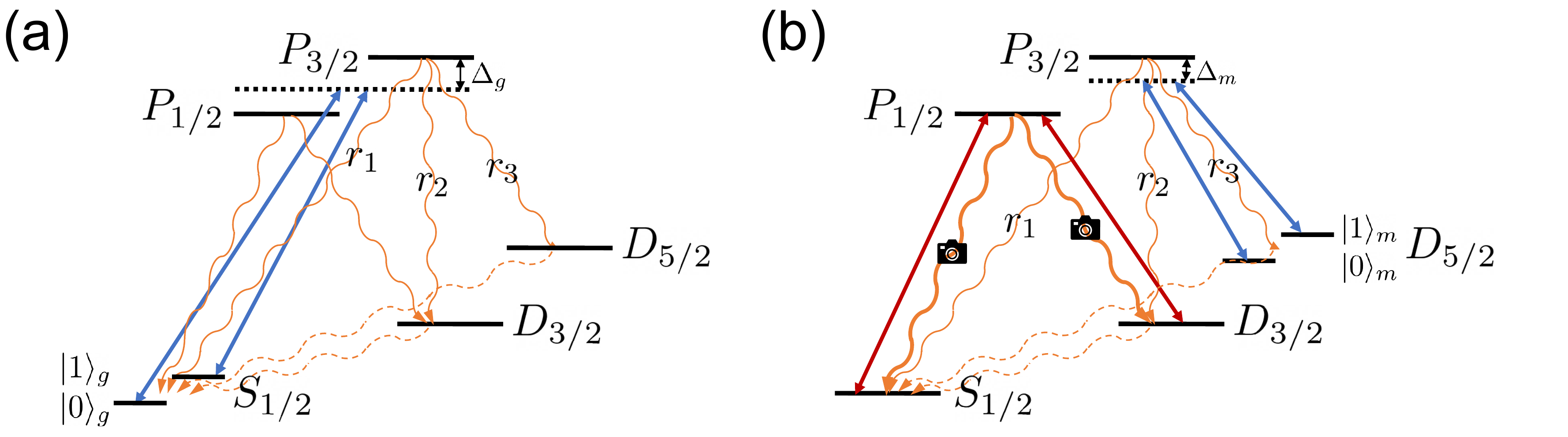}
\caption{(a) The laser-based gate operation (blue) on the $S_{1/2}$ ground qubit. (b) The laser-based gate operation (blue) and leakage detection (red) on the $D_{5/2}$ metastable qubit. The orange curvy arrows show the paths of spontaneous decay. The gate operation on the ground (metastable) qubit uses two Raman beams detuned from the $S_{1/2}$ $(D_{5/2}) \rightarrow P_{3/2}$ transition by $-\Delta_g$ $(-\Delta_m)$. When a $P_{3/2}$ state decays, the state decays to one of the $S_{1/2}$, $D_{3/2}$, and $D_{5/2}$ manifolds with probability $r_1$, $r_2$, and $r_3$, respectively. For the metastable qubit, leakage detection detects the decay to the $S_{1/2}$ and $D_{3/2}$ manifolds using lasers that are resonant to the $S_{1/2} \rightarrow P_{1/2}$ and $D_{3/2} \rightarrow P_{1/2}$ transitions, which cause photons to scatter from the $P_{1/2}$ state and get collected.}
\label{fig:fig1}
\end{figure*}
%%%%%%%%%%%%%%%%%%%%%%%%%%%%%%%%%%%%

Laser-based gate operations on ground (metastable) qubits are performed using the two-photon Raman transition, where the laser frequencies are detuned from the $S_{1/2}$ $(D_{5/2}) \rightarrow P_{3/2}$ transition, as described in Fig.~\ref{fig:fig1}. Here we define the detuning $\Delta_g$ ($\Delta_m$) as the laser frequency minus the frequency difference between the $S_{1/2}$ ($D_{5/2}$) and $P_{3/2}$ manifolds. Apart from the ``technical'' sources of gate error due to noise in the experimental system, a fundamental source of gate error is the spontaneous scattering of the atomic state from the short-lived $P$ states. During ground-qubit gates, both the $P_{1/2}$ and the $P_{3/2}$ states contribute to the two-photon transition as well as gate error, while for metastable-qubit gates, only the $P_{3/2}$ states contribute, as transition between the $D_{5/2}$ and $P_{1/2}$ states is forbidden.

When an ion that is initially in the $P_{3/2}$ state decays, the state falls to one of the $S_{1/2}$, $D_{3/2}$, and $ D_{5/2}$ manifolds with probability $r_1$, $r_2$, and $r_3$, respectively ($r_1 + r_2 + r_3 = 1$), where these probabilities are known as the \textit{resonant branching fractions}. Typically, $r_1$ is several times larger than $r_2$ and $r_3$. 

For ground (metastable) qubits, if the atomic state decays to either qubit level of the$S_{1/2}$ ($D_{5/2}$) manifold, the resulting gate error can be described as a \textit{Pauli} error. On the other hand, if the atomic state decays to the $D_{3/2}$ manifold, or the $D_{5/2}$ ($S_{1/2}$) manifold, or the hyperfine states of the $S_{1/2}$ ($D_{5/2}$) manifold other than the qubit states, the resulting gate error is a \textit{leakage}. 

We now describe how the majority of the leakage can be detected when metastable qubits are used, similarly to the scheme proposed in Ref.~\cite{Wu22}. Specifically, whenever the atomic state has decayed to either the $S_{1/2}$ or the $D_{3/2}$ manifold, the state can be detected using lasers that induce fluorescence on cycling transitions resonant to $S_{1/2} \rightarrow P_{1/2}$ and to $D_{3/2} \rightarrow P_{1/2}$, as described in Fig.~\ref{fig:fig1}(b). Unlike a typical qubit-state detection scheme where the $\ket{1}$ state is selectively optically cycled between $\ket{1}$ and appropriate sublevels in the $P_{1/2}$ manifold, this leakage detection can be performed using broadband lasers (such as in hyperfine-repumped laser cooling) such that all hyperfine levels in the $S_{1/2}$ and $D_{3/2}$ manifolds are cycled to $P_{1/2}$. 

In the rare event of detecting leakage, the qubit is reset to either $\ket{0}_m$ or $\ket{1}_m$, with probability $1/2$ each. This effectively replaces the leaked state to the maximally mixed state $I/2$ in the qubit subspace, which completes converting leakage to \textit{erasure}. Resetting the metastable qubit can be performed by the standard ground-qubit state preparation followed by coherent electric quadrupole transition. This has recently been experimentally demonstrated with high fidelity in less than \mbox{1 $\mu$s} using Yb$^+$ ions~\cite{Yang22}.  

Note that as transition between the $P_{1/2}$ and $D_{5/2}$ states is forbidden, the photons fluoresced from the $P_{1/2}$ manifold during leakage detection and ground-qubit state preparation are not resonant to any transition with the metastable-qubit states involved. This allows the erasure conversion to be performed on an ion without destroying the qubit states of the nearby ions with high probability. For ground qubits, an analogous erasure-conversion scheme of detecting leakage to the $D_{3/2}$ and $D_{5/2}$ manifolds will destroy the ground-qubit states of the nearby ions, as both the $P_{1/2}$ and the $P_{3/2}$ states decay to $S_{1/2}$ states with high probability. 

In the scheme described above, leakage to $D_{5/2}$ states other than the qubit states remains undetected. Such leakage can be handled by selectively pumping the $D_{5/2}$ hyperfine states except for $\ket{0}_m$ and $\ket{1}_m$ to the $S_{1/2}$ manifold through $P_{3/2}$. With high probability, the atomic state eventually decays to either the $S_{1/2}$ or the $D_{3/2}$ manifold, which then can be detected as described above. However, this requires the laser polarization to be aligned with high precision such that the qubit states are not accidentally pumped~\cite{Brown18}. Therefore, we defer a careful analysis on whether such process is feasible and classify leakage to other $D_{5/2}$ states as undetected leakage when the erasure-conversion scheme is used.

%%%%%%%%%%%%%%%%%%%%%%%%%
%%%%%%%SECTION 3%%%%%%%%%
%%%%%%%%%%%%%%%%%%%%%%%%%
\section{Two-qubit-gate Error model} \label{sec:sec3}

%%%%%%%%%%%%%%%FIGURE 2%%%%%%%%%%%%%%
\begin{figure}[ht]
\includegraphics[width=\linewidth]{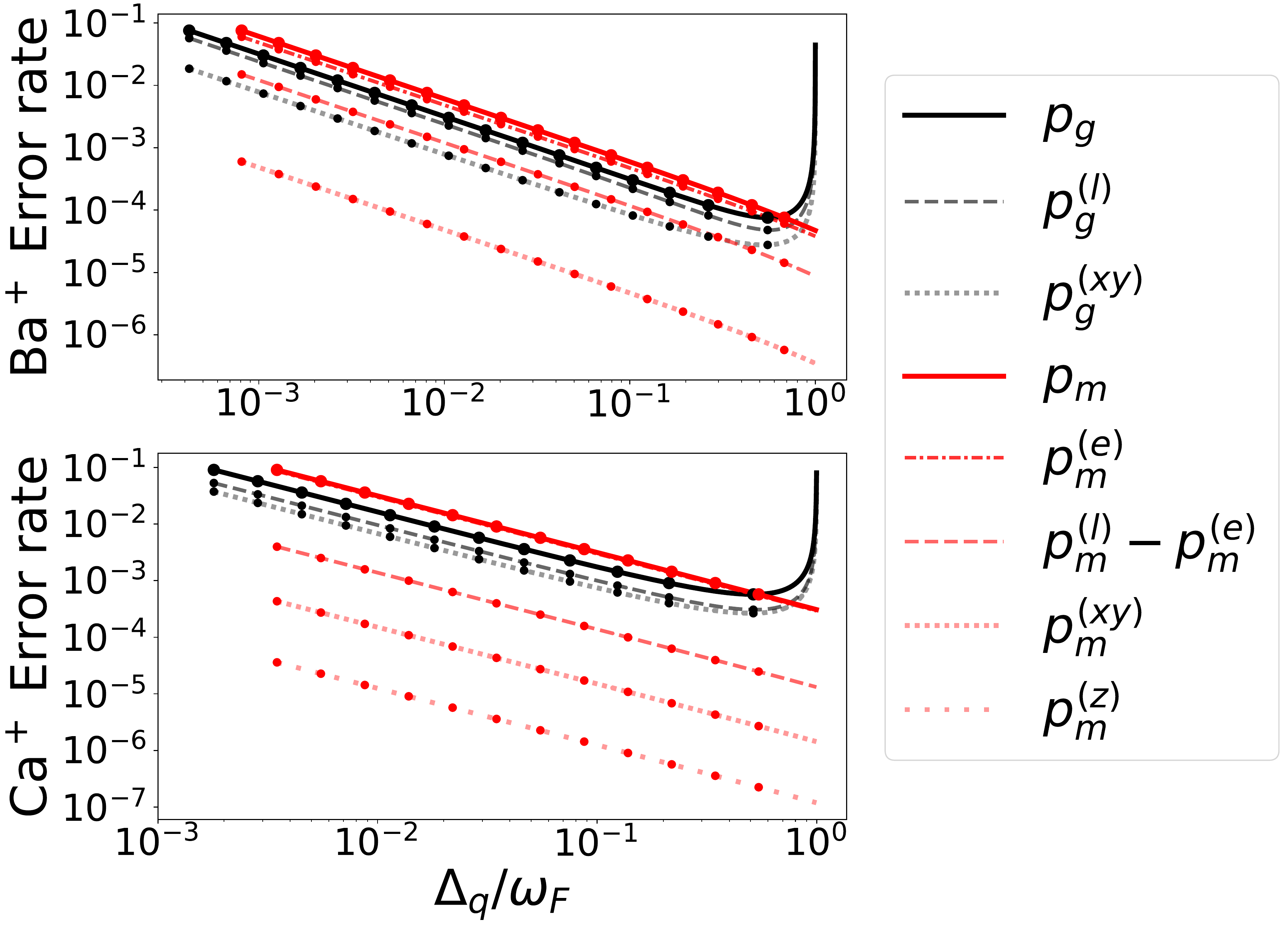}
\caption{The various types of error rates of the Ba$^+$ (top) and Ca$^+$ (bottom) qubits as the detuning $\Delta_q$ ($q=g,m$) from the $P_{3/2}$ manifold is varied. $\omega_F$ is the frequency difference between the $P_{1/2}$ and $P_{3/2}$ manifolds. For the metastable Ba$^+$ (Ca$^+$) qubit, $I=1/2$ (7/2) and $F_0 = 2$ (5), as shown in Table~\ref{tab:coeffs}. For ground (metastable) qubits, $p_g^{(l)}$ ($p_m^{(e)}$) most significantly contributes to $p_g$ ($p_m$). For Ca$^+$, the $p_m$ and $p_m^{(e)}$ curves are barely distinguishable. Note also that metastable qubits require larger detuning than ground qubits in order to have the same total error rate.}
\label{fig:fig2}
\end{figure}
%%%%%%%%%%%%%%%%%%%%%%%%%%%%%%%%%%%%

In this section, we describe the error model that we use for comparing the logical performance of ground and metastable qubits. The source of gate errors that we consider here is the spontaneous decay of excited states, which can cause various types of errors. When excited states decay back to one of the qubit states, either a bit flip or a phase flip occurs. When excited states decay to any other state, a leakage error occurs. Finally, for metastable qubits, when the state after decay is outside the $D_{5/2}$ manifold, such leakage can be converted to an erasure. 

Figure~\ref{fig:fig2} shows the various types of error rates of the Ba$^+$ and Ca$^+$ qubits as the lasers' detuning from the $P_{3/2}$ manifold is varied. Here, subscripts $g$ and $m$ denote the ground and metastable qubit, respectively, and $p_q^{(xy)}$, $p_q^{(z)}$, $p_q^{(l)}$, $p_q^{(e)}$, and $p_q$ \mbox{($q=g,m$)} denote the rate of bit flip, phase flip, leakage, erasure, and total error, respectively, for each qubit on which a two-qubit gate is applied. Up to SubSec.~\ref{subsec:2C}, we provide qualitative explanations on how these error rates are calculated from atomic physics, following the discussion in Refs.~\cite{Moore23, Uys10}. The quantitative derivations are deferred to Appendix~\ref{app:B}. 

In our model, the only controllable parameter that determines the error rates is the laser detuning from the transition to excited states. In reality, the laser power is also important, as the gate time is determined by both the detuning and the laser power. In SubSec.~\ref{subsec:2D}, we provide methods of comparing ground and metastable qubits with a fixed gate time, such that the errors due to technical noise are upper bounded to the same amount.

\subsection{Definitions}

In order to compare the gate error rates between ground and metastable qubits, we need to define several quantities. First, we define the maximal one-photon Rabi frequency of the transition between a state in the manifold of the qubit states, denoted with subscript \mbox{$q \in \{g = S_{1/2},\: m = D_{5/2}\}$}, and an excited $P$ state \mbox{($L=1$)}, as
\begin{equation} \label{eq:g}
    g_{q} := \frac{E_q}{2 \hbar} \mu_{q},
\end{equation}
where
\begin{equation} \label{eq:mu}
    \mu_{q} := \sqrt{k_q} \left| \langle L=1 || T^{(1)}({\vec{d}}) || L_q \rangle \right|.
\end{equation}
Here, $E_q$ is the electric field amplitude of the laser used for the qubit in manifold $q$, $\mu_{q}$ is the largest dipole-matrix element of transition between a state in manifold $q$ and a $P$ state, and $T^{(1)}({\vec{d}})$ is the dipole tensor operator of rank 1. Also, $k_q$ is a coefficient that relates $g_{q}$ to the \textit{orbital} dipole transition-matrix element, which is calculated in Appendix~\ref{app:A} using the Wigner-$3j$ and $6j$ coefficients.

Next, in order to obtain the scattering rates that lead to various types of errors, we first define the decay rate of the manifold of the excited states, denoted with subscript $e \in \{P_{1/2}, P_{3/2}\}$, to the final manifold, denoted with subscript $f \in \{S_{1/2}, D_{3/2}, D_{5/2}\}$, as
\begin{align} \label{eq:gamma}
    \gamma_{e,f} &:= \frac{\omega_{e,f}^3}{3\pi c^3 \hbar \epsilon_0}
    \sum_{F_f, M_f} \left| \langle L_e, J_e; F_e, M_e | \vec{d} | L_f, J_f ; F_f, M_f \rangle \right|^2 \nonumber \\
    &= \frac{\alpha_{e,f} \omega_{e,f}^3}{3\pi c^3 \hbar \epsilon_0}
    \left| \langle L=1 || T^{(1)}({\vec{d}}) || L_f \rangle \right|^2,
\end{align}
where $\omega_{e,f}$ is the frequency difference between the manifolds of the excited and final states. Also, $\alpha_{e,f}$ is a coefficient that relates $\gamma_{e,f}$ to the \textit{orbital} dipole transition-matrix element, which is calculated in Appendix~\ref{app:A} using the Wigner-$3j$ and $6j$ coefficients. Note that $\gamma_{e,f}$ does not depend on $F_e$ and $M_e$ as the frequency differences between hyperfine states of the same manifold are ignored. 

The laser-based gate operations use two-photon Raman beams of frequency $\omega_L$ that are detuned from the transition between manifolds $e$ and $q$ by $-\Delta_{e,q}$. In such case, the decay rate is given by \cite{Moore23}
\begin{equation} \label{eq:gammaprime}
\gamma'_{e,f} := \gamma_{e,f} \left(\frac{\omega_{e,f} - \Delta_{e,q}}{\omega_{e,f}} \right)^3 = \gamma_{e,f} \left(\frac{\omega_L - \omega_{f,q}}{\omega_{e,f}} \right)^3,
\end{equation}
where $\omega_{f,q}$ is the energy of manifold $f$ minus the energy of manifold $q$. Note that the numerator of the cubed factor does not depend on the choice of the manifold $e$ of the excited states. 

While manifold $e$ can be either $P_{1/2}$ or $P_{3/2}$, Eqs. (\ref{eq:gamma}) and (\ref{eq:gammaprime}) remove the dependence of $\gamma'_{e,f}/\alpha_{e,f}$ on $e$. This allows us to calculate the rates of scattering from the states of both $P_{1/2}$ and $P_{3/2}$ manifolds only using $\gamma'_{f} / \alpha_f$, where we define
\begin{gather}
    \alpha_f := \alpha_{P_{3/2}, f}, \quad 
    \gamma'_f := \gamma'_{P_{3/2}, f}. \label{eq:alphaf}
\end{gather}
Specifically, combining Eq. (\ref{eq:gammaprime}) and the branching fractions of $P_{3/2}$ states, $\gamma'_f$ is given by
\begin{equation} \label{eq:gammaprimef}
\gamma'_f = 
\begin{cases}
    (1 - \Delta_q / \omega_{P_{3/2}, S_{1/2}})^3 \times r_1 \gamma, & f = S_{1/2},\\
    (1 - \Delta_q / \omega_{P_{3/2}, D_{3/2}})^3 \times r_2 \gamma, & f = D_{3/2},\\
    (1 - \Delta_q / \omega_{P_{3/2}, D_{5/2}})^3 \times r_3 \gamma, & f =  D_{5/2},\\
\end{cases}
\end{equation}
where $\gamma$ is the total decay rate of a $P_{3/2}$ state, and $\Delta_q$ is the detuning defined as the laser frequency minus the frequency difference between manifold $q$ and the $P_{3/2}$ manifold. For the detunings considered in this paper, $\Delta_q / \omega_{P_{3/2}, f}$ is at most approximately 0.1, so $r_i \gamma$ ($i=1,2,3$) is a reasonably close upper bound for the corresponding $\gamma'_f$. 

Table~\ref{tab:ka} shows the values of $k_q$ ($\alpha_f$) for the manifolds of various qubit (final) state, and their derivations can be found in Appendix~\ref{app:A}. 

\begin{table}
\caption{\label{tab:ka} The values of the coefficient $k_q$ ($\alpha_f$), for the manifolds of various qubit (final) states.}
\begin{ruledtabular}
\begin{tabular}{ccccccc}
 & $q = S_{1/2}$ & 
 $q = D_{5/2}$ & & 
 $f = S_{1/2}$ & 
 $f = D_{3/2}$ &
 $f = D_{5/2}$ \\ \hline
 $k_q$ & 1/3 & 1/5 & $\alpha_f$ & 1/3 & 1/30 & 3/10 \\
\end{tabular}
\end{ruledtabular}
\end{table}

 \subsection{How errors arise from spontaneous scattering} \label{subsec:2B}

Spontaneous scattering of the short-lived $P_{1/2}$ and $P_{3/2}$ states is the fundamental source of errors for laser-based gates. The type of error (phase flip, bit flip, leakage, or erasure) depends on to which atomic state the short-lived states decay. 

Rayleigh and Raman scattering are the two types of spontaneous scattering. Rayleigh scattering is the elastic case where the scattered photons and the atom do not exchange energy or angular momentum. Raman scattering is the inelastic case where the photons and the atom exchange energy, thus changing the internal state of the atom. For Rayleigh scattering, the error occurs to the qubit only when the scattering rates differ between the two qubit states, which we call effective Rayleigh scattering. This results in the dephasing of the qubit, or \textit{phase-flip} ($\hat{Z}$) error. For Raman scattering, either \textit{bit-flip} ($\hat{X}$ or $\hat{Y}$) or \textit{leakage} error occurs. Finally, for metastable qubits, if the atomic state after Raman scattering is in either the $S_{1/2}$ or the $D_{3/2}$ manifold, the leakage can be detected and converted to \textit{erasure}, as described in Sec.~\ref{sec:sec2}.

We note in passing that for ground qubits, physically converting leakage to Pauli errors may be considered. Specifically, leakage of ground qubits to the $D_{3/2}$ and $D_{5/2}$ manifolds can be pumped back to the $S_{1/2}$ manifold, and leaked states in the $S_{1/2}$ manifold can be selectively pumped to the qubit states. While the former process is straightforward, the latter process suffers when the laser polarization is imperfect and unwanted (qubit) states are pumped~\cite{Brown18}. Therefore, we assume for simplicity that for ground qubits, all leakage during gates remains as leakage. 

The scattering rates during the two-photon Raman transition can be calculated using the Kramers-Heisenberg formula, as outlined in Ref.~\cite{Uys10}. In this section, we only introduce the scaling behavior and we defer the quantitative equations to Appendix~\ref{app:B}. The contribution of each excited state $\ket{J}$ in manifold $e$ to the rate of scattering from qubit state $\ket{i}$ in manifold $q$ to final state $\ket{j}$ in manifold $f$ is given by
\begin{equation} \label{eq:Gammaij}
    \Gamma_{i,j,J} = C_{i,j,J} \frac{k_q \gamma'_f}{\alpha_f} \left(\frac{g_q}{\Delta_{e,q}} \right)^2, 
\end{equation}
where $C_{i,j,J}$ is a proportionality constant that depends on the hyperfine structure of the atom and the polarization of the laser beams. By summing up over all excited and final states appropriately, the scattering rate that leads to each type of gate error can be calculated. 

For ground qubits, both the $P_{1/2}$ and $P_{3/2}$ states contribute to the scattering rates. Thus, the rates consist of terms that are proportional to both $(\omega_F - \Delta_g)^{-2}$ and $\Delta_g^{-2}$, where $\omega_F$ is the frequency difference between the $P_{1/2}$ and $P_{3/2}$ manifolds. Meanwhile, for metastable qubits, only the $P_{3/2}$ states contribute, as the $D_{5/2}$ states do not transition to $P_{1/2}$ states. Thus, the scattering rates are directly proportional to $\Delta_m^{-2}$. This explains why, in Fig.~\ref{fig:fig2}, as $\Delta_q / \omega_F$ approaches 1, the error rates of ground qubits increase but those of metastable qubits continue to decrease.

\subsection{Two-qubit-gate error rates and erasure-conversion rate} \label{subsec:2C}

\begin{table*}
\caption{\label{tab:coeffs}
The values of the nuclear spin $I$, the hyperfine total angular-momentum number $F_0$ of $\ket{0}_m$, the lifetime $\tau_m$ of the metastable state~\cite{Zhang20, Kreuter04}, the lifetime $\gamma^{-1}$ of the $P_{3/2}$ state~\cite{Zhang20, Jin93}, the geometric coefficient $c_0$ for the qubit-state Rabi frequency, the branching fractions $r_1$, $r_2$, and $r_3$ of the $P_{3/2}$ state~\cite{Zhang20, Gerritsma08}, and the zero-detuning erasure-conversion rate $R_e^{(0)}$ (in bold type) of two metastable qubits chosen as examples. 
}
\begin{ruledtabular}
\begin{tabular}{cccccccccc}
 Isotope & $I$ & $F_0$ & $\tau_m$ (s) & $\gamma^{-1}$ (ns) & $c_0$ & $r_1$ & $r_2$ & $r_3$ & $R_e^{(0)}$ \\ 
\hline
$^{133} {\rm Ba}^+$ & $1/2$ & 2 & 30.14 & 6.2615 & 1/10 & 0.7417 & 0.0280 & 0.2303 
& \textbf{0.7941}\\
$^{43} {\rm Ca}^+$ & $7/2$ & 5 & 1.16 & 6.924 & $\sqrt{7/220}$ & 0.9347 & 0.0066 & 0.0587
& \textbf{0.9509}\\
\end{tabular}
\end{ruledtabular}
\end{table*}

The error rates are obtained by the scattering rate times the gate time, which is determined by the Rabi frequency of the qubit-state transition. For ground qubits, both $S_{1/2} \rightarrow P_{1/2}$ and $S_{1/2} \rightarrow P_{3/2}$ transitions, detuned by $\omega_F - \Delta_g$ and $-\Delta_g$, respectively, contribute to the two-photon Raman transition and similarly to the scattering rate. When the two Raman beams are both linearly polarized and mutually perpendicular, the Rabi frequency of the ground qubit is given by~\cite{Wineland03, Ozeri07}
\begin{equation} \label{eq:Rabiground}
\Omega_g = \frac{g_g^2}{3} \left|\frac{\omega_F}{(\omega_F - \Delta_g) \Delta_g} \right|,
\end{equation}
where $g_g$ is the maximal one-photon Rabi frequency of a $S_{1/2}$ state.

For metastable qubits, only the $D_{5/2} \rightarrow P_{3/2}$ transition, detuned by $-\Delta_m$, contributes to the Raman transition. The Rabi frequency of the metastable qubit is given by
\begin{equation}\label{eq:Rabimeta}
\Omega_m = \frac{c_0 g_m^2}{|\Delta_m|},
\end{equation}
where $g_m$ is the maximal one-photon Rabi frequency of a $D_{5/2}$ state, and $c_0$ is a geometric coefficient determined by $I$ and $F_0$. 

The gate time for a two-qubit gate following the M\o{}lmer-S\o{}rensen (MS) scheme~\cite{Sorensen99} is typically given by
\begin{equation}\label{eq:gatetime}
t_{\rm gate} = \frac{\pi \sqrt{K}}{2 \eta \Omega},
\end{equation}
where $\Omega$ is the qubit-state Rabi frequency, $\eta$ is the Lamb-Dicke parameter, and $K$ is the number of revolutions of the trajectory of the ions in phase space~\cite{Ozeri07}. 

The error rates of each qubit on which a two-qubit gate is applied can be obtained by multiplying the corresponding sums of scattering rates in Eq.~(\ref{eq:Gammaij}) and the gate time in Eq.~(\ref{eq:gatetime}). Importantly, the $g_q^2$ factors are canceled out, which removes the dependence of the error rates on the electric field amplitude. Thus, the error rates can be expressed as functions of only the detuning $\Delta_q$, as shown in Fig.~\ref{fig:fig2}. Here, $\eta = 0.05$ and $K=1$ are fixed to their typical values~\footnote{While we use a fixed value of $\eta$ for ground and metastable qubits in our calculations, in reality the Lamb-Dicke parameter is proportional to $2 \omega_L \sin(\theta/2)/c \times \sqrt{\hbar/2M\omega}$ (up to a normalization coefficient that depends on the length of ion chain), where $\theta$ ($0 \leq \theta \leq \pi)$ is the angle between the two Raman beams, $c$ is the speed of light, $M$ is the ion mass, and $\omega$ is the motional-mode frequency. As the laser frequency $\omega_L$ is smaller for metastable qubits than for ground qubits (by a factor of roughly 1.3 for Ba$^+$ and 2.2 for Ca$^+$), fixing $\eta$ between metastable and ground qubits may require using larger $\theta$ or smaller $\omega$ for metastable qubits than ground qubits. Alternatively, metastable qubits may require additional laser power to compensate the effects of smaller $\eta$~\cite{Moore23}.}. The quantitative equations for the error rates of various types can be found in Appendix~\ref{app:B}. 

We note that single-qubit-gate error rates can also be similarly obtained by inserting $\pi/2\Omega$ into the gate time. We do not consider single-qubit-gate errors in this paper as two-qubit-gate errors are more than an order of magnitude larger, due to the additional factor $\sqrt{K}/\eta$. 

An important metric for metastable qubits is the ratio of the erasure rate to the total error rate
\begin{equation}\label{eq:Re}
    R_e := p_m^{(e)} / p_m,
\end{equation}
which we denote as the \textit{erasure-conversion rate}, following the terminology of Ref.~\cite{Wu22}. An intuitive guess of $R_e$ from Fig.~\ref{fig:fig1} would be $r_1 + r_2$, the branching fraction to the $S_{1/2}$ and $D_{3/2}$ manifolds; however, $R_e$ is slightly larger than $r_1 + r_2$ for two reasons. 

First, while $\gamma'_f = r_i \gamma$ for the corresponding $i$ when $\Delta_m = 0$, as $\Delta_m$ increases, $\gamma'_{D_{5/2}}$ decreases faster than $\gamma'_{S_{1/2}}$ and $\gamma'_{D_{3/2}}$ [see Eq.~(\ref{eq:gammaprimef})], which leads to larger $R_e$. To set a constant lower bound on $R_e$, we define the zero-detuning erasure-conversion rate
\begin{equation}\label{eq:Re0}
    R_e^{(0)} := \lim_{\Delta_m / \omega_F \rightarrow 0} R_e.
\end{equation}
For the detunings considered in this paper, $R_e$ is larger than $R_e^{(0)}$ by at most approximately $0.01$ for the Ba$^+$ qubit and $0.001$ for the Ca$^+$ qubit. 

Second, scattering to the qubit subspace of the $D_{5/2}$ manifold does not always cause an error. When the scattering is elastic, the phase-flip rate is not proportional to the Rayleigh-scattering rate but to the difference between the Rayleigh-scattering rate of the two qubit states, as explained in Sec.~\ref{subsec:2B}. 
Therefore, we expect $R_e^{(0)}$ to be slightly larger than $r_1 + r_2$ as well. The quantitative equation for $R_e^{(0)}$ can be found in Appendix~\ref{app:B}. 

Table~\ref{tab:coeffs} shows the values of various parameters relevant to the Rabi frequency and the scattering rates for two metastable qubits chosen as examples. In this paper, for each of the two species Ba$^+$ and Ca$^+$, we choose the isotope and $F_0$, the hyperfine total angular-momentum number of $\ket{0}_m$, such that the qubit-splitting frequency is the largest among the candidates shown in Ref.~\cite{Allcock21}. The chosen isotopes and $F_0$ values are $^{133} {\rm Ba}^+$, $F_0=2$ and $^{43} {\rm Ca}^+$, $F_0=5$. 

The zero-detuning erasure-conversion rate is 0.7941 (0.9509) for the Ba$^+$ (Ca$^+$) metastable qubit, which is slightly larger than $r_1 + r_2$. For both species, a large portion of the errors can be converted to erasures, which significantly improves the logical performance, as shown in Sec.~\ref{sec:sec5}.

\subsection{Comparison of ground and metastable qubits}  \label{subsec:2D}

To compare the logical performance of ground and metastable qubits, we consider the following three cases:
\begin{itemize}
    \item Case \rom{1}: $p_g = p_m$
    \item Case \rom{2}: $\Omega_g = \Omega_m$ and $E_g = E_m$
    \item Case \rom{3}: $\Omega_g = \Omega_m$ and $\Delta_g = \Delta_m$
\end{itemize}
where $E_g$ ($E_m$) is the electric field amplitude of the laser used for ground (metastable) qubits.

In case \rom{1}, the total error rate is fixed between ground and metastable qubits, as in Ref.~\cite{Wu22}. Here, we expect metastable qubits to outperform ground qubits, as a significant portion ($R_e$) of the gate errors of the metastable qubits are erasures, which is more favorable than Pauli errors for QEC.

However, such a comparison does not reflect an important disadvantage of metastable qubits. Namely, the transition between metastable and excited states is significantly weaker than that between ground and excited states, i.e., \mbox{$\mu_m \ll \mu_g$}. Therefore, given the same laser power, gates on metastable qubits require either smaller detuning [see Eq.~(\ref{eq:Rabimeta})] or a longer gate time [see Eq.~(\ref{eq:gatetime})]. Note that $R_e$ being close to one, which is an advantage of metastable qubits, also comes from the fact that \mbox{$\mu_m \ll \mu_g$}, as we see in detail below. 

If we completely ignore noise in the experimental system, we can simply use a sufficiently longer gate time for metastable qubits than for ground qubits, such that the total error rates due to spontaneous scattering match. However, this is unrealistic, especially given that the dominant sources of two-qubit-gate errors in the current state-of-the-art trapped-ion systems are motional heating and motional dephasing~\cite{Wang20, Cetina22, Kang23ff}, the effects of which build up with gate time. 

Therefore, we also compare ground and metastable qubits with fixed Rabi frequency, i.e., $\Omega_g = \Omega_m$. This requires smaller detuning for metastable qubits, which leads to larger gate error due to spontaneous scattering for metastable qubits. However, the gate time is fixed, so the gate error due to technical noise is upper bounded to the same amount.

We note that with metastable qubits and the erasure-conversion scheme, decreasing the detuning $\Delta_m$ is in some sense converting Pauli errors to erasures (and small amount of undetected leakage), as the gate time decreases at the cost of a larger spontaneous-scattering rate. Given the magnitude of technical noise, an optimal amount of detuning should exist, where the optimum is determined by the \textit{logical} error rate (for a similar approach, see Ref.~\cite{Jandura22}). On the other hand, for ground qubits, $\Delta_g$ needs to be set such that the Raman beams are far detuned from both $P_{1/2}$ and $P_{3/2}$ manifolds, as the detrimental leakage errors cannot be converted to erasures. 

To fix the Rabi frequencies, we first express the ratio $\mu_m/\mu_g$ using variables with experimentally known values. From Eqs. (\ref{eq:mu}) and (\ref{eq:gamma}), we have
\begin{equation} \label{eq:muratio}
    \bigg( \frac{\mu_m}{\mu_g} \bigg)^2 = \frac{k_m \alpha_g \gamma_{e,m}}{k_g \alpha_m \gamma_{e,g}} \bigg(\frac{\omega_{e,g}}{\omega_{e,m}} \bigg)^3 
    = \frac{2 r_3}{3 r_1} \bigg(\frac{\omega_{e,g}}{\omega_{e,m}} \bigg)^3 ,
\end{equation}
where $\omega_{e,g}$ ($\omega_{e,m}$) is the frequency difference between the $P_{3/2}$ and $S_{1/2}$ ($D_{5/2}$) manifolds. Note that $r_3/r_1$ is proportional to $(\mu_m/\mu_g)^2$, which shows that large $R_e$ stems from $\mu_m \ll \mu_g$. 

Then, the Rabi-frequency ratio $\Omega_m / \Omega_g$ is obtained by Eqs. (\ref{eq:g}), (\ref{eq:Rabiground}), and (\ref{eq:Rabimeta}) as
\begin{align} \label{eq:Rabiratio}
    \frac{\Omega_m}{\Omega_g} &= 3c_0 \left| \frac{(\omega_F - \Delta_g) \Delta_g}{\omega_F \Delta_m} \right| \bigg( \frac{\mu_m E_m}{\mu_g E_g} \bigg)^2, \nonumber\\
    &= 2c_0 \frac{r_3}{r_1} \left| \frac{(\omega_F - \Delta_g) \Delta_g}{\omega_F \Delta_m} \right| \bigg(\frac{\omega_{e,g}}{\omega_{e,m}} \bigg)^3 \bigg( \frac{E_m}{E_g} \bigg)^2.
\end{align}

The condition $\Omega_m / \Omega_g = 1$ is used for comparing ground and metastable qubits with fixed gate time. There are two additional choices of fixing variables: $E_g = E_m$ (case \rom{2}) and $\Delta_g = \Delta_m$ (case \rom{3}). In case \rom{2}, the ratio of detunings is given by
\begin{equation} \label{eq:case2}
    {\rm case \: \rom{2} :} \quad \frac{\Delta_m}{\Delta_g} = 2c_0 \frac{r_3}{r_1} \left|1 - \frac{\Delta_g}{\omega_F} \right| \bigg(\frac{\omega_{e,g}}{\omega_{e,m}} \bigg)^3,
\end{equation}
which is typically in the order of $10^{-1}$ unless $\Delta_g$ is close to $\omega_F$. In case \rom{3}, the ratio of electric field amplitudes is given by
\begin{equation} \label{eq:case3}
    {\rm case \: \rom{3} :} \quad \frac{E_m}{E_g} = \sqrt{\frac{r_1}{2c_0 r_3}} \left|1 - \frac{\Delta_g}{\omega_F} \right|^{-1/2} \bigg(\frac{\omega_{e,m}}{\omega_{e,g}} \bigg)^{3/2},
\end{equation}
which is typically several times larger than one unless \mbox{$\Delta_g=\Delta_m$} is close to $\omega_F$. 

We note that case \rom{3}, where $E_m$ is larger than $E_g$, is experimentally motivated, as the limitation on laser power is often imposed by material loss in optical devices such as mirrors and waveguides~\cite{Brown21}. Such loss is less severe with a longer laser wavelength. As the $D_{5/2} \rightarrow P_{3/2}$ transition has a longer wavelength than the $S_{1/2} \rightarrow P_{3/2}$ transition, we expect that using metastable qubits allows significantly higher laser power for gate operations than using ground qubits. The laser power required to achieve a typical Rabi frequency is estimated for both ground and metastable qubits in Appendix~\ref{app:C}. 

%%%%%%%%%%%%%%%%%%%%%%%%%
%%%%%%%SECTION 4%%%%%%%%%
%%%%%%%%%%%%%%%%%%%%%%%%%

\section{Surface-code simulation} \label{sec:sec4}

It is well established that erasures, or errors with known locations, are more favorable than other types of errors for quantum~\cite{Grassl97, Bennett97, Lu08} and classical codes, as the decoder can use the information of the erasure locations. Thus, we expect the advantage of the erasure conversion of metastable qubits to be valid for all QEC codes; however, it is certainly valuable to estimate how much the QEC performance, such as the circuit-level threshold, of a particular code is improved by erasure conversion.

\begin{figure}[ht]
\includegraphics[width=\linewidth]{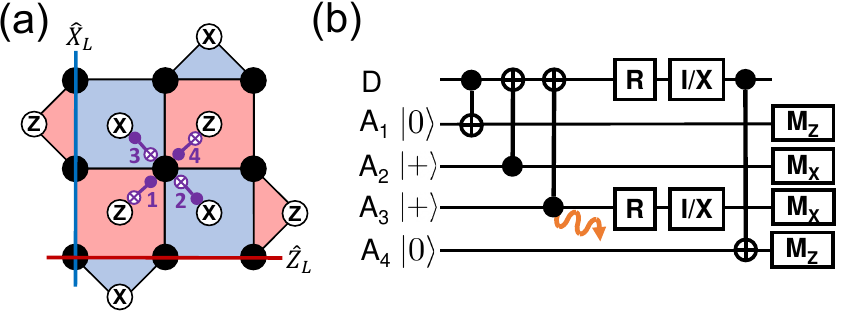}
\caption{(a) The layout of the rotated surface code of distance $d=3$. The black (white) circles represent data (syndrome) qubits. The red (blue) plaquettes represent $Z$ ($X$) stabilizers. The red horizontal (blue vertical) line represents the logical $\hat{Z}_L$ ($\hat{X}_L$) operator. The CNOT gates between a data qubit and corresponding syndrome qubits are applied in the numbered order. (b) The quantum circuit for a single round of error correction. Here, we show an example where a leakage is detected during the third CNOT gate. In such a case, both the data qubit and the third syndrome qubit are converted to the maximally mixed state by erasure conversion. $R$ denotes resetting the qubit to $\ket{0}$. $I/X$ denotes that either the $\hat{I}$ or the $\hat{X}$ gate is applied with probability 1/2 each. $M_Z$ ($M_X$) denotes qubit-state measurement in the $Z$ ($X$) basis.}
\label{fig:fig3}
\end{figure}

In this paper, we choose to simulate the surface code~\cite{Raussendorf07, Fowler09, Fowler12} (for a detailed review, see Ref.~\cite{Fowler12}). In particular, we consider the rotated surface code~\cite{Tomita14}, which uses slightly fewer qubits than the standard surface code. Figure~\ref{fig:fig3}(a) shows the rotated surface code of distance $d=3$ consisting of $2d^2-1=17$ qubits. The logical qubit is encoded in $d^2$ data qubits (black circles) and the $Z$ and $X$ stabilizers (the red and blue plaquettes, respectively) are measured using $d^2-1$ syndrome qubits (white circles). The logical operator $\hat{Z}_L$ ($\hat{X}_L$) is the product of the $\hat{Z}$ ($\hat{X}$) operators of $d$ data qubits across a horizontal (vertical) line. The measured stabilizers are used by a decoder in inferring the locations and types ($\hat{X}$, $\hat{Y}$, or $\hat{Z}$) of errors.

%%%%%%%%%%%%%%%FIGURE 3%%%%%%%%%%%%%%
%%%%%%%%%%%%%%%%%%%%%%%%%%%%%%%%%%%%

The surface code is a viable candidate for QEC in an experimental system, as it has a high (approximately 1\%) circuit-level threshold and it can be implemented using only nearest-neighbor interactions on a two-dimensional layout. Recently, there has been wide experimental success in demonstrating fault-tolerant memory of a single logical qubit encoded in the surface code using superconducting qubits~\cite{Krinner22, Zhao22, Google23}. 

We note that for trapped-ion systems, the use of nearest-neighbor interactions is not required. Thus, recent QEC experiments in trapped-ion systems have used the Bacon-Shor code~\cite{Egan21}, the five-qubit code~\cite{RyanAnderson22}, and the color code~\cite{RyanAnderson21, RyanAnderson22, Postler22}. For distance \mbox{$d=3$}, these codes allow fewer physical qubits and gate operations~\cite{Debroy20}. However, simulating the error-correction threshold using these codes can be complicated, as the family of each code of various distances is defined using code concatenation (for an example of the color code, see Ref.~\cite{Steane97}). Meanwhile, the family of rotated surface codes is straightforwardly defined on square lattices of various sizes, which leads to feasible threshold simulations.

Figure~\ref{fig:fig3}(b) shows the circuit for a single round of error correction for surface codes. First, the syndrome qubits for measuring the $Z$ ($X$) stabilizers are initialized to state $\ket{0}$ ($\ket{+}$). Then, four CNOT gates are performed between the data qubit and the syndrome qubits, in the correct order. Finally, the syndrome qubits are measured in the respective basis to provide the error syndromes. This circuit is performed on all data qubits in parallel. As the measurements can be erroneous as well, typically $d$ rounds of error correction are consecutively performed for a distance-$d$ code. 

The most probable set of errors that could have caused the observed syndromes is inferred by a decoder run by a classical computer. Among the various efficient decoders for surface codes~\cite{Dennis02, Delfosse21, Huang20faulttolerant}, we choose the minimum-weight perfect-matching (MWPM) decoder, which finds the error chain of minimum weight using Edmonds' algorithm~\cite{Dennis02}. The errors are corrected by simply keeping track of the Pauli frame, thus not requiring any physical gate operations~\cite{Fowler12}. 

Erasure conversion on metastable qubits is performed by replacing both qubits with the maximally mixed state whenever leakage is detected during a two-qubit gate. In the actual implementation, this can be done by resetting both qubits to $\ket{0}_m$ and then performing a $\hat{X}$ gate with probability $1/2$, as shown for the case of the leakage of the third syndrome qubit in Fig.~\ref{fig:fig3}(b). The data qubit is also erased as a leakage error may propagate to the other qubit during a two-qubit gate (for details, see Appendix~\ref{app:D}). The decoder uses the information on the erasure locations by setting the weight of erased data qubits to zero, which decreases the weights of error chains consisting of the erased data qubits.

We note that resetting to $\ket{0}_m$ instead of the maximally mixed state may be sufficient for erasure conversion. In this paper, we choose to reset to the maximally mixed state, as this can be described by a (maximally depolarizing) Pauli channel that is easy to simulate at the circuit level.

To evaluate the QEC performance, we simulate the logical memory of rotated surface codes in the $Z$ basis. Specifically, we initialize the data qubits to $\ket{0}$, perform $d$ rounds of error correction for the distance-$d$ code, and then measure the data qubits in the $Z$ basis. Whether a logical error has occurred is determined by comparing the measurement of the $\hat{Z}_L$ operator [see Fig.~\ref{fig:fig3}(a)] and the expected value of $\hat{Z}_L$ after decoding. This is repeated many times to determine the logical error rate. 

The circuit-level simulations in this paper are performed using STIM~\cite{Gidney21}, a software package for fast simulations of quantum stabilizer circuits. The error syndromes generated by the circuit simulations are decoded using PyMatching~\cite{Higgott22}, a software package that executes the MWPM decoder. In particular, STIM allows the simulation of erasures and feeding of the location information into the decoder~\cite{stim_erasure}. 

%%%%%%%%%%%%%%%%%%%%%%%%%
%%%%%%%SECTION 5%%%%%%%%%
%%%%%%%%%%%%%%%%%%%%%%%%%

\section{Results} \label{sec:sec5}

%%%%%%%%%%%%%%%FIGURE 4%%%%%%%%%%%%%%
\begin{figure*}[ht]
\includegraphics[width=\linewidth]{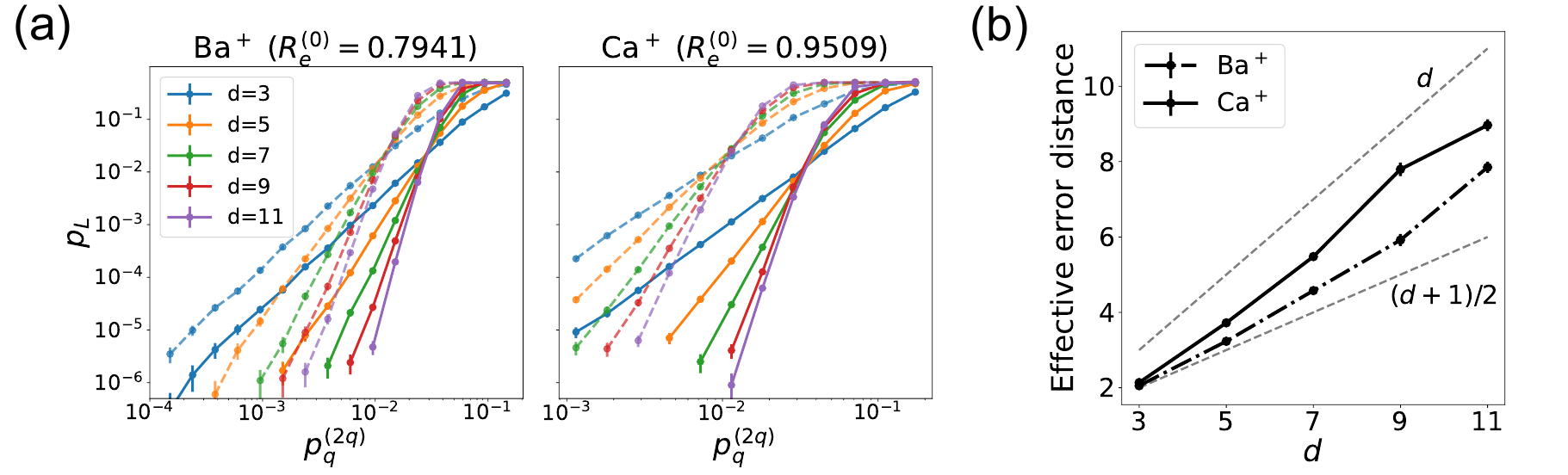}
\caption{(a) The logical error rates of ground (dashed) and metastable (solid) qubits for various code distances $d$ and two-qubit-gate error rates $p_q^{(2q)} = 2p_q-p_q^2$ ($q=g,m$), where \mbox{$p_g = p_m$} as in case \rom{1}. The error rates marked as points in Fig.~\ref{fig:fig2} are used in the simulations. The thresholds, determined by intersections of the curves for $d=5,\cdots,11$, are 1.36\%, 2.97\%, 1.22\%, and 3.42\% for ground Ba$^+$, metastable Ba$^+$, ground Ca$^+$, and metastable Ca$^+$ qubits, respectively. The error bars represent the 95\% confidence interval. (b) The effective error distances of metastable qubits~\cite{Wu22}, measured from the slope of points below the threshold in (a). The effective error distance is $(d+1)/2$ for pure Pauli errors ($R_e=0$) and $d$ for pure erasures ($R_e=1$), which are shown as dashed lines.}
\label{fig:fig4}
\end{figure*}
%%%%%%%%%%%%%%%%%%%%%%%%%%%%%%%%%%%%

In this section, we compare the logical performance of ground and metastable qubits on the surface code, under the three cases in Sec.~\ref{subsec:2D}. For each two-qubit-gate error $p^{(2q)}_q := 2p_q - p^2_q$ ($q=g,m$), the composition of various types of errors is given by Fig.~\ref{fig:fig2}. For ground (metastable) qubits, the undetected-leakage (erasure) rate takes up the largest portion of the total error rate.

First, we use the error model where the spontaneous scattering during the two-qubit gate is the only source of error. Both the undetected leakage and the erasure of rate $p$ are simulated as depolarizing error, i.e., Pauli error randomly chosen from $\{\hat{I}, \hat{X}, \hat{Y}, \hat{Z}\}$ with probability $p/4$ each. In the simulations, the only difference between undetected leakage and erasure is that the decoder knows and uses the locations of erasures but not the locations of undetected leakage. The errors during single-qubit gates, state preparation, idling, and measurement are not considered.

During a two-qubit gate, an error on one of the two qubits may propagate to the other qubit. For the circuit-level simulations, we use a detailed error model that includes the propagation of Pauli~\cite{Schwerdt22} and leakage errors (for details, see Appendix~\ref{app:D}).

Figure~\ref{fig:fig4}(a) shows the logical error rates of ground (dashed) and metastable (solid) qubits for various code distances and two-qubit-gate error rates. Here, \mbox{$p_g = p_m$} as in case \rom{1}. As expected from Ref.~\cite{Wu22}, using erasure conversion on metastable qubits significantly improves the threshold, which leads to a dramatic reduction in the logical error rates, compared to using ground qubits. The thresholds for our error model and the MWPM decoder are determined by intersections of the curves for $d=$5, 7, 9, and 11. For the Ba$^+$ ion, which has $R_e^{(0)} = 0.7941$, the threshold improves from 1.36\% to 2.97\% when metastable qubits are used. For the Ca$^+$ ion, which has $R_e^{(0)} = 0.9509$, the threshold improves from 1.22\% to 3.42\%. 

In additional to a higher threshold, erasure conversion also results in a steeper decrease of the logical error rate as the physical error rate decreases. Below the threshold, the logical error rate is proportional to $[p_q^{(2q)}]^\xi$, where $\xi$ is the \textit{effective error distance} or the slope of the logical error-rate curve. For odd $d$, $\xi = (d+1)/2$ for pure Pauli errors ($R_e=0$) and $\xi=d$ for pure erasures ($R_e=1$). Therefore, $\xi$ is expected to increase from $(d+1)/2$ to $d$ as $R_e$ increases from 0 to 1~\cite{Wu22}. Indeed, Fig.~\ref{fig:fig4}(b) shows that the effective error distance of metastable qubits is larger than $(d+1)/2$, where that of Ca$^+$ is larger than that of Ba$^+$. Meanwhile, the effective error distance of ground qubits is always $(d+1)/2$, as erasure conversion is not feasible. Thus, the advantage of metastable qubits over ground qubits grows even larger as the physical error rate decreases further from the threshold. 

While these results are promising, the disadvantages of metastable qubits are not yet reflected. The shorter lifetimes of metastable qubits are not considered. Also, Fig.~\ref{fig:fig2} shows that in order to achieve a fixed error rate (case \rom{1}), metastable qubits require larger detuning than ground qubits, which leads to a lower Rabi frequency and a longer gate time. 

%%%%%%%%%%%%%%%FIGURE 5%%%%%%%%%%%%%%
\begin{figure*}[ht]
\includegraphics[width=\linewidth]{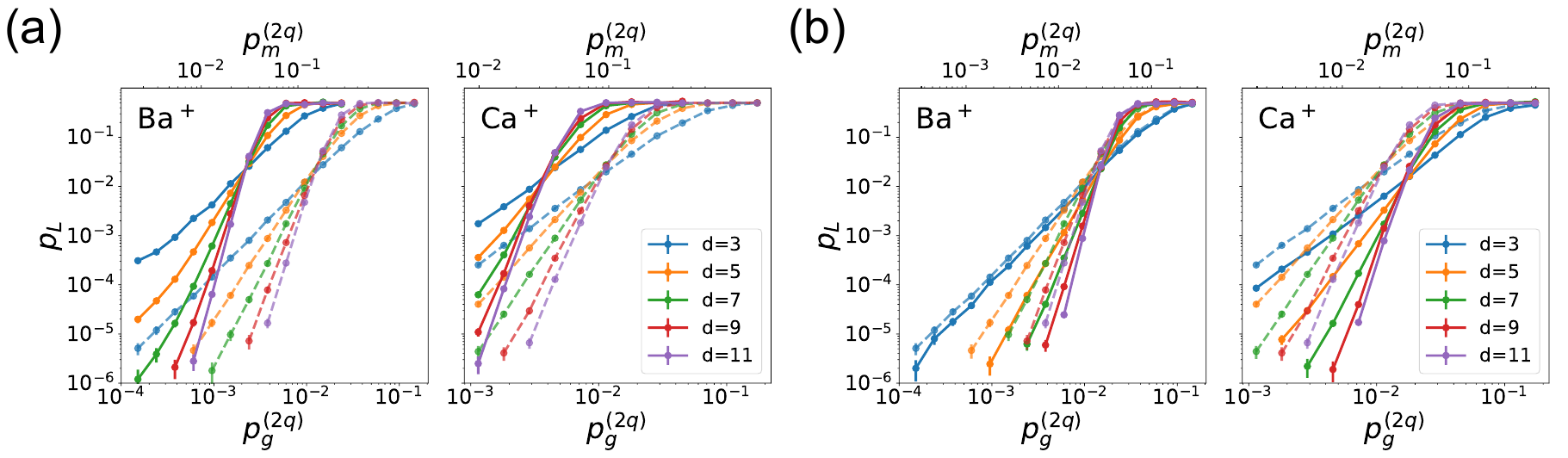}
\caption{The logical error rates of ground (dashed) and metastable (solid) qubits for various code distances $d$ and two-qubit-gate error rates. Idling (measurement) errors of fixed rate are added for metastable (both) qubits. The thresholds with respect to $p_q^{(2q)}$, determined by intersections of the curves for $d=5,\cdots,11$, are 1.36\%, 2.84\%, 1.21\%, and 3.15\% for ground Ba$^+$, metastable Ba$^+$, ground Ca$^+$, and metastable Ca$^+$ qubits, respectively. The error bars represent the 95\% confidence interval. (a) The two-qubit-gate error rates $p^{(2q)}_g$ and $p^{(2q)}_m$ of ground and metastable qubits, respectively, are aligned with respect to case \rom{2}: $\Omega_g = \Omega_m$ and $E_g=E_m$. From the left to the right of the region where the logical error rates of both the ground and metastable qubits are plotted, the ratio of detunings $\Delta_m / \Delta_g$ given by Eq.~(\ref{eq:case2}) is varied from 0.0681 to 0.152 for Ba$^+$ and from 0.111 to 0.229 for Ca$^+$. For case \rom{2}, ground qubits outperform metastable qubits. (b) The two-qubit-gate error rates $p^{(2q)}_g$ and $p^{(2q)}_m$ of the ground and metastable qubits are aligned with respect to case \rom{3}: $\Delta_g=\Delta_m$. To match the Rabi frequencies ($\Omega_g = \Omega_m$), from left to right of the plotted region, the ratio of electric field amplitudes $E_m / E_g$ given by Eq.~(\ref{eq:case3}) is varied from 3.83 to 2.56 for Ba$^+$ and from 3.00 to 2.09 for Ca$^+$. For case \rom{3}, metastable qubits outperform ground qubits. For both (a) and (b), the top horizontal axes are not exactly at log scale.}
\label{fig:fig5}
\end{figure*}
%%%%%%%%%%%%%%%%%%%%%%%%%%%%%%%%%%%%

In the following simulations, we add idling errors of metastable qubits into the error model. Specifically, for a surface-code cycle time $T$, we assume that \textit{all} metastable qubits have an additional probability of erasure $p^{\rm (idle)}(T)/4$ [see Eq.~(\ref{eq:idleerror})] before each layer of CNOT gates, where the factor of $1/4$ comes from the fact that each error-correction round consists of four layers of CNOT gates. Here, we assume \mbox{$T = 3$ ms}, considering the time required for the state-of-the-art sideband cooling~\cite{Rasmusson21}. For the Ba$^+$ (Ca$^+$) ion, the idling error rate is fixed at $p^{\rm (idle)}(T)/4 = 2.49 \times 10^{-5}$ ($6.46 \times 10^{-4}$). The QEC performance of metastable qubits for various values of the idling-error rates is discussed in Appendix~\ref{app:E}. We also add measurement errors of fixed rate $10^{-4}$, which is roughly the state-of-the-art for trapped ions~\cite{Christensen20, Ransford21}, for both ground and metastable qubits. The addition of idling and measurement errors of fixed rates reduces the thresholds only slightly, as the two-qubit-gate error dominates for the range of $p_q^{(2q)}$ simulated here.

Figure~\ref{fig:fig5}(a) shows the comparison result for \mbox{case \rom{2}}. The top and bottom horizontal axes represent the two-qubit-gate error rates of the ground and metastable qubits, respectively. For each $p^{(2q)}_g$, the vertically aligned $p^{(2q)}_m$ is determined by the ratio of detunings $\Delta_m / \Delta_g$ that satisfies the conditions $\Omega_g = \Omega_m$ and $E_g = E_m$ [see Eq.~(\ref{eq:case2})]. As a result, from the lowest to highest simulated $p^{(2q)}_m$, $\Delta_m / \Delta_g$ is varied from 0.0681 to 0.152 for the Ba$^+$ ion and from 0.111 to 0.229 for the Ca$^+$ ion. Thus, each $p^{(2q)}_m$ is compared with a $p^{(2q)}_g$ that is an order of magnitude lower. In this case, for both Ba$^+$ and Ca$^+$ ions, ground qubits outperform metastable qubits, despite having lower thresholds. The effects of idling and measurement errors are negligible in the range of two-qubit-gate errors simulated here. 

Figure~\ref{fig:fig5}(b) shows the comparison result for case \rom{3}. Here, $p^{(2q)}_g$ and $p^{(2q)}_m$ that are vertically aligned correspond to the same detuning ($\Delta_g = \Delta_m$). Thus, each $p^{(2q)}_m$ is compared with a $p^{(2q)}_g$ that is at most two times lower, which can be overcome by the improvement in the threshold due to erasure conversion. For both Ba$^+$ and Ca$^+$ ions, metastable qubits outperform ground qubits. The advantage of having a larger effective error distance shows up especially for smaller gate-error rates and larger code distances. The advantage of metastable qubits is greater for Ca$^+$, due to the larger erasure-conversion rate. Again, the effects of idling and measurement errors are negligible. 

For case \rom{3}, to achieve both $\Omega_g = \Omega_m$ and $\Delta_g = \Delta_m$, from the lowest to the highest simulated error rates, $E_m/E_g$ is varied from 3.83 to 2.56 for the Ba$^+$ ion and from 3.00 to 2.09 for the Ca$^+$ ion [see Eq.~(\ref{eq:case3})]. Therefore, in order to achieve the advantage shown in Fig.~\ref{fig:fig5}(b), metastable qubits require higher laser power than ground qubits (for typical values of the laser power, see Appendix~\ref{app:C}). We emphasize that it is reasonable to assume that metastable qubits allow higher laser power for gate operations, as material loss due to lasers is less severe for a longer laser wavelength. In reality, the advantage of metastable qubits may depend on the achievable laser powers. 

%%%%%%%%%%%%%%%%%%%%%%%%%
%%%%%%%SECTION 6%%%%%%%%%
%%%%%%%%%%%%%%%%%%%%%%%%%

\section{Discussion and outlook} \label{sec:sec6}

\subsection{Comparison with the Rydberg-atom platform}

As our erasure-conversion scheme on metastable trapped-ion qubits is motivated by that on metastable Rydberg-atom qubits~\cite{Wu22}, in this section we provide a discussion on comparing the two platforms, as well as suggesting the uniqueness of erasure conversion on trapped-ion qubits.  

The two-qubit gates on Rydberg-atom qubits are performed by coupling the qubit state $\ket{1}$ to the Rydberg state $\ket{r}$ with Rabi frequency $\Omega$. The van der Waals interaction between the two atoms prevents them from being simultaneously excited to $\ket{r}$, a phenomenon known as the Rydberg blockade. This acquires a phase on the two-qubit state $\ket{11}$ that is different from the phase of $\ket{01}$ and $\ket{10}$, which, if carefully controlled, leads to a fully entangling two-qubit gate~\cite{Jaksch00, Lukin01}.

The fundamental sources of two-qubit-gate errors are the spontaneous decay of the Rydberg states and the finite Rydberg-blockade strength~\cite{Zhang12, Graham19}. It has been argued that the effects of the finite Rydberg-blockade strength can be compensated by tuning the laser parameters~\cite{Levine19}. Thus, we focus only on the spontaneous decay of the Rydberg states. This is similar to the fact that the spontaneous decay of the excited $P$ states is the fundamental source of errors for laser-based gates on trapped ions. For both platforms, the state-of-the-art gates are dominated by other technical sources of errors in the experimental system~\cite{Wang20, Cetina22, Kang23ff, Graham19, Levine19}. 

The crucial difference between the two platforms is that for trapped ions, the $P$ states are only virtually occupied as the lasers are far detuned from the transition to the $P$ states, while for Rydberg atoms, the Rydberg states need to be occupied for a sufficiently long time in order to acquire the entangling phases. Thus, for Rydberg atoms, the gate-error rate due to the spontaneous decay is reduced by minimizing the time that the atoms spend in the Rydberg states, denoted as $t_R$. 

The state-of-the-art method directly minimizes the error rate by finding the $t_R$-optimal pulse of length \mbox{$t_{\rm gate} > t_R$} using quantum optimal control techniques~\cite{Jandura22Time}. The minimal two-qubit-gate error rate is given by
\begin{equation} \label{eq:Rydberg}
    p^{(2q)} = \gamma_R t_R = 2.947 \times \frac{\gamma_R}{\Omega_{\rm max}},
\end{equation}
where $\gamma_R$ is the lifetime of the Rydberg states and $\Omega_{\rm max}$ is the maximal Rabi frequency of the time-optimal pulse~\cite{Jandura22Time}. Using \mbox{$1/\gamma_R = 540$ $\mu$s} and \mbox{$\Omega_{\rm max} = 10$ MHz} as suggested by Ref.~\cite{Jandura22Time}, the error rate is $8.7 \times 10^{-5}$, which is on par with the minimum two-qubit-gate error rate of ground Ba$^+$ ion qubits ($1.5 \times 10^{-4}$) over $\Delta_g \in [0, \omega_F]$. 

We note that with limited laser power, the lower bound of the two-qubit-gate error rate of the Rydberg atoms is determined by the maximum achievable $\Omega_{\rm max}$. For trapped ions, the error rate of ground (metastable) qubits can in principle be reduced to an arbitrarily small value by increasing the detuning from the $P_{1/2}$ ($P_{3/2}$) manifold~\cite{Moore23}, at the cost of a lower Rabi frequency.

In Eq.~(\ref{eq:Rydberg}), the gate-error rate decreases as the Rabi frequency increases. Meanwhile, for trapped ions, the gate-error rate due to the spontaneous scattering does not explicitly depend on the Rabi frequency (see Appendix~\ref{app:B}). Instead, as the detuning $\Delta_q$ decreases, both the Rabi frequency and the gate-error rate increase. This trade-off is central to cases \rom{2} and \rom{3} of our comparison between ground and metastable trapped-ion qubits. 

Also, as noted in Sec.~\ref{subsec:2D}, in the presence of technical errors that increase with the gate time, decreasing the detuning $\Delta_m$ of metastable qubits converts the technical errors into erasures (and a small amount of undetected leakage). This is a method of erasure conversion that is unique to the trapped-ion platform. Exploiting this trade-off for minimizing the \textit{logical} error rate is an opportunity for metastable trapped-ion qubits. 

For both platforms, it is desirable to increase the Rabi frequency by using higher laser power. This reduces the technical errors for trapped ions and both the technical and the fundamental errors for Rydberg atoms. For trapped ions, the erasure conversion of metastable qubits is performed at the cost of a higher laser-power requirement for achieving a fixed Rabi frequency, as the inherent matrix element of the transition from the metastable states to the $P$ states is weaker than from the ground states. For Rydberg alkaline-earth-like atoms, the trade-off is more complicated, as the Rydberg excitation from the ground qubit requires two-photon transitions via $^3P_1$~\cite{Ma22} but only a single photon from the metastable $^3P_0$ qubit state~\cite{Wu22}. The complication is that if we include the time of state preparation and then Rydberg excitation, the metastable qubit will require more laser power to create the Rydberg excitation in the same amount of time, but in the active processing of quantum information the state-preparation stage is rare, so the metastable qubit will use $^1S \rightarrow {^3P}$ laser less often and therefore less laser power overall. A more careful comparison of the laser-power requirements between ground and metastable Rydberg-atom qubits is beyond the scope of this paper.

\subsection{Outlook}

While the simulation results in this work may serve as a proof of principle, they have several limitations that lead to future directions. First, undetected-leakage errors are simulated as depolarizing errors. In reality, the fault-tolerant handling of leakage errors requires significant overhead, such as leakage-reducing operations~\cite{Wu02, Byrd04, Byrd05} and circuits~\cite{Fowler13, Ghosh15, Suchara15, Brown18, Brown19, Brown20}. In particular, the trade offs of using leakage-reducing circuits on ground trapped-ion qubits have been discussed in Refs.~\cite{Brown18, Brown19}. After implementing such overhead, the effects of undetected leakage errors are equivalent to those of depolarizing errors of a larger rate, in terms of the threshold and error distances~\cite{Brown18, Brown19}; thus, the effects of undetected leakage in our simulations may be considered as a lower bound of the actual effects. When the cost of handling leakage is considered, we expect the advantage of using metastable qubits to be significantly greater, as undetected leakage is the dominant type of error for ground qubits but not for metastable qubits equipped with erasure conversion.

Second, we do not consider errors due to miscalibration of the physical parameters, which may cause overrotation errors. This is important because (i) overrotation error is often larger than stochastic error for state-of-the-art two-qubit gates~\cite{Cetina22, Kang23ff} and (ii) calibrating physical parameters to high precision may require a large number of shots and a long experiment time~\cite{Kang23mode}. 

Notably, overrotation errors can be converted to erasures using certified quantum gates on metastable qubits~\cite{Campbell20}. In this gate scheme, auxiliary states outside the qubit subspace are used, such that overrotation causes residual occupation of the auxiliary states, which can be detected by optical pumping. While Ref.~\cite{Campbell20} considers a heralded gate of success probability smaller than one, in the perspective of QEC, this is essentially erasure conversion. More work needs to be done on generalizing certified quantum gates to other encodings of qubit states and to more commonly used two-qubit-gate schemes such as the MS scheme. 

Finally, we only consider the examples of $D_{5/2}$ metastable qubits, which have a limited lifetime (see Table~\ref{tab:coeffs}). Meanwhile, the metastable qubit of the Yb$^+$ ion encoded in the $F_{7/2}^o$ manifold is even more promising, as its lifetime ranges from several days to years~\cite{Allcock21}. Laser-based gate operations on such a metastable qubit may use excited states that are more ``exotic'' than the $P$ states. We hope that this work motivates future research on utilizing these exotic states, as well as measuring their properties, such as their decay rates and branching fractions.

\section{Conclusion} \label{sec:sec7}

In this paper, we show that trapped-ion metastable qubits equipped with erasure conversion can outperform ground qubits in fault-tolerant logical memory using surface codes. Even when the Rabi frequency is fixed between ground and metastable qubits, the logical error rates of the metastable qubits are lower when reasonably higher laser power is allowed (such that detuning from the $P_{3/2}$ manifold is the same). We hope that this paper motivates further research in using metastable trapped-ion qubits for scalable and fault-tolerant quantum computing~\cite{Allcock21}. Also, the methodology of using our detailed error model for comparing ground and metastable qubits may be applied back to the Rydberg-atom platform~\cite{Wu22}.

\begin{acknowledgments}
M.K. thanks Craig Gidney for his advice on using STIM and thanks Shilin Huang for helpful discussions.  W.C.C. acknowledges support from NSF grant no.\ PHY-2207985 and ARO grant no.\ W911NF-20-1-0037. M.K. and K.R.B acknowledge support from the Office of the Director of National Intelligence, Intelligence Advanced Research Projects Activity through ARO grant no. W911NF-16-1-0082 and the NSF-sponsored Quantum Leap Challenge Institute for Robust Quantum Simulation grant no.\ OMA-2120757.
\end{acknowledgments}

\appendix

%%%%%%%%%%%%%%%%%%APPENDIX A%%%%%%%%%%%%%%%%%%%%
\section{Derivations of $k_q$ and $\alpha_f$} \label{app:A}

In this appendix we calculate the coefficients $k_q$ and $\alpha_f$, introduced in (\ref{eq:mu}), (\ref{eq:gamma}), and (\ref{eq:alphaf}), using the Wigner $3j$ and $6j$ symbols. We mainly use two relations. First is the relation between the hyperfine transition element and the fine-structure-level transition element, given by~\cite{Zare88}
\begin{widetext}
\begin{align}
    \langle L_a, J_a ; F_a, M_a | T^{(1)}_\lambda(\vec{d}) | L_b J_b ; F_b, M_b \rangle 
    &= (-1)^{F_a - M_a + J_a + I + F_b + 1} \sqrt{(2F_a+1)(2F_b+1)} 
    \begin{pmatrix} F_a & 1 & F_b \\ -M_a & s & M_b \end{pmatrix} \nonumber \\
    &\quad \times \begin{Bmatrix} J_a & F_a & I \\ F_b & J_b & 1 \end{Bmatrix} 
    \langle L_a, J_a || T^{(1)}(\vec{d}) || L_b, J_b \rangle, \label{eq:ME1}
\end{align}
where $\lambda$ is the photon polarization. Next is the relation between the fine-structure-level transition element and the orbital transition element, given by~\cite{Zare88}
\begin{equation}
    \langle L_a, J_a || T^{(1)}(\vec{d}) || L_b, J_b \rangle = 
    (-1)^{(L_a + \frac{1}{2} + J_b + 1)} \sqrt{(2J_a+1)(2J_b+1)}
    \begin{Bmatrix} L_a & J_a & S \\ J_b & L_b & 1 \end{Bmatrix}
    \langle L_a || T^{(1)}(\vec{d}) || L_b \rangle, \label{eq:ME2}
\end{equation}
where $S = 1/2$ is the electron spin. 

We first calculate $k_q$. The largest dipole-matrix element of transition between a state in manifold $q$ and a $P$ state is given by
\begin{equation}
    \mu_q := \langle 1, \frac{3}{2} ; I + \frac{3}{2}, I + \frac{3}{2} | T^{(1)}_{\frac{3}{2} - J_q}(\vec{d}) | L_q, J_q ; I + J_q, I + J_q \rangle,
\end{equation}
where $I$ is the nuclear spin, which we assume to be a half integer. Applying (\ref{eq:ME1}) and (\ref{eq:ME2}) sequentially, we obtain
\begin{align}
    \mu_q &= (-1)^{J_q + 2I + \frac{1}{2}} \sqrt{(2I + 4) (2I + 2J_q + 1)}
    \begin{pmatrix} I + \frac{3}{2} & 1 & I + J_q \\ -I - \frac{3}{2} & \frac{3}{2} - J_q & I + J_q \end{pmatrix} \nonumber \\
    &\quad \times \begin{Bmatrix} \frac{3}{2} & I + \frac{3}{2} & I \\ I + J_q & J_q & 1 \end{Bmatrix}
    \langle L=1, J=\frac{3}{2} || T^{(1)}(\vec{d}) || L_q, J_q \rangle \nonumber \\
    &= (-1)^{2J_q + 2I + 1}
    \sqrt{(2I + 4) (2I + 2J_q + 1)(4)(2J_q+1)} \nonumber \\
    &\quad \times 
    \begin{pmatrix} I + \frac{3}{2} & 1 & I + J_q \\ -I - \frac{3}{2} & \frac{3}{2} - J_q & I + J_q \end{pmatrix}
    \begin{Bmatrix} \frac{3}{2} & I + \frac{3}{2} & I \\ I + J_q & J_q & 1 \end{Bmatrix}
    \begin{Bmatrix} 1 & \frac{3}{2} & \frac{1}{2} \\ J_q & L_q & 1 \end{Bmatrix}
    \langle L=1 || T^{(1)}(\vec{d}) || L_q \rangle. \label{eq:muq_app}
\end{align}

For ground qubits ($q=g$), inserting $L_q=0$ and $J_q=\frac{1}{2}$ to (\ref{eq:muq_app}) gives
\begin{align}
    \mu_g &= (-1)^{2I} \sqrt{(2I+4)(2I+2)(4)(2)} \nonumber \\
     &\quad \times \begin{pmatrix} I + \frac{3}{2} & 1 & I + \frac{1}{2} \\ -I - \frac{3}{2} & 1 & I + \frac{1}{2} \end{pmatrix}
     \begin{Bmatrix} \frac{3}{2} & I + \frac{3}{2} & I \\ I + \frac{1}{2} & \frac{1}{2} & 1 \end{Bmatrix}
     \begin{Bmatrix} 1 & \frac{3}{2} & \frac{1}{2} \\ \frac{1}{2} & 0 & 1 \end{Bmatrix}
    \langle L=1 || T^{(1)}(\vec{d}) || L_g=0 \rangle \nonumber \\
    &= (-1)^{2I} \sqrt{(2I+4)(2I+2)(4)(2)} \times \frac{1}{\sqrt{2I+4}}
     \frac{(-1)^{2I+1}}{\sqrt{(2I+2)(4)}}
     \frac{(-1)}{\sqrt{6}}
    \langle L=1 || T^{(1)}(\vec{d}) || L_g=0 \rangle \nonumber \\
    &= \frac{1}{\sqrt{3}} \langle L=1 || T^{(1)}(\vec{d}) || L_g=0 \rangle.
\end{align}
Therefore, from (\ref{eq:mu}), we have $k_g = 1/3$. 

Similarly, for metastable qubits ($q = m$), inserting $L_q=2$ and $J_q=5/2$ to (\ref{eq:muq_app}) gives
\begin{align}
    \mu_m &= (-1)^{2I} \sqrt{(2I+4)(2I+6)(4)(6)} \nonumber \\
     &\quad \times \begin{pmatrix} I + \frac{3}{2} & 1 & I + \frac{5}{2} \\ -I - \frac{3}{2} & -1 & I + \frac{5}{2} \end{pmatrix}
     \begin{Bmatrix} \frac{3}{2} & I + \frac{3}{2} & I \\ I + \frac{5}{2} & \frac{5}{2} & 1 \end{Bmatrix}
     \begin{Bmatrix} 1 & \frac{3}{2} & \frac{1}{2} \\ \frac{5}{2} & 2 & 1 \end{Bmatrix}
    \langle L=1 || T^{(1)}(\vec{d}) || L_m=2 \rangle \nonumber \\
    &= (-1)^{2I} \sqrt{(2I+4)(2I+6)(4)(6)} \times \frac{1}{\sqrt{2I+6}}
     \frac{(-1)^{2I+1}}{\sqrt{(2I+4)(6)}}
     \frac{(-1)}{\sqrt{20}}
    \langle L=1 || T^{(1)}(\vec{d}) || L_m=2 \rangle \nonumber \\
    &= \frac{1}{\sqrt{5}} \langle L=1 || T^{(1)}(\vec{d}) || L_m=2 \rangle.
\end{align}
From (\ref{eq:mu}), we have $k_m = 1/5$. 

Now we calculate $\alpha_f$. Before we start, we introduce two useful identities for Wigner $3j$ and $6j$ symbols:
\begin{gather}
    \sum_{m_1, m_2} \begin{pmatrix} j_1 & j_2 & j_3 \\ m_1 & m_2 & m_3 \end{pmatrix}
     \begin{pmatrix} j_1 & j_2 & j'_3 \\ m_1 & m_2 & m'_3 \end{pmatrix}
     = \frac{\delta_{j_3, j'_3} \delta_{m_3, m'_3}}{2j_3+1}, \label{eq:sumrule1} \\
     \sum_{j_3} (2j_3+1) \begin{Bmatrix} j_1 & j_2 & j_3 \\ m_1 & m_2 & m_3 \end{Bmatrix}
     \begin{Bmatrix} j_1 & j_2 & j_3 \\ m_1 & m_2 & m'_3 \end{Bmatrix}
     = \frac{\delta_{m_3, m'_3}}{2m_3+1}. \label{eq:sumrule2}
\end{gather}
We start with the left-hand side of the first equation of (\ref{eq:gamma}) and apply the Wigner-Eckart theorem, which leads to
\begin{align}
    \sum_{F_f, M_f} |\langle L_e, J_e; F_e, M_e | \vec{d} | L_f, J_f ; F_f, M_f \rangle |^2
    &= \sum_{F_f} \left| \langle L_e, J_e; F_e || T^{(1)}(\vec{d}) || L_f, J_f ; F_f \rangle \right|^2
    \sum_{M_f} \left| \begin{pmatrix} F_f & 1 & F_e \\ -M_f & M_f - M_e & M_e \end{pmatrix} \right|^2 \nonumber \\
    &= \sum_{F_f} \left| \langle L_e, J_e; F_e || T^{(1)}(\vec{d}) || L_f, J_f ; F_f \rangle \right|^2
    \sum_{M_f, s} \left| \begin{pmatrix} F_f & 1 & F_e \\ -M_f & s & M_e \end{pmatrix} \right|^2 \nonumber \\
    &= \frac{1}{2F_e+1} \sum_{F_f} \left| \langle L_e, J_e; F_e || T^{(1)}(\vec{d}) || L_f, J_f ; F_f \rangle \right|^2,
\end{align}
where the last equality uses (\ref{eq:sumrule1}). Applying (\ref{eq:ME2}) with $J \rightarrow F$, $L \rightarrow J$, and $S \rightarrow I$ gives
\begin{align}
    \sum_{F_f, M_f} |\langle L_e, J_e; F_e, M_e | \vec{d} | L_f, J_f ; F_f, M_f \rangle |^2
    &= \frac{1}{2F_e+1}
    \sum_{F_f} (2F_e+1)(2F_f+1) 
    \left| \begin{Bmatrix} J_e & F_e & I \\ F_f & J_f & 1 \end{Bmatrix} \right|^2
    \left| \langle L_e, J_e || T^{(1)}(\vec{d}) || L_f, J_f \rangle \right|^2 \nonumber \\
    &=  \left| \langle L_e, J_e || T^{(1)}(\vec{d}) || L_f, J_f \rangle \right|^2
    \sum_{F_f} (2F_f+1) \left| \begin{Bmatrix} J_e & F_e & I \\ F_f & J_f & 1 \end{Bmatrix} \right|^2 \nonumber\\
    &= \frac{1}{2J_e+1} \left| \langle L_e, J_e || T^{(1)}(\vec{d}) || L_f, J_f \rangle \right|^2,
\end{align}
where the last equality uses (\ref{eq:sumrule2}). Applying (\ref{eq:ME2}) gives
\begin{align}
    \sum_{F_f, M_f} |\langle L_e, J_e; F_e, M_e | \vec{d} | L_f, J_f ; F_f, M_f \rangle |^2
    &= (2J_f+1) 
    \left|\begin{Bmatrix} L_e & J_e & \frac{1}{2} \\ J_f & L_f & 1 \end{Bmatrix} \right|^2
    \left| \langle L_e || T^{(1)}(\vec{d}) || L_f \rangle \right|^2 \nonumber\\
\end{align}
Therefore, from (\ref{eq:gamma}), we obtain
\begin{equation}
\alpha_{e,f} = (2J_f+1) \left|\begin{Bmatrix} L_e & J_e & \frac{1}{2} \\ J_f & L_f & 1 \end{Bmatrix} \right|^2.    
\end{equation}
Finally, inserting $L_e = 1$ and $J_e = 3/2$ gives the values of $\alpha_f$ shown in Table~\ref{tab:ka}. 

\end{widetext}

%%%%%%%%%%%%%%%%%%APPENDIX B%%%%%%%%%%%%%%%%%%%%
\section{Derivations of two-qubit-gate error rates} \label{app:B}

In this appendix we derive the various types of two-qubit-gate error rates in Fig.~\ref{fig:fig2}. Specifically, we show how the proportionality constants $C_{i,j,J}$ in (\ref{eq:Gammaij}) are calculated, which leads to equations for the error rates and the erasure-conversion rate. 

Generalizing the notation of Ref.~\cite{Uys10}, we define the scattering amplitude from $\ket{i}$ in manifold $q$ to $\ket{j}$ in manifold $f$ via excited state $\ket{J}$ in manifold $e$ as~\cite{Cohen98}
\begin{equation}
    A^{i \rightarrow j}_{J,\lambda} = -\frac{b_\lambda \langle j | \vec{d} | J \rangle \langle J | \vec{d} | i \rangle}{\Delta_{e,q} \mu_f \mu_q},
\end{equation}
where $b_\lambda$ is the normalized amplitude of the polarization component $\hat{\epsilon}_\lambda$ of the Raman beam ($\lambda = 1, 0, -1$). Note that $A^{i \rightarrow j}_{J,\lambda} \neq 0$ only if the magnetic quantum number of $\ket{J}$ is equal to that of $\ket{i}$ plus $\lambda$. 

As explained in Sec.~\ref{subsec:2B}, phase-flip errors are caused by the effective Rayleigh scattering, and bit-flip and leakage errors are caused by the Raman scattering. The scattering rates of various types can be calculated using the Kramers-Heisenberg formula. First, the effective Rayleigh scattering rate during the qubit's Raman transition is given by~\cite{Uys10}
\begin{equation}\label{eq:GammaZ}
    \Gamma^{(z)} = k_q g_q^2 \frac{\gamma'_q}{\alpha_q} \sum_\lambda \left(\sum_J A^{1 \rightarrow 1}_{J,\lambda} - \sum_{J'} A^{0 \rightarrow 0}_{J',\lambda} \right)^2,
\end{equation}
where 0 and 1 in the superscript denote the qubit states. Next, the rate of Raman scattering that leads to bit-flip errors is given by \cite{Uys10, Ozeri07}
\begin{equation}\label{eq:GammaXY}
    \Gamma^{(xy)} = k_q g_q^2 \frac{\gamma'_q}{\alpha_q} \sum_\lambda \left[ 
    \bigg(\sum_J A^{0 \rightarrow 1}_{J,\lambda} \bigg)^2 + \bigg( \sum_{J'} A^{1 \rightarrow 0}_{J',\lambda} \bigg)^2
    \right], 
\end{equation} 
and the rate of Raman scattering that leads to leakage errors is given by
\begin{equation}\label{eq:GammaL}
    \Gamma^{(l)} = k_q g_q^2 \sum_{j \neq 0,1} \frac{\gamma'_f}{\alpha_f} \sum_\lambda \left[ 
    \bigg(\sum_J A^{0 \rightarrow j}_{J,\lambda} \bigg)^2 + \bigg( \sum_{J'} A^{1 \rightarrow j}_{J',\lambda} \bigg)^2 \right],
\end{equation}
where $j \neq 0,1$ denotes that $\ket{j}$ is not a qubit state. Finally, for $D_{5/2}$ metastable qubits with erasure conversion, the rate of Raman scattering that leads to erasures is given by
\begin{equation}\label{eq:GammaE}
    \Gamma^{(e)} = k_q g_q^2 \sum_{j \notin D_{5/2}} \frac{\gamma'_f}{\alpha_f} \sum_\lambda \left[ 
    \bigg(\sum_J A^{0 \rightarrow j}_{J,\lambda} \bigg)^2 + \bigg( \sum_{J'} A^{1 \rightarrow j}_{J',\lambda} \bigg)^2 \right],
\end{equation}
where $j \notin D_{5/2}$ denotes that $\ket{j}$ is not in the $D_{5/2}$ manifold. 

Now we perform the sums in (\ref{eq:GammaZ})-(\ref{eq:GammaE}). For notational convenience, we define the detuning-dependent branching fractions~\cite{Moore23}
\begin{align} \label{eq:rprime}
    r'_1 := (1 - \Delta_q / \omega_{P_{3/2}, S_{1/2}})^3 \times r_1, \nonumber \\
    r'_2 := (1 - \Delta_q / \omega_{P_{3/2}, D_{3/2}})^3 \times r_2, \\
    r'_3 := (1 - \Delta_q / \omega_{P_{3/2}, D_{5/2}})^3 \times r_3, \nonumber
\end{align}
such that $\gamma'_f = r'_i \gamma$ for the corresponding $i$ for each manifold $f$ [see (\ref{eq:gammaprimef})]. Also, we assume that the Raman beams are linearly polarized and mutually perpendicular. Then for ground qubits, the scattering rates are given by
\begin{align}
    \Gamma^{(z)}_g &= 0, \label{eq:GammaZg} \\
    \Gamma^{(xy)}_g &= \frac{2}{9} r'_1 \gamma g_g^2 \frac{\omega_F^2}{(\omega_F - \Delta_g)^2 \Delta_g^2}, \label{eq:GammaXYg} \\
    \Gamma^{(l)}_g &= \frac{2}{9} \gamma \big(\frac{g_g}{\Delta_g}\big)^2 \Bigg[
    r'_1 \frac{\omega_F^2}{(\omega_F - \Delta_g)^2} \nonumber \\
    &\quad\quad\quad\quad + 6r'_2 \frac{\omega_F^2  - 2\omega_F\Delta_g  + 6 \Delta_g^2}{(\omega_F - \Delta_g)^2}
    + 6r'_3 \Bigg], \label{eq:GammaLg}
\end{align}
where we replace the subscripts $q$ with $g$. Note that (\ref{eq:GammaZg})-(\ref{eq:GammaLg}) are valid for any value of $I$. Similarly, for metastable qubits,
\begin{align}
&\Gamma^{(z)}_m = c_z r'_3 \gamma \big(\frac{g_m}{\Delta_m} \big)^2, \label{eq:GammaZm}\\
&\Gamma^{(xy)}_m = c_{xy} r'_3 \gamma \big(\frac{g_m}{\Delta_m} \big)^2, \label{eq:GammaXYm}\\
&\Gamma^{(l)}_m = (c_1 r'_1 + c_2 r'_2 + c_l r'_3) \gamma \big(\frac{g_m}{\Delta_m} \big)^2, \label{eq:GammaLm}\\
&\Gamma^{(e)}_m = (c_1 'r_1 + c_2 r'_2) \gamma \big(\frac{g_m}{\Delta_m} \big)^2,\label{eq:GammaEm}
\end{align}
where we replace the subscripts $q$ with $m$. Here, $c_z$, $c_{xy}$, $c_1$, $c_2$, and $c_l$ are geometric coefficients determined by $I$ and $F_0$. Note that $c_{xy}$ ($c_l$) comes from the Raman-scattering rates to the $D_{5/2}$ states within (outside) the qubit subspace. Also, $c_1$ ($c_2$) comes from the Raman-scattering rates to the $S_{1/2}$ ($D_{3/2}$) states. The values of these coefficients for the metastable Ba$^+$ and Ca$^+$ qubits considered in this paper can be found in Table \ref{tab:coeffs2}.

\begin{table}
\caption{\label{tab:coeffs2}
Values of the geometric coefficients used in (\ref{eq:GammaZm})-(\ref{eq:GammaEm}) for two metastable qubits chosen as examples. 
}
\begin{ruledtabular}
\begin{tabular}{cccccccccccccc}
 Metastable qubit & $c_z$ & $c_{xy}$ & $c_1$ & $c_2$ & $c_l$ \\ 
\hline
$^{133} {\rm Ba}^+$, $F_0=2$ & 0 & 1/75 & 2/5 & 2/5 & 1/3\\
$^{43} {\rm Ca}^+$, $F_0=5$ & 0.0035 & 7/165 & 29/55 & 29/55 & 0.3904\\
\end{tabular}
\end{ruledtabular}
\end{table}

We note in passing that the approximations used to derive (\ref{eq:GammaZg}) - (\ref{eq:GammaEm}) are valid when $|\Delta_g|$ and $|\Delta_m|$ are much larger than the hyperfine-splitting frequencies and much smaller than the laser frequency. For extremely far-detuned illumination, other effects, such as coupling to higher excited states, may need to be taken into account (see \textit{e.g.} Ref.~\cite{Moore23} for a more thorough treatment).

The error rates of each qubit on which two-qubit gate is applied follow straightforwardly by multiplying the gate time $t_{\rm gate}$ in (\ref{eq:gatetime}) to the scattering rates. Note that as $t_{\rm gate}$ is inversely proportional to $\Omega$ in (\ref{eq:Rabiground}) and (\ref{eq:Rabimeta}), the $g_q^2$ factor is cancelled out. The error rates of each ground qubit are given by 
\begin{align}
    &p^{(z)}_g = 0, \\
    &p^{(xy)}_g = \frac{\pi \sqrt{K}}{3 \eta} r'_1 \gamma
    \left|\frac{\omega_F}{(\omega_F - \Delta_g) \Delta_g}  \right|, \\
    &p^{(l)}_g = \frac{\pi \sqrt{K}}{3 \eta} \frac{\gamma}{|\Delta_g|} \Bigg( 
    r'_1 \left|\frac{\omega_F}{\omega_F - \Delta_g}  \right| \nonumber\\
    &\quad\quad + 6r'_2 \left|\frac{\omega_F^2  - 2\omega_F\Delta_g  + 6 \Delta_g^2}{(\omega_F - \Delta_g) \omega_F} \right| 
    + 6r'_3 \left| \frac{\omega_F - \Delta_g}{\omega_F} \right| \Bigg),\\
    &p_g = p^{(z)}_g + p^{(xy)}_g + p^{(l)}_g.
\end{align}
Similarly, the error rates of each metastable qubit are given by 
\begin{align}
&p^{(z)}_m = \frac{\pi \sqrt{K}}{2 \eta} \frac{c_z r'_3}{c_0} \frac{\gamma}{|\Delta_m|}, \label{eq:pmz}\\
&p^{(xy)}_m = \frac{\pi \sqrt{K}}{2 \eta} \frac{c_{xy} r'_3}{c_0} \frac{\gamma}{|\Delta_m|}, \\
&p^{(l)}_m = \frac{\pi \sqrt{K}}{2 \eta} \frac{c_1 r'_1 + c_2 r'_2 + c_l r'_3}{c_0} \frac{\gamma}{|\Delta_m|},\\
&p^{(e)}_m = \frac{\pi \sqrt{K}}{2 \eta} \frac{c_1 r'_1 + c_2 r'_2}{c_0} \frac{\gamma}{|\Delta_m|},\\
&p_m = p^{(z)}_m + p^{(xy)}_m + p^{(l)}_m. \label{eq:pm}
\end{align}

For metastable qubits, We define $c_3 := c_z + c_{xy} + c_l$. Then, the erasure-conversion rate defined in (\ref{eq:Re}) is given by
\begin{equation}\label{eq:Re2}
R_e := \frac{p^{(e)}_m}{p_m} = \frac{c_1 r'_1 + c_2 r'_2}{c_1 r'_1 + c_2 r'_2 + c_3 r'_3},
\end{equation}
and the zero-detuning erasure-conversion rate defined in (\ref{eq:Re0}) is given by
\begin{equation}\label{eq:Re02}
    R_e^{(0)} := \lim_{\frac{\Delta_m}{\omega_F} \rightarrow 0} R_e =  \frac{c_1 r_1 + c_2 r_2}{c_1 r_1 + c_2 r_2 + c_3 r_3}.
\end{equation}
As explained in Sec.~\ref{subsec:2C}, $R_e$ is slightly larger than $R_e^{(0)}$ for nonzero $\Delta_m$, and $R_e^{(0)}$ is slightly larger than $r_1 + r_2$. This completes the derivation of the error rates and the erasure-conversion rate used in the main text.

%%%%%%%%%%%%%%%%%%APPENDIX C%%%%%%%%%%%%%%%%%%%%
\section{Laser power estimation} \label{app:C}

In this appendix we provide estimates of the typical laser power for two-qubit gates on both ground and metastable qubits. Following Ref.~\cite{Ozeri07}, the laser power required for achieving a given qubit-state Rabi frequency can be calculated. 

Assuming that the laser beams are Gaussian, the electric-field amplitude $E_q$ at the center of the beam is related to the power $P_q$ of each Raman beam by
\begin{equation}
    E_q^2 = \frac{4P_q}{\pi c \epsilon_0 w_0^2}
\end{equation}
where $c$ is the speed of light, $\epsilon_0$ is the vacuum permittivity, and $w_0$ is the beam waist at the ion's position~\cite{Ozeri07}. Also, from (\ref{eq:g})-(\ref{eq:gamma}), we have the relation
\begin{equation}
    \frac{g_q^2}{\gamma_q} = \frac{3k_q \pi \epsilon_0 c^3 E_q^2}{4 \alpha_q \hbar \omega_{e,q}^3} = \frac{3k_q c^2 P_q}{\alpha_q \hbar \omega_{e,q}^3 w_0^2}.
\end{equation}
Then, as the qubit-state Rabi frequency is proportional to $g_q^2$, the relation between the laser beam power and the Rabi frequency can be found. For ground qubits ($q=g$), from (\ref{eq:Rabiground}) and $\gamma_g = r_1 \gamma$,
\begin{equation}
    P_g = \frac{\alpha_g \hbar \omega_{e,g}^3 w_0^2}{k_g c^2 r_1 \gamma \omega_F}  |\Delta_g (\omega_F - \Delta_g)| \Omega_g.
\end{equation}
Similarly, for metastable qubits ($q=m$), from (\ref{eq:Rabimeta}) and $\gamma_m = r_3 \gamma$,
\begin{equation}
    P_m = \frac{\alpha_m \hbar \omega_{e,m}^3 w_0^2}{3 c_0 k_m c^2 r_3 \gamma} |\Delta_m| \Omega_m.
\end{equation} 

%%%%%%%%%%%%%%%FIGURE power%%%%%%%%%%
\begin{figure}[ht]
\includegraphics[width=\linewidth]{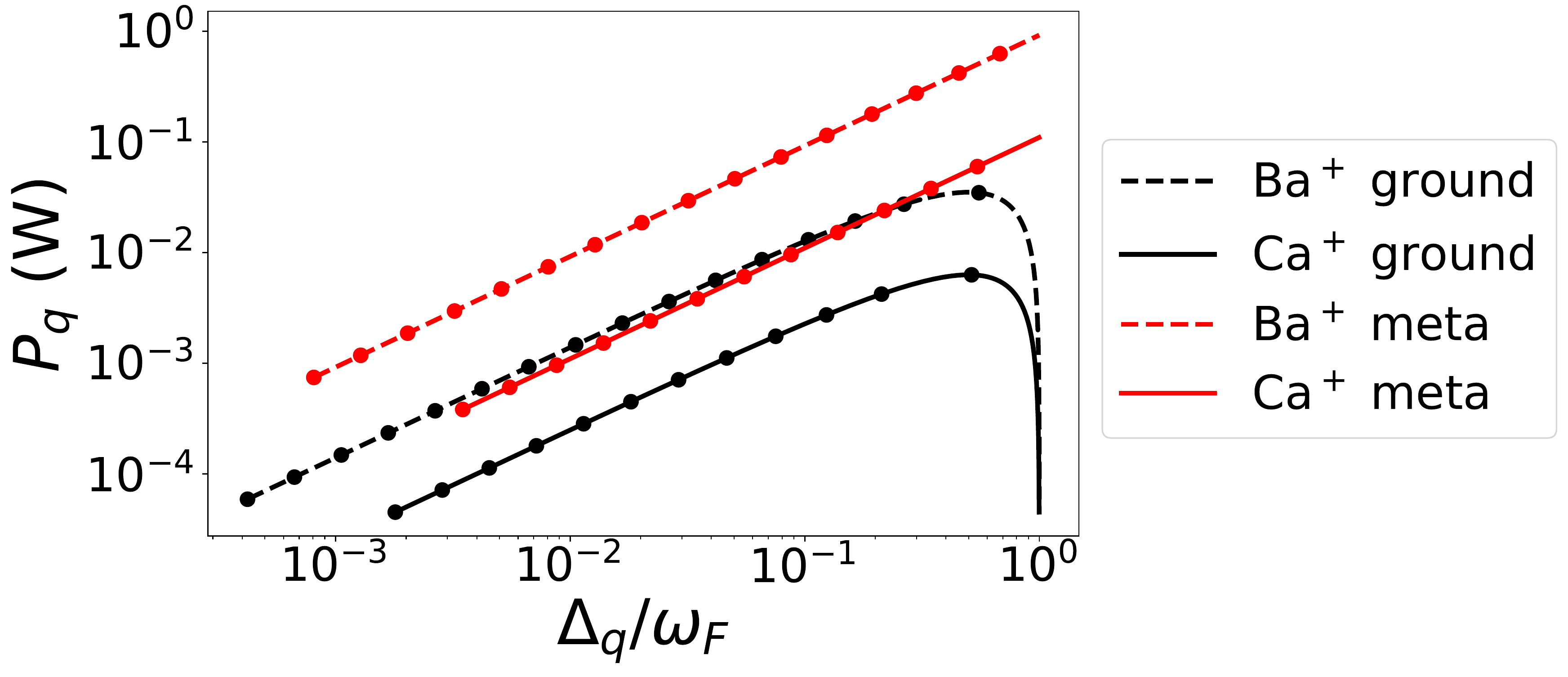}
\caption{Laser power of each Raman beam as the detuning $\Delta_q$ ($q=g,m$) from the $P_{3/2}$ manifold is varied. The points are marked at same detunings as the points in Fig.~\ref{fig:fig2} and \ref{fig:fig4}(a). For metastable Ba$^+$ (Ca$^+$) qubit, $I=1/2$ (7/2) and $F_0 = 2$ (5), as shown in Table~\ref{tab:coeffs}. Here we assume \mbox{$\Omega_q = 2\pi \times 0.25$ MHz} \mbox{($t_{\rm gate} = 20$ $\mu$s)} and \mbox{$w_0 = 20$ $\mu$m.} }
\label{fig:fig_power}
\end{figure}
%%%%%%%%%%%%%%%%%%%%%%%%%%%%%%%%%%%%

Figure~\ref{fig:fig_power} shows the laser power of each Raman beam for various detunings, for a typical Rabi frequency \mbox{$\Omega_q = 2\pi \times 0.25$ MHz} (which gives \mbox{$t_{\rm gate} = 20$ $\mu$s} for \mbox{$\eta=0.05$} and \mbox{$K=1$}) and beam waist \mbox{$w_0 = 20$ $\mu$m}, following Ref.~\cite{Ozeri07}. Combined with Fig.~\ref{fig:fig2}, the laser power is higher at detuning that yields lower gate-error rate. As expected from Case \rom{3}, for fixed Rabi frequency and detuning, the metastable qubit requires an order of magnitude higher laser power when $\Delta_q$ is not too close to $\omega_F$. 

In practice, the laser power is often limited by material loss in mirrors and waveguides~\cite{Brown21}, which is less severe for metastable qubits than ground qubits as the laser wavelength is longer. We expect our results provide a guideline to future experiments on whether the achievable laser power is large enough for metastable qubits to have advantage over ground qubits by erasure conversion.

%%%%%%%%%%%%%%%%%%APPENDIX D%%%%%%%%%%%%%%%%%%%%
\section{Error propagation during two-qubit gates} \label{app:D}

In order to accurately evaluate the circuit-level performance of the surface code, we use a detailed model on how an error on one qubit due to the spontaneous scattering propagate to the other qubit during a CNOT gate. 

\begin{figure}[!htb]
\includegraphics[width=\linewidth]{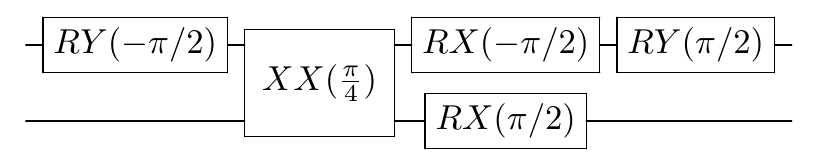}
\caption{Circuit diagram of a CNOT gate (upper: control, lower: target) decomposed into native gates for trapped ions. Here, $RX(\theta) = \exp(-i \frac{\theta}{2} X)$ and $RY(\theta) = \exp(-i \frac{\theta}{2} Y)$ are the single-qubit gates and $XX(\theta) = \exp(-i \theta X \otimes X)$ is the MS gate, where $X$ and $Y$ are single-qubit Pauli operations. When the MS gate is omitted, the circuit becomes $RZ(\pi/2) \otimes RX(\pi/2)$.}
\label{fig:cnotcircuit}
\end{figure}

As the MS gate is a widely-used native two-qubit gate for trapped ions, we first decompose the CNOT gate into MS and single-qubit gates using the circuit in Fig.~\ref{fig:cnotcircuit}~\cite{Maslov17}. As single-qubit gates are assumed to be perfect, we analyze how errors propagate during an MS gate. The errors are then altered as they go through the following single-qubit gates by the standard rules.  

First, a bit-flip ($X$) error due to the spontaneous scattering \textit{during} a MS gate does not propagate, regardless of at which point during gate time it occurred. This is because an $X$ error commutes through $XX(\theta)$ for any $\theta$.

Next, for the propagation of a phase-flip ($Z$) error, we start with two extreme cases: error completely before and after a MS gate. A $Z \otimes I$ phase flip occurring before the MS gate is equivalent to a $(X \otimes X) (Z \otimes I)$ error occurring after the MS gate~\cite{Schwerdt22}. A phase flip after the MS gate trivially does not propagate. Thus, for a general case where a phase-flip error occurs \textit{during} a MS gate, we expect that additional bit-flip error occurs to one of or both of the two qubits. 

To find the average rate of additional bit-flip error, we use a master-equation simulation following the method in Ref.~\cite{Schwerdt22}. Specifically, we prepare the initial state $|00\rangle$, run a perfect $XX(-\pi/4)$ gate, run a $XX(\pi/4)$ gate with a phase flip injected at time $t$ ($0 \leq t \leq t_{\rm gate}$), and then measure the population of the $\ket{1}$ state of the qubits. The $\ket{1}$-state population indicates the additional bit-flip error rate. The populations are averaged over many values of $t$, drawn from a uniform distribution over the gate duration. According to our simulation results, when a phase flip occurs to one of the qubits during a MS gate, additional $X \otimes I$, $I \otimes X$, or $X \otimes X$ error occurs with probability $r$, $r$, and $1/2-r$, respectively, where $r=0.1349$.

Finally, we consider the propagation of a leakage error. If a leakage happens at time $t < t_{\rm gate}$ after a MS gate started, the two-qubit rotation $XX(\theta)$ is not fully performed up to $\theta = \pi/4$. Thus, the effect of a leakage can be described by two qubits undergoing a partial $XX$ rotation followed by the leaked qubit being traced out. 

To start, we describe how the partial rotation angle evolves over time. During a MS gate, a normal mode of the ion chain's collective motion is briefly excited by laser beams near-resonant to the sideband transition. Here we denote $\delta$ as the detuning from the sideband transition. Then, the action of a MS gate up to time $t$ ($0 \leq t \leq t_{\rm gate}$) in the subspace of the two qubits is given by $\exp(-i \theta(t) X \otimes X)$, where~\cite{Wu18}
\begin{equation}
    \theta(t) = \frac{\delta t - \sin \delta t}{\delta t_{\rm gate} - \sin \delta t_{\rm gate}} \times \frac{\pi}{4}
\end{equation}
such that $\theta(0) = 0$ and $\theta(t_{\rm gate}) = \pi/4$. There is an additional condition that the motional mode needs to be completely disentangled from the qubits at the end of the gate, which is satisfied when $\delta t_{\rm gate} = 2K\pi$. We choose $K=1$ as in the main text and obtain
\begin{equation}
    \theta(t) = \frac{\pi t}{4 t_{\rm gate}} - \frac{1}{8}\sin\left(2\pi \frac{t}{t_{\rm gate}} \right).
\end{equation}
We note that even when one of the qubit is leaked during the gate, the $\delta t_{\rm gate} = 2K\pi$ condition guarantees that the other qubit is completely disentangled from the motional mode after the gate. 

Now we describe the action of tracing out the leaked qubit after an incomplete $XX$ rotation. Such channel is complicated to write in a general form that is valid for all initial states. However, for the surface code, in the absence of errors, the syndrome qubit is always in the $\ket{0}$ state right before a MS gate is applied. If the syndrome qubit is leaked at time $t$, it can be straightforwardly shown that the channel on the data qubit is a bit-flip channel with probability $\sin^2 \theta(t)$, regardless of its initial state. If the data qubit is leaked at time $t$, the channel on the syndrome qubit is also a bit-flip channel with probability $\sin^2 \theta(t)$, where here we additionally use that in the subspace of the pair of data and syndrome qubits, the data qubit is in the maximally mixed state. 

Therefore, for both data and syndrome qubits, a leakage in one of the qubits at time $t$ results in a bit flip with probability $\sin^2 \theta(t)$ on the other qubit. Assuming again that the distribution of probability that a leakage occurs is uniform over the gate duration, the average probability of bit-flip propagation is given by
\begin{equation}
    \frac{1}{t_{\rm gate}} \int_0^{t_{\rm gate}} \sin^2 \theta(t) dt = 0.2078.
\end{equation}
Thus, whenever one of the two qubits is leaked (and becomes a maximally mixed state in our simulations), the other qubit undergoes a bit flip with probability 0.2078. 

This concludes the model for error propagation during a MS gate. Then, the errors go through the single-qubit gates following the MS gate in Fig.~\ref{fig:cnotcircuit}, leading to the error model for a CNOT gate. Specifically, the rates of two-qubit Pauli errors $IX$, $IY$, ..., $ZZ$ for each CNOT gate can be calculated in terms of $p_q^{(xy)}$, $p_q^{(z)}$, $p_q^{(l)}$, and $p_q^{(e)}$ ($q = g,m$), which are fed into the circuit simulated by Stim~\cite{Gidney21}. 

A caveat here is that the two-qubit Pauli errors are always appended after the execution of each CNOT gate. In reality, for the case of a leakage error, the MS gate is replaced by an incomplete $XX$ rotation that directly becomes the propagated error. Thus, the two-qubit Pauli errors are ideally appended after a $RZ(\pi/2) \otimes RX(\pi/2)$ gate rather than a CNOT gate (see Fig.~\ref{fig:cnotcircuit}). However, switching between CNOT and $RZ(\pi/2) \otimes RX(\pi/2)$ gates probabilistically is challenging to simulate efficiently. 

We justify always appending the two-qubit Pauli errors after the CNOT gate by the following argument. The CNOT gate can be expressed as
\begin{equation}
    \text{CNOT} = I \otimes |+\rangle \langle +| + Z \otimes |-\rangle \langle -|
    = |0\rangle \langle 0| \otimes I + |1\rangle \langle 1| \otimes X. \nonumber
\end{equation}
In our simulations, a leaked qubit is in the maximally mixed state. Thus, when the target qubit is leaked, the CNOT gate becomes either $I$ or $Z$ on the control qubit with probability 1/2 each. This can be approximated to $RZ(\pi/2)$ on the control qubit on average. Similarly, when the control qubit is leaked, the CNOT gate becomes either $I$ or $X$ on the target qubit with probability 1/2 each, and this can be approximated to $RX(\pi/2)$ on the target qubit on average.

%%%%%%%%%%%%%%%%%%APPENDIX E%%%%%%%%%%%%%%%%%%%%
\section{Effects of metastable qubits' idling errors on QEC} \label{app:E}

Here we simulate how the idling errors of metastable qubits affect the QEC performance. Due to the finite lifetime of the $D_{5/2}$ states, metastable qubits can spontaneously decay to the $S_{1/2}$ manifold during idling. Such leakage errors can always be converted to erasures, using the erasure-conversion scheme described in Sec.~\ref{sec:sec2}. The idling-error rate is given in (\ref{eq:idleerror}), where the upper bound of time $t$ is determined by the surface-code cycle time $T$, which can be considered as the clock time for fault-tolerant quantum computation. 

We consider the example of metastable Ca$^+$ qubit in Table~\ref{tab:coeffs}, as its lifetime \mbox{1.16 s} is significantly shorter than the metastable Ba$^+$ qubit's \mbox{30.14 s}. In the surface-code simulations, similarly to Sec.~\ref{sec:sec5}, we assume that all qubits can be erased, each with probability $p^{\rm (idle)}(T)/4$ before each layer of CNOT gates, where the factor $1/4$ is from the four layers of CNOT gates per cycle. Various values of surface-code cycle time $T$ ranging from \mbox{3 ms} to \mbox{1.19 s} are considered, such that the idling-error rate $p^{\rm (idle)}(T)/4$ is varied from $6.46 \times 10^{-4}$ to $0.161$. We also add two-qubit-gate errors of fixed rate $p_m^{(2q)} = 1.14 \times 10^{-3}$ and measurement errors of fixed rate $10^{-4}$, where the former is the lowest value considered in Fig.~\ref{fig:fig4}(a) for Ca$^+$ qubits.  

%%%%%%%%%%%%%%%FIGURE 5%%%%%%%%%%%%%%
\begin{figure}[ht]
\includegraphics[width=8.6cm]{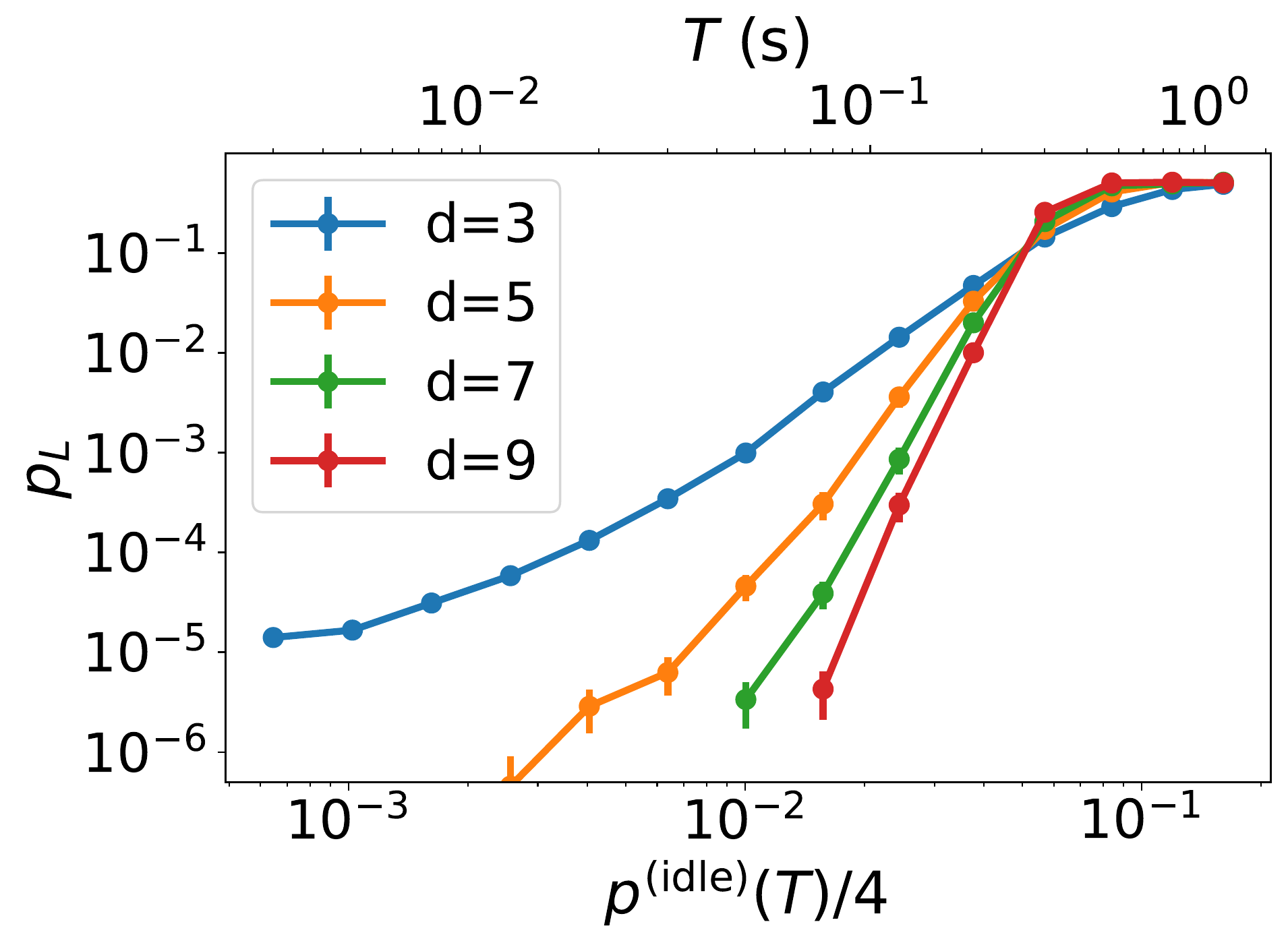}
\caption{Logical error rates of metastable Ca$^+$ qubits for various code distances $d$ and idling-error rates $p^{\rm (idle)}(T)/4$, where $T$ is the surface-code cycle time, represented in the top horizontal axis. The factor $1/4$ comes from the four layers of CNOT gates per cycle. Two-qubit-gate errors and measurement errors of fixed rate $1.14 \times 10^{-3}$ and $10^{-4}$, respectively, are added. The error bars represent the 95\% confidence interval. The threshold idling-error rate is 5.14\%, which corresponds to \mbox{$T = 0.267$ s}. We expect that significantly shorter cycle time, thus lower idling-error rate, is achievable.}
\label{fig:fig_loss}
\end{figure}
%%%%%%%%%%%%%%%%%%%%%%%%%%%%%%%%%%%%

Figure~\ref{fig:fig_loss} shows the logical error rates $p_L$ for various code distances $d$ and idling-error rates $p^{\rm (idle)}(T)/4$. The threshold idling-error rate is 5.14\%, which is higher than the threshold two-qubit-gate error rate 3.42\%, as idling errors are completely converted to erasures. A distance-$d$ code is guaranteed to correct $d-1$ idling errors per cycle, as indicated by the slopes of the logical-error curves. 

The top horizontal axis represents the surface-code cycle time $T$. While Fig.~\ref{fig:fig_loss} plots the logical error rates for cycle times as long as \mbox{$\sim 1$ s} to show the threshold, such long cycle time is impractical for quantum computation. For relatively feasible cycle times, say, \mbox{$T < 10$ ms}, $p_L$ quickly decreases below $10^{-6}$ for $d \geq 5$. For $d=3$, $p_L$ converges to a nonzero value as the idling-error rate decreases, which indicates that the effects of the two-qubit-gate errors of fixed rate $p_m^{(2q)} = 1.14 \times 10^{-3}$ dominate the effects of the idling errors when \mbox{$T \lesssim 3$ ms}. Therefore, for reasonably short surface-code cycle times, we expect that the metastable qubits' idling errors have significantly smaller impact on the QEC performance than the gate errors. For metastable qubits with longer lifetime, such as Ba$^+$ and Yb$^+$, the idling errors will be even more negligible.  

\bibliography{bib}% Produces the bibliography via BibTeX.

\begin{thebibliography}{88}%
\makeatletter
\providecommand \@ifxundefined [1]{%
 \@ifx{#1\undefined}
}%
\providecommand \@ifnum [1]{%
 \ifnum #1\expandafter \@firstoftwo
 \else \expandafter \@secondoftwo
 \fi
}%
\providecommand \@ifx [1]{%
 \ifx #1\expandafter \@firstoftwo
 \else \expandafter \@secondoftwo
 \fi
}%
\providecommand \natexlab [1]{#1}%
\providecommand \enquote  [1]{``#1''}%
\providecommand \bibnamefont  [1]{#1}%
\providecommand \bibfnamefont [1]{#1}%
\providecommand \citenamefont [1]{#1}%
\providecommand \href@noop [0]{\@secondoftwo}%
\providecommand \href [0]{\begingroup \@sanitize@url \@href}%
\providecommand \@href[1]{\@@startlink{#1}\@@href}%
\providecommand \@@href[1]{\endgroup#1\@@endlink}%
\providecommand \@sanitize@url [0]{\catcode `\\12\catcode `\$12\catcode
  `\&12\catcode `\#12\catcode `\^12\catcode `\_12\catcode `\%12\relax}%
\providecommand \@@startlink[1]{}%
\providecommand \@@endlink[0]{}%
\providecommand \url  [0]{\begingroup\@sanitize@url \@url }%
\providecommand \@url [1]{\endgroup\@href {#1}{\urlprefix }}%
\providecommand \urlprefix  [0]{URL }%
\providecommand \Eprint [0]{\href }%
\providecommand \doibase [0]{https://doi.org/}%
\providecommand \selectlanguage [0]{\@gobble}%
\providecommand \bibinfo  [0]{\@secondoftwo}%
\providecommand \bibfield  [0]{\@secondoftwo}%
\providecommand \translation [1]{[#1]}%
\providecommand \BibitemOpen [0]{}%
\providecommand \bibitemStop [0]{}%
\providecommand \bibitemNoStop [0]{.\EOS\space}%
\providecommand \EOS [0]{\spacefactor3000\relax}%
\providecommand \BibitemShut  [1]{\csname bibitem#1\endcsname}%
\let\auto@bib@innerbib\@empty
%</preamble>
\bibitem [{\citenamefont {Knill}\ and\ \citenamefont
  {Laflamme}(1997)}]{Knill97}%
  \BibitemOpen
  \bibfield  {author} {\bibinfo {author} {\bibfnamefont {E.}~\bibnamefont
  {Knill}}\ and\ \bibinfo {author} {\bibfnamefont {R.}~\bibnamefont
  {Laflamme}},\ }\bibfield  {title} {\bibinfo {title} {Theory of quantum
  error-correcting codes},\ }\href {https://doi.org/10.1103/PhysRevA.55.900}
  {\bibfield  {journal} {\bibinfo  {journal} {Phys. Rev. A}\ }\textbf {\bibinfo
  {volume} {55}},\ \bibinfo {pages} {900} (\bibinfo {year} {1997})}\BibitemShut
  {NoStop}%
\bibitem [{\citenamefont {Mirrahimi}\ \emph {et~al.}(2014)\citenamefont
  {Mirrahimi}, \citenamefont {Leghtas}, \citenamefont {Albert}, \citenamefont
  {Touzard}, \citenamefont {Schoelkopf}, \citenamefont {Jiang},\ and\
  \citenamefont {Devoret}}]{Mirrahimi14}%
  \BibitemOpen
  \bibfield  {author} {\bibinfo {author} {\bibfnamefont {M.}~\bibnamefont
  {Mirrahimi}}, \bibinfo {author} {\bibfnamefont {Z.}~\bibnamefont {Leghtas}},
  \bibinfo {author} {\bibfnamefont {V.~V.}\ \bibnamefont {Albert}}, \bibinfo
  {author} {\bibfnamefont {S.}~\bibnamefont {Touzard}}, \bibinfo {author}
  {\bibfnamefont {R.~J.}\ \bibnamefont {Schoelkopf}}, \bibinfo {author}
  {\bibfnamefont {L.}~\bibnamefont {Jiang}},\ and\ \bibinfo {author}
  {\bibfnamefont {M.~H.}\ \bibnamefont {Devoret}},\ }\bibfield  {title}
  {\bibinfo {title} {Dynamically protected cat-qubits: a new paradigm for
  universal quantum computation},\ }\href@noop {} {\bibfield  {journal}
  {\bibinfo  {journal} {New Journal of Physics}\ }\textbf {\bibinfo {volume}
  {16}},\ \bibinfo {pages} {045014} (\bibinfo {year} {2014})}\BibitemShut
  {NoStop}%
\bibitem [{\citenamefont {Ofek}\ \emph {et~al.}(2016)\citenamefont {Ofek},
  \citenamefont {Petrenko}, \citenamefont {Heeres}, \citenamefont {Reinhold},
  \citenamefont {Leghtas}, \citenamefont {Vlastakis}, \citenamefont {Liu},
  \citenamefont {Frunzio}, \citenamefont {Girvin}, \citenamefont {Jiang} \emph
  {et~al.}}]{Ofek16}%
  \BibitemOpen
  \bibfield  {author} {\bibinfo {author} {\bibfnamefont {N.}~\bibnamefont
  {Ofek}}, \bibinfo {author} {\bibfnamefont {A.}~\bibnamefont {Petrenko}},
  \bibinfo {author} {\bibfnamefont {R.}~\bibnamefont {Heeres}}, \bibinfo
  {author} {\bibfnamefont {P.}~\bibnamefont {Reinhold}}, \bibinfo {author}
  {\bibfnamefont {Z.}~\bibnamefont {Leghtas}}, \bibinfo {author} {\bibfnamefont
  {B.}~\bibnamefont {Vlastakis}}, \bibinfo {author} {\bibfnamefont
  {Y.}~\bibnamefont {Liu}}, \bibinfo {author} {\bibfnamefont {L.}~\bibnamefont
  {Frunzio}}, \bibinfo {author} {\bibfnamefont {S.}~\bibnamefont {Girvin}},
  \bibinfo {author} {\bibfnamefont {L.}~\bibnamefont {Jiang}}, \emph {et~al.},\
  }\bibfield  {title} {\bibinfo {title} {Extending the lifetime of a quantum
  bit with error correction in superconducting circuits},\ }\href@noop {}
  {\bibfield  {journal} {\bibinfo  {journal} {Nature}\ }\textbf {\bibinfo
  {volume} {536}},\ \bibinfo {pages} {441} (\bibinfo {year}
  {2016})}\BibitemShut {NoStop}%
\bibitem [{\citenamefont {Puri}\ \emph {et~al.}(2020)\citenamefont {Puri},
  \citenamefont {St-Jean}, \citenamefont {Gross}, \citenamefont {Grimm},
  \citenamefont {Frattini}, \citenamefont {Iyer}, \citenamefont {Krishna},
  \citenamefont {Touzard}, \citenamefont {Jiang}, \citenamefont {Blais},
  \citenamefont {Flammia},\ and\ \citenamefont {Girvin}}]{Puri20}%
  \BibitemOpen
  \bibfield  {author} {\bibinfo {author} {\bibfnamefont {S.}~\bibnamefont
  {Puri}}, \bibinfo {author} {\bibfnamefont {L.}~\bibnamefont {St-Jean}},
  \bibinfo {author} {\bibfnamefont {J.~A.}\ \bibnamefont {Gross}}, \bibinfo
  {author} {\bibfnamefont {A.}~\bibnamefont {Grimm}}, \bibinfo {author}
  {\bibfnamefont {N.~E.}\ \bibnamefont {Frattini}}, \bibinfo {author}
  {\bibfnamefont {P.~S.}\ \bibnamefont {Iyer}}, \bibinfo {author}
  {\bibfnamefont {A.}~\bibnamefont {Krishna}}, \bibinfo {author} {\bibfnamefont
  {S.}~\bibnamefont {Touzard}}, \bibinfo {author} {\bibfnamefont
  {L.}~\bibnamefont {Jiang}}, \bibinfo {author} {\bibfnamefont
  {A.}~\bibnamefont {Blais}}, \bibinfo {author} {\bibfnamefont {S.~T.}\
  \bibnamefont {Flammia}},\ and\ \bibinfo {author} {\bibfnamefont {S.~M.}\
  \bibnamefont {Girvin}},\ }\bibfield  {title} {\bibinfo {title}
  {Bias-preserving gates with stabilized cat qubits},\ }\href
  {https://doi.org/10.1126/sciadv.aay5901} {\bibfield  {journal} {\bibinfo
  {journal} {Science Advances}\ }\textbf {\bibinfo {volume} {6}},\ \bibinfo
  {pages} {eaay5901} (\bibinfo {year} {2020})}\BibitemShut {NoStop}%
\bibitem [{\citenamefont {Cong}\ \emph {et~al.}(2022)\citenamefont {Cong},
  \citenamefont {Levine}, \citenamefont {Keesling}, \citenamefont {Bluvstein},
  \citenamefont {Wang},\ and\ \citenamefont {Lukin}}]{Cong22}%
  \BibitemOpen
  \bibfield  {author} {\bibinfo {author} {\bibfnamefont {I.}~\bibnamefont
  {Cong}}, \bibinfo {author} {\bibfnamefont {H.}~\bibnamefont {Levine}},
  \bibinfo {author} {\bibfnamefont {A.}~\bibnamefont {Keesling}}, \bibinfo
  {author} {\bibfnamefont {D.}~\bibnamefont {Bluvstein}}, \bibinfo {author}
  {\bibfnamefont {S.-T.}\ \bibnamefont {Wang}},\ and\ \bibinfo {author}
  {\bibfnamefont {M.~D.}\ \bibnamefont {Lukin}},\ }\bibfield  {title} {\bibinfo
  {title} {Hardware-efficient, fault-tolerant quantum computation with rydberg
  atoms},\ }\href {https://doi.org/10.1103/PhysRevX.12.021049} {\bibfield
  {journal} {\bibinfo  {journal} {Phys. Rev. X}\ }\textbf {\bibinfo {volume}
  {12}},\ \bibinfo {pages} {021049} (\bibinfo {year} {2022})}\BibitemShut
  {NoStop}%
\bibitem [{\citenamefont {Aliferis}\ and\ \citenamefont
  {Preskill}(2008)}]{Aliferis08}%
  \BibitemOpen
  \bibfield  {author} {\bibinfo {author} {\bibfnamefont {P.}~\bibnamefont
  {Aliferis}}\ and\ \bibinfo {author} {\bibfnamefont {J.}~\bibnamefont
  {Preskill}},\ }\bibfield  {title} {\bibinfo {title} {Fault-tolerant quantum
  computation against biased noise},\ }\href
  {https://doi.org/10.1103/PhysRevA.78.052331} {\bibfield  {journal} {\bibinfo
  {journal} {Phys. Rev. A}\ }\textbf {\bibinfo {volume} {78}},\ \bibinfo
  {pages} {052331} (\bibinfo {year} {2008})}\BibitemShut {NoStop}%
\bibitem [{\citenamefont {Li}\ \emph {et~al.}(2019)\citenamefont {Li},
  \citenamefont {Miller}, \citenamefont {Newman}, \citenamefont {Wu},\ and\
  \citenamefont {Brown}}]{Li19}%
  \BibitemOpen
  \bibfield  {author} {\bibinfo {author} {\bibfnamefont {M.}~\bibnamefont
  {Li}}, \bibinfo {author} {\bibfnamefont {D.}~\bibnamefont {Miller}}, \bibinfo
  {author} {\bibfnamefont {M.}~\bibnamefont {Newman}}, \bibinfo {author}
  {\bibfnamefont {Y.}~\bibnamefont {Wu}},\ and\ \bibinfo {author}
  {\bibfnamefont {K.~R.}\ \bibnamefont {Brown}},\ }\bibfield  {title} {\bibinfo
  {title} {2d compass codes},\ }\href
  {https://doi.org/10.1103/PhysRevX.9.021041} {\bibfield  {journal} {\bibinfo
  {journal} {Phys. Rev. X}\ }\textbf {\bibinfo {volume} {9}},\ \bibinfo {pages}
  {021041} (\bibinfo {year} {2019})}\BibitemShut {NoStop}%
\bibitem [{\citenamefont {Guillaud}\ and\ \citenamefont
  {Mirrahimi}(2019)}]{Guillaud19}%
  \BibitemOpen
  \bibfield  {author} {\bibinfo {author} {\bibfnamefont {J.}~\bibnamefont
  {Guillaud}}\ and\ \bibinfo {author} {\bibfnamefont {M.}~\bibnamefont
  {Mirrahimi}},\ }\bibfield  {title} {\bibinfo {title} {Repetition cat qubits
  for fault-tolerant quantum computation},\ }\href
  {https://doi.org/10.1103/PhysRevX.9.041053} {\bibfield  {journal} {\bibinfo
  {journal} {Phys. Rev. X}\ }\textbf {\bibinfo {volume} {9}},\ \bibinfo {pages}
  {041053} (\bibinfo {year} {2019})}\BibitemShut {NoStop}%
\bibitem [{\citenamefont {Huang}\ and\ \citenamefont
  {Brown}(2020)}]{Huang20compass}%
  \BibitemOpen
  \bibfield  {author} {\bibinfo {author} {\bibfnamefont {S.}~\bibnamefont
  {Huang}}\ and\ \bibinfo {author} {\bibfnamefont {K.~R.}\ \bibnamefont
  {Brown}},\ }\bibfield  {title} {\bibinfo {title} {Fault-tolerant compass
  codes},\ }\href {https://doi.org/10.1103/PhysRevA.101.042312} {\bibfield
  {journal} {\bibinfo  {journal} {Phys. Rev. A}\ }\textbf {\bibinfo {volume}
  {101}},\ \bibinfo {pages} {042312} (\bibinfo {year} {2020})}\BibitemShut
  {NoStop}%
\bibitem [{\citenamefont {Bonilla~Ataides}\ \emph {et~al.}(2021)\citenamefont
  {Bonilla~Ataides}, \citenamefont {Tuckett}, \citenamefont {Bartlett},
  \citenamefont {Flammia},\ and\ \citenamefont {Brown}}]{Bonilla21}%
  \BibitemOpen
  \bibfield  {author} {\bibinfo {author} {\bibfnamefont {J.~P.}\ \bibnamefont
  {Bonilla~Ataides}}, \bibinfo {author} {\bibfnamefont {D.~K.}\ \bibnamefont
  {Tuckett}}, \bibinfo {author} {\bibfnamefont {S.~D.}\ \bibnamefont
  {Bartlett}}, \bibinfo {author} {\bibfnamefont {S.~T.}\ \bibnamefont
  {Flammia}},\ and\ \bibinfo {author} {\bibfnamefont {B.~J.}\ \bibnamefont
  {Brown}},\ }\bibfield  {title} {\bibinfo {title} {The {XZZX} surface code},\
  }\href@noop {} {\bibfield  {journal} {\bibinfo  {journal} {Nat. Commun}\
  }\textbf {\bibinfo {volume} {12}},\ \bibinfo {pages} {2172} (\bibinfo {year}
  {2021})}\BibitemShut {NoStop}%
\bibitem [{\citenamefont {Darmawan}\ \emph {et~al.}(2021)\citenamefont
  {Darmawan}, \citenamefont {Brown}, \citenamefont {Grimsmo}, \citenamefont
  {Tuckett},\ and\ \citenamefont {Puri}}]{Darmawan21}%
  \BibitemOpen
  \bibfield  {author} {\bibinfo {author} {\bibfnamefont {A.~S.}\ \bibnamefont
  {Darmawan}}, \bibinfo {author} {\bibfnamefont {B.~J.}\ \bibnamefont {Brown}},
  \bibinfo {author} {\bibfnamefont {A.~L.}\ \bibnamefont {Grimsmo}}, \bibinfo
  {author} {\bibfnamefont {D.~K.}\ \bibnamefont {Tuckett}},\ and\ \bibinfo
  {author} {\bibfnamefont {S.}~\bibnamefont {Puri}},\ }\bibfield  {title}
  {\bibinfo {title} {Practical quantum error correction with the {XZZX} code
  and kerr-cat qubits},\ }\href {https://doi.org/10.1103/PRXQuantum.2.030345}
  {\bibfield  {journal} {\bibinfo  {journal} {PRX Quantum}\ }\textbf {\bibinfo
  {volume} {2}},\ \bibinfo {pages} {030345} (\bibinfo {year}
  {2021})}\BibitemShut {NoStop}%
\bibitem [{\citenamefont {Dua}\ \emph {et~al.}(2022)\citenamefont {Dua},
  \citenamefont {Kubica}, \citenamefont {Jiang}, \citenamefont {Flammia},\ and\
  \citenamefont {Gullans}}]{Dua22}%
  \BibitemOpen
  \bibfield  {author} {\bibinfo {author} {\bibfnamefont {A.}~\bibnamefont
  {Dua}}, \bibinfo {author} {\bibfnamefont {A.}~\bibnamefont {Kubica}},
  \bibinfo {author} {\bibfnamefont {L.}~\bibnamefont {Jiang}}, \bibinfo
  {author} {\bibfnamefont {S.~T.}\ \bibnamefont {Flammia}},\ and\ \bibinfo
  {author} {\bibfnamefont {M.~J.}\ \bibnamefont {Gullans}},\ }\bibfield
  {title} {\bibinfo {title} {Clifford-deformed surface codes},\ }\href@noop {}
  {\bibfield  {journal} {\bibinfo  {journal} {arXiv preprint arXiv:2201.07802}\
  } (\bibinfo {year} {2022})}\BibitemShut {NoStop}%
\bibitem [{\citenamefont {Xu}\ \emph {et~al.}(2023)\citenamefont {Xu},
  \citenamefont {Mannucci}, \citenamefont {Seif}, \citenamefont {Kubica},
  \citenamefont {Flammia},\ and\ \citenamefont {Jiang}}]{Xu23}%
  \BibitemOpen
  \bibfield  {author} {\bibinfo {author} {\bibfnamefont {Q.}~\bibnamefont
  {Xu}}, \bibinfo {author} {\bibfnamefont {N.}~\bibnamefont {Mannucci}},
  \bibinfo {author} {\bibfnamefont {A.}~\bibnamefont {Seif}}, \bibinfo {author}
  {\bibfnamefont {A.}~\bibnamefont {Kubica}}, \bibinfo {author} {\bibfnamefont
  {S.~T.}\ \bibnamefont {Flammia}},\ and\ \bibinfo {author} {\bibfnamefont
  {L.}~\bibnamefont {Jiang}},\ }\bibfield  {title} {\bibinfo {title} {Tailored
  {XZZX} codes for biased noise},\ }\href
  {https://doi.org/10.1103/PhysRevResearch.5.013035} {\bibfield  {journal}
  {\bibinfo  {journal} {Phys. Rev. Res.}\ }\textbf {\bibinfo {volume} {5}},\
  \bibinfo {pages} {013035} (\bibinfo {year} {2023})}\BibitemShut {NoStop}%
\bibitem [{\citenamefont {Grassl}\ \emph {et~al.}(1997)\citenamefont {Grassl},
  \citenamefont {Beth},\ and\ \citenamefont {Pellizzari}}]{Grassl97}%
  \BibitemOpen
  \bibfield  {author} {\bibinfo {author} {\bibfnamefont {M.}~\bibnamefont
  {Grassl}}, \bibinfo {author} {\bibfnamefont {T.}~\bibnamefont {Beth}},\ and\
  \bibinfo {author} {\bibfnamefont {T.}~\bibnamefont {Pellizzari}},\ }\bibfield
   {title} {\bibinfo {title} {Codes for the quantum erasure channel},\ }\href
  {https://doi.org/10.1103/PhysRevA.56.33} {\bibfield  {journal} {\bibinfo
  {journal} {Phys. Rev. A}\ }\textbf {\bibinfo {volume} {56}},\ \bibinfo
  {pages} {33} (\bibinfo {year} {1997})}\BibitemShut {NoStop}%
\bibitem [{\citenamefont {Bennett}\ \emph {et~al.}(1997)\citenamefont
  {Bennett}, \citenamefont {DiVincenzo},\ and\ \citenamefont
  {Smolin}}]{Bennett97}%
  \BibitemOpen
  \bibfield  {author} {\bibinfo {author} {\bibfnamefont {C.~H.}\ \bibnamefont
  {Bennett}}, \bibinfo {author} {\bibfnamefont {D.~P.}\ \bibnamefont
  {DiVincenzo}},\ and\ \bibinfo {author} {\bibfnamefont {J.~A.}\ \bibnamefont
  {Smolin}},\ }\bibfield  {title} {\bibinfo {title} {Capacities of quantum
  erasure channels},\ }\href {https://doi.org/10.1103/PhysRevLett.78.3217}
  {\bibfield  {journal} {\bibinfo  {journal} {Phys. Rev. Lett.}\ }\textbf
  {\bibinfo {volume} {78}},\ \bibinfo {pages} {3217} (\bibinfo {year}
  {1997})}\BibitemShut {NoStop}%
\bibitem [{\citenamefont {Lu}\ \emph {et~al.}(2008)\citenamefont {Lu},
  \citenamefont {Gao}, \citenamefont {Zhang}, \citenamefont {Zhou},
  \citenamefont {Yang},\ and\ \citenamefont {Pan}}]{Lu08}%
  \BibitemOpen
  \bibfield  {author} {\bibinfo {author} {\bibfnamefont {C.-Y.}\ \bibnamefont
  {Lu}}, \bibinfo {author} {\bibfnamefont {W.-B.}\ \bibnamefont {Gao}},
  \bibinfo {author} {\bibfnamefont {J.}~\bibnamefont {Zhang}}, \bibinfo
  {author} {\bibfnamefont {X.-Q.}\ \bibnamefont {Zhou}}, \bibinfo {author}
  {\bibfnamefont {T.}~\bibnamefont {Yang}},\ and\ \bibinfo {author}
  {\bibfnamefont {J.-W.}\ \bibnamefont {Pan}},\ }\bibfield  {title} {\bibinfo
  {title} {Experimental quantum coding against qubit loss error},\ }\href
  {https://doi.org/10.1073/pnas.0800740105} {\bibfield  {journal} {\bibinfo
  {journal} {Proceedings of the National Academy of Sciences}\ }\textbf
  {\bibinfo {volume} {105}},\ \bibinfo {pages} {11050} (\bibinfo {year}
  {2008})}\BibitemShut {NoStop}%
\bibitem [{\citenamefont {Alber}\ \emph {et~al.}(2001)\citenamefont {Alber},
  \citenamefont {Beth}, \citenamefont {Charnes}, \citenamefont {Delgado},
  \citenamefont {Grassl},\ and\ \citenamefont {Mussinger}}]{Alber01}%
  \BibitemOpen
  \bibfield  {author} {\bibinfo {author} {\bibfnamefont {G.}~\bibnamefont
  {Alber}}, \bibinfo {author} {\bibfnamefont {T.}~\bibnamefont {Beth}},
  \bibinfo {author} {\bibfnamefont {C.}~\bibnamefont {Charnes}}, \bibinfo
  {author} {\bibfnamefont {A.}~\bibnamefont {Delgado}}, \bibinfo {author}
  {\bibfnamefont {M.}~\bibnamefont {Grassl}},\ and\ \bibinfo {author}
  {\bibfnamefont {M.}~\bibnamefont {Mussinger}},\ }\bibfield  {title} {\bibinfo
  {title} {Stabilizing distinguishable qubits against spontaneous decay by
  detected-jump correcting quantum codes},\ }\href
  {https://doi.org/10.1103/PhysRevLett.86.4402} {\bibfield  {journal} {\bibinfo
   {journal} {Phys. Rev. Lett.}\ }\textbf {\bibinfo {volume} {86}},\ \bibinfo
  {pages} {4402} (\bibinfo {year} {2001})}\BibitemShut {NoStop}%
\bibitem [{\citenamefont {Vala}\ \emph {et~al.}(2005)\citenamefont {Vala},
  \citenamefont {Whaley},\ and\ \citenamefont {Weiss}}]{Vala05}%
  \BibitemOpen
  \bibfield  {author} {\bibinfo {author} {\bibfnamefont {J.}~\bibnamefont
  {Vala}}, \bibinfo {author} {\bibfnamefont {K.~B.}\ \bibnamefont {Whaley}},\
  and\ \bibinfo {author} {\bibfnamefont {D.~S.}\ \bibnamefont {Weiss}},\
  }\bibfield  {title} {\bibinfo {title} {Quantum error correction of a qubit
  loss in an addressable atomic system},\ }\href
  {https://doi.org/10.1103/PhysRevA.72.052318} {\bibfield  {journal} {\bibinfo
  {journal} {Phys. Rev. A}\ }\textbf {\bibinfo {volume} {72}},\ \bibinfo
  {pages} {052318} (\bibinfo {year} {2005})}\BibitemShut {NoStop}%
\bibitem [{\citenamefont {Wu}\ \emph {et~al.}(2022)\citenamefont {Wu},
  \citenamefont {Kolkowitz}, \citenamefont {Puri},\ and\ \citenamefont
  {Thompson}}]{Wu22}%
  \BibitemOpen
  \bibfield  {author} {\bibinfo {author} {\bibfnamefont {Y.}~\bibnamefont
  {Wu}}, \bibinfo {author} {\bibfnamefont {S.}~\bibnamefont {Kolkowitz}},
  \bibinfo {author} {\bibfnamefont {S.}~\bibnamefont {Puri}},\ and\ \bibinfo
  {author} {\bibfnamefont {J.}~\bibnamefont {Thompson}},\ }\bibfield  {title}
  {\bibinfo {title} {Erasure conversion for fault-tolerant quantum computing in
  alkaline earth rydberg atom arrays},\ }\href
  {https://doi.org/10.1038/s41467-022-32094-6} {\bibfield  {journal} {\bibinfo
  {journal} {Nature Communications}\ }\textbf {\bibinfo {volume} {13}},\
  \bibinfo {pages} {4657} (\bibinfo {year} {2022})}\BibitemShut {NoStop}%
\bibitem [{\citenamefont {Kubica}\ \emph {et~al.}(2022)\citenamefont {Kubica},
  \citenamefont {Haim}, \citenamefont {Vaknin}, \citenamefont {Brand{\~a}o},\
  and\ \citenamefont {Retzker}}]{Kubica22}%
  \BibitemOpen
  \bibfield  {author} {\bibinfo {author} {\bibfnamefont {A.}~\bibnamefont
  {Kubica}}, \bibinfo {author} {\bibfnamefont {A.}~\bibnamefont {Haim}},
  \bibinfo {author} {\bibfnamefont {Y.}~\bibnamefont {Vaknin}}, \bibinfo
  {author} {\bibfnamefont {F.}~\bibnamefont {Brand{\~a}o}},\ and\ \bibinfo
  {author} {\bibfnamefont {A.}~\bibnamefont {Retzker}},\ }\bibfield  {title}
  {\bibinfo {title} {Erasure qubits: Overcoming the {$T_1$} limit in
  superconducting circuits},\ }\href@noop {} {\bibfield  {journal} {\bibinfo
  {journal} {arXiv preprint arXiv:2208.05461}\ } (\bibinfo {year}
  {2022})}\BibitemShut {NoStop}%
\bibitem [{\citenamefont {Ghosh}\ \emph {et~al.}(2013)\citenamefont {Ghosh},
  \citenamefont {Fowler}, \citenamefont {Martinis},\ and\ \citenamefont
  {Geller}}]{Ghosh13}%
  \BibitemOpen
  \bibfield  {author} {\bibinfo {author} {\bibfnamefont {J.}~\bibnamefont
  {Ghosh}}, \bibinfo {author} {\bibfnamefont {A.~G.}\ \bibnamefont {Fowler}},
  \bibinfo {author} {\bibfnamefont {J.~M.}\ \bibnamefont {Martinis}},\ and\
  \bibinfo {author} {\bibfnamefont {M.~R.}\ \bibnamefont {Geller}},\ }\bibfield
   {title} {\bibinfo {title} {Understanding the effects of leakage in
  superconducting quantum-error-detection circuits},\ }\href
  {https://doi.org/10.1103/PhysRevA.88.062329} {\bibfield  {journal} {\bibinfo
  {journal} {Phys. Rev. A}\ }\textbf {\bibinfo {volume} {88}},\ \bibinfo
  {pages} {062329} (\bibinfo {year} {2013})}\BibitemShut {NoStop}%
\bibitem [{\citenamefont {Wu}\ \emph {et~al.}(2002)\citenamefont {Wu},
  \citenamefont {Byrd},\ and\ \citenamefont {Lidar}}]{Wu02}%
  \BibitemOpen
  \bibfield  {author} {\bibinfo {author} {\bibfnamefont {L.-A.}\ \bibnamefont
  {Wu}}, \bibinfo {author} {\bibfnamefont {M.~S.}\ \bibnamefont {Byrd}},\ and\
  \bibinfo {author} {\bibfnamefont {D.~A.}\ \bibnamefont {Lidar}},\ }\bibfield
  {title} {\bibinfo {title} {Efficient universal leakage elimination for
  physical and encoded qubits},\ }\href
  {https://doi.org/10.1103/PhysRevLett.89.127901} {\bibfield  {journal}
  {\bibinfo  {journal} {Phys. Rev. Lett.}\ }\textbf {\bibinfo {volume} {89}},\
  \bibinfo {pages} {127901} (\bibinfo {year} {2002})}\BibitemShut {NoStop}%
\bibitem [{\citenamefont {Byrd}\ \emph {et~al.}(2004)\citenamefont {Byrd},
  \citenamefont {Wu},\ and\ \citenamefont {Lidar}}]{Byrd04}%
  \BibitemOpen
  \bibfield  {author} {\bibinfo {author} {\bibfnamefont {M.~S.}\ \bibnamefont
  {Byrd}}, \bibinfo {author} {\bibfnamefont {L.-A.}\ \bibnamefont {Wu}},\ and\
  \bibinfo {author} {\bibfnamefont {D.~A.}\ \bibnamefont {Lidar}},\ }\bibfield
  {title} {\bibinfo {title} {Overview of quantum error prevention and leakage
  elimination},\ }\href {https://doi.org/10.1080/09500340408231803} {\bibfield
  {journal} {\bibinfo  {journal} {Journal of Modern Optics}\ }\textbf {\bibinfo
  {volume} {51}},\ \bibinfo {pages} {2449} (\bibinfo {year}
  {2004})}\BibitemShut {NoStop}%
\bibitem [{\citenamefont {Byrd}\ \emph {et~al.}(2005)\citenamefont {Byrd},
  \citenamefont {Lidar}, \citenamefont {Wu},\ and\ \citenamefont
  {Zanardi}}]{Byrd05}%
  \BibitemOpen
  \bibfield  {author} {\bibinfo {author} {\bibfnamefont {M.~S.}\ \bibnamefont
  {Byrd}}, \bibinfo {author} {\bibfnamefont {D.~A.}\ \bibnamefont {Lidar}},
  \bibinfo {author} {\bibfnamefont {L.-A.}\ \bibnamefont {Wu}},\ and\ \bibinfo
  {author} {\bibfnamefont {P.}~\bibnamefont {Zanardi}},\ }\bibfield  {title}
  {\bibinfo {title} {Universal leakage elimination},\ }\href
  {https://doi.org/10.1103/PhysRevA.71.052301} {\bibfield  {journal} {\bibinfo
  {journal} {Phys. Rev. A}\ }\textbf {\bibinfo {volume} {71}},\ \bibinfo
  {pages} {052301} (\bibinfo {year} {2005})}\BibitemShut {NoStop}%
\bibitem [{\citenamefont {Fowler}(2013)}]{Fowler13}%
  \BibitemOpen
  \bibfield  {author} {\bibinfo {author} {\bibfnamefont {A.~G.}\ \bibnamefont
  {Fowler}},\ }\bibfield  {title} {\bibinfo {title} {Coping with qubit leakage
  in topological codes},\ }\href {https://doi.org/10.1103/PhysRevA.88.042308}
  {\bibfield  {journal} {\bibinfo  {journal} {Phys. Rev. A}\ }\textbf {\bibinfo
  {volume} {88}},\ \bibinfo {pages} {042308} (\bibinfo {year}
  {2013})}\BibitemShut {NoStop}%
\bibitem [{\citenamefont {Ghosh}\ and\ \citenamefont {Fowler}(2015)}]{Ghosh15}%
  \BibitemOpen
  \bibfield  {author} {\bibinfo {author} {\bibfnamefont {J.}~\bibnamefont
  {Ghosh}}\ and\ \bibinfo {author} {\bibfnamefont {A.~G.}\ \bibnamefont
  {Fowler}},\ }\bibfield  {title} {\bibinfo {title} {Leakage-resilient approach
  to fault-tolerant quantum computing with superconducting elements},\ }\href
  {https://doi.org/10.1103/PhysRevA.91.020302} {\bibfield  {journal} {\bibinfo
  {journal} {Phys. Rev. A}\ }\textbf {\bibinfo {volume} {91}},\ \bibinfo
  {pages} {020302(R)} (\bibinfo {year} {2015})}\BibitemShut {NoStop}%
\bibitem [{\citenamefont {Suchara}\ \emph {et~al.}(2015)\citenamefont
  {Suchara}, \citenamefont {Cross},\ and\ \citenamefont
  {Gambetta}}]{Suchara15}%
  \BibitemOpen
  \bibfield  {author} {\bibinfo {author} {\bibfnamefont {M.}~\bibnamefont
  {Suchara}}, \bibinfo {author} {\bibfnamefont {A.~W.}\ \bibnamefont {Cross}},\
  and\ \bibinfo {author} {\bibfnamefont {J.~M.}\ \bibnamefont {Gambetta}},\
  }\bibfield  {title} {\bibinfo {title} {Leakage suppression in the toric
  code},\ }in\ \href {https://doi.org/10.1109/ISIT.2015.7282629} {\emph
  {\bibinfo {booktitle} {2015 IEEE International Symposium on Information
  Theory (ISIT)}}}\ (\bibinfo {year} {2015})\ pp.\ \bibinfo {pages}
  {1119--1123}\BibitemShut {NoStop}%
\bibitem [{\citenamefont {Brown}\ and\ \citenamefont {Brown}(2018)}]{Brown18}%
  \BibitemOpen
  \bibfield  {author} {\bibinfo {author} {\bibfnamefont {N.~C.}\ \bibnamefont
  {Brown}}\ and\ \bibinfo {author} {\bibfnamefont {K.~R.}\ \bibnamefont
  {Brown}},\ }\bibfield  {title} {\bibinfo {title} {Comparing zeeman qubits to
  hyperfine qubits in the context of the surface code: $^{174}\mathrm{Yb}^{+}$
  and $^{171}\mathrm{Yb}^{+}$},\ }\href
  {https://doi.org/10.1103/PhysRevA.97.052301} {\bibfield  {journal} {\bibinfo
  {journal} {Phys. Rev. A}\ }\textbf {\bibinfo {volume} {97}},\ \bibinfo
  {pages} {052301} (\bibinfo {year} {2018})}\BibitemShut {NoStop}%
\bibitem [{\citenamefont {Brown}\ and\ \citenamefont {Brown}(2019)}]{Brown19}%
  \BibitemOpen
  \bibfield  {author} {\bibinfo {author} {\bibfnamefont {N.~C.}\ \bibnamefont
  {Brown}}\ and\ \bibinfo {author} {\bibfnamefont {K.~R.}\ \bibnamefont
  {Brown}},\ }\bibfield  {title} {\bibinfo {title} {Leakage mitigation for
  quantum error correction using a mixed qubit scheme},\ }\href
  {https://doi.org/10.1103/PhysRevA.100.032325} {\bibfield  {journal} {\bibinfo
   {journal} {Phys. Rev. A}\ }\textbf {\bibinfo {volume} {100}},\ \bibinfo
  {pages} {032325} (\bibinfo {year} {2019})}\BibitemShut {NoStop}%
\bibitem [{\citenamefont {Brown}\ \emph {et~al.}(2019)\citenamefont {Brown},
  \citenamefont {Newman},\ and\ \citenamefont {Brown}}]{Brown20}%
  \BibitemOpen
  \bibfield  {author} {\bibinfo {author} {\bibfnamefont {N.~C.}\ \bibnamefont
  {Brown}}, \bibinfo {author} {\bibfnamefont {M.}~\bibnamefont {Newman}},\ and\
  \bibinfo {author} {\bibfnamefont {K.~R.}\ \bibnamefont {Brown}},\ }\bibfield
  {title} {\bibinfo {title} {Handling leakage with subsystem codes},\
  }\href@noop {} {\bibfield  {journal} {\bibinfo  {journal} {New Journal of
  Physics}\ }\textbf {\bibinfo {volume} {21}},\ \bibinfo {pages} {073055}
  (\bibinfo {year} {2019})}\BibitemShut {NoStop}%
\bibitem [{\citenamefont {Brown}\ \emph {et~al.}(2016)\citenamefont {Brown},
  \citenamefont {Kim},\ and\ \citenamefont {Monroe}}]{Brown16}%
  \BibitemOpen
  \bibfield  {author} {\bibinfo {author} {\bibfnamefont {K.~R.}\ \bibnamefont
  {Brown}}, \bibinfo {author} {\bibfnamefont {J.}~\bibnamefont {Kim}},\ and\
  \bibinfo {author} {\bibfnamefont {C.}~\bibnamefont {Monroe}},\ }\bibfield
  {title} {\bibinfo {title} {Co-designing a scalable quantum computer with
  trapped atomic ions},\ }\href@noop {} {\bibfield  {journal} {\bibinfo
  {journal} {npj Quantum Information}\ }\textbf {\bibinfo {volume} {2}},\
  \bibinfo {pages} {1} (\bibinfo {year} {2016})}\BibitemShut {NoStop}%
\bibitem [{\citenamefont {Schindler}\ \emph {et~al.}(2011)\citenamefont
  {Schindler}, \citenamefont {Barreiro}, \citenamefont {Monz}, \citenamefont
  {Nebendahl}, \citenamefont {Nigg}, \citenamefont {Chwalla}, \citenamefont
  {Hennrich},\ and\ \citenamefont {Blatt}}]{Schindler11}%
  \BibitemOpen
  \bibfield  {author} {\bibinfo {author} {\bibfnamefont {P.}~\bibnamefont
  {Schindler}}, \bibinfo {author} {\bibfnamefont {J.~T.}\ \bibnamefont
  {Barreiro}}, \bibinfo {author} {\bibfnamefont {T.}~\bibnamefont {Monz}},
  \bibinfo {author} {\bibfnamefont {V.}~\bibnamefont {Nebendahl}}, \bibinfo
  {author} {\bibfnamefont {D.}~\bibnamefont {Nigg}}, \bibinfo {author}
  {\bibfnamefont {M.}~\bibnamefont {Chwalla}}, \bibinfo {author} {\bibfnamefont
  {M.}~\bibnamefont {Hennrich}},\ and\ \bibinfo {author} {\bibfnamefont
  {R.}~\bibnamefont {Blatt}},\ }\bibfield  {title} {\bibinfo {title}
  {Experimental repetitive quantum error correction},\ }\href@noop {}
  {\bibfield  {journal} {\bibinfo  {journal} {Science}\ }\textbf {\bibinfo
  {volume} {332}},\ \bibinfo {pages} {1059} (\bibinfo {year}
  {2011})}\BibitemShut {NoStop}%
\bibitem [{\citenamefont {Nigg}\ \emph {et~al.}(2014)\citenamefont {Nigg},
  \citenamefont {Mueller}, \citenamefont {Martinez}, \citenamefont {Schindler},
  \citenamefont {Hennrich}, \citenamefont {Monz}, \citenamefont
  {Martin-Delgado},\ and\ \citenamefont {Blatt}}]{Nigg14}%
  \BibitemOpen
  \bibfield  {author} {\bibinfo {author} {\bibfnamefont {D.}~\bibnamefont
  {Nigg}}, \bibinfo {author} {\bibfnamefont {M.}~\bibnamefont {Mueller}},
  \bibinfo {author} {\bibfnamefont {E.~A.}\ \bibnamefont {Martinez}}, \bibinfo
  {author} {\bibfnamefont {P.}~\bibnamefont {Schindler}}, \bibinfo {author}
  {\bibfnamefont {M.}~\bibnamefont {Hennrich}}, \bibinfo {author}
  {\bibfnamefont {T.}~\bibnamefont {Monz}}, \bibinfo {author} {\bibfnamefont
  {M.~A.}\ \bibnamefont {Martin-Delgado}},\ and\ \bibinfo {author}
  {\bibfnamefont {R.}~\bibnamefont {Blatt}},\ }\bibfield  {title} {\bibinfo
  {title} {Quantum computations on a topologically encoded qubit},\ }\href@noop
  {} {\bibfield  {journal} {\bibinfo  {journal} {Science}\ }\textbf {\bibinfo
  {volume} {345}},\ \bibinfo {pages} {302} (\bibinfo {year}
  {2014})}\BibitemShut {NoStop}%
\bibitem [{\citenamefont {Linke}\ \emph {et~al.}(2017)\citenamefont {Linke},
  \citenamefont {Gutierrez}, \citenamefont {Landsman}, \citenamefont {Figgatt},
  \citenamefont {Debnath}, \citenamefont {Brown},\ and\ \citenamefont
  {Monroe}}]{Linke17}%
  \BibitemOpen
  \bibfield  {author} {\bibinfo {author} {\bibfnamefont {N.~M.}\ \bibnamefont
  {Linke}}, \bibinfo {author} {\bibfnamefont {M.}~\bibnamefont {Gutierrez}},
  \bibinfo {author} {\bibfnamefont {K.~A.}\ \bibnamefont {Landsman}}, \bibinfo
  {author} {\bibfnamefont {C.}~\bibnamefont {Figgatt}}, \bibinfo {author}
  {\bibfnamefont {S.}~\bibnamefont {Debnath}}, \bibinfo {author} {\bibfnamefont
  {K.~R.}\ \bibnamefont {Brown}},\ and\ \bibinfo {author} {\bibfnamefont
  {C.}~\bibnamefont {Monroe}},\ }\bibfield  {title} {\bibinfo {title}
  {Fault-tolerant quantum error detection},\ }\href@noop {} {\bibfield
  {journal} {\bibinfo  {journal} {Science advances}\ }\textbf {\bibinfo
  {volume} {3}},\ \bibinfo {pages} {e1701074} (\bibinfo {year}
  {2017})}\BibitemShut {NoStop}%
\bibitem [{\citenamefont {Egan}\ \emph {et~al.}(2021)\citenamefont {Egan},
  \citenamefont {Debroy}, \citenamefont {Noel}, \citenamefont {Risinger},
  \citenamefont {Zhu}, \citenamefont {Biswas}, \citenamefont {Newman},
  \citenamefont {Li}, \citenamefont {Brown}, \citenamefont {Cetina},\ and\
  \citenamefont {Monroe}}]{Egan21}%
  \BibitemOpen
  \bibfield  {author} {\bibinfo {author} {\bibfnamefont {L.}~\bibnamefont
  {Egan}}, \bibinfo {author} {\bibfnamefont {D.~M.}\ \bibnamefont {Debroy}},
  \bibinfo {author} {\bibfnamefont {C.}~\bibnamefont {Noel}}, \bibinfo {author}
  {\bibfnamefont {A.}~\bibnamefont {Risinger}}, \bibinfo {author}
  {\bibfnamefont {D.}~\bibnamefont {Zhu}}, \bibinfo {author} {\bibfnamefont
  {D.}~\bibnamefont {Biswas}}, \bibinfo {author} {\bibfnamefont
  {M.}~\bibnamefont {Newman}}, \bibinfo {author} {\bibfnamefont
  {M.}~\bibnamefont {Li}}, \bibinfo {author} {\bibfnamefont {K.~R.}\
  \bibnamefont {Brown}}, \bibinfo {author} {\bibfnamefont {M.}~\bibnamefont
  {Cetina}},\ and\ \bibinfo {author} {\bibfnamefont {C.}~\bibnamefont
  {Monroe}},\ }\bibfield  {title} {\bibinfo {title} {Fault-tolerant control of
  an error-corrected qubit},\ }\href
  {https://doi.org/10.1038/s41586-021-03928-y} {\bibfield  {journal} {\bibinfo
  {journal} {Nature}\ }\textbf {\bibinfo {volume} {598}},\ \bibinfo {pages}
  {281} (\bibinfo {year} {2021})}\BibitemShut {NoStop}%
\bibitem [{\citenamefont {Ryan-Anderson}\ \emph {et~al.}(2021)\citenamefont
  {Ryan-Anderson}, \citenamefont {Bohnet}, \citenamefont {Lee}, \citenamefont
  {Gresh}, \citenamefont {Hankin}, \citenamefont {Gaebler}, \citenamefont
  {Francois}, \citenamefont {Chernoguzov}, \citenamefont {Lucchetti},
  \citenamefont {Brown}, \citenamefont {Gatterman}, \citenamefont {Halit},
  \citenamefont {Gilmore}, \citenamefont {Gerber}, \citenamefont {Neyenhuis},
  \citenamefont {Hayes},\ and\ \citenamefont {Stutz}}]{RyanAnderson21}%
  \BibitemOpen
  \bibfield  {author} {\bibinfo {author} {\bibfnamefont {C.}~\bibnamefont
  {Ryan-Anderson}}, \bibinfo {author} {\bibfnamefont {J.~G.}\ \bibnamefont
  {Bohnet}}, \bibinfo {author} {\bibfnamefont {K.}~\bibnamefont {Lee}},
  \bibinfo {author} {\bibfnamefont {D.}~\bibnamefont {Gresh}}, \bibinfo
  {author} {\bibfnamefont {A.}~\bibnamefont {Hankin}}, \bibinfo {author}
  {\bibfnamefont {J.~P.}\ \bibnamefont {Gaebler}}, \bibinfo {author}
  {\bibfnamefont {D.}~\bibnamefont {Francois}}, \bibinfo {author}
  {\bibfnamefont {A.}~\bibnamefont {Chernoguzov}}, \bibinfo {author}
  {\bibfnamefont {D.}~\bibnamefont {Lucchetti}}, \bibinfo {author}
  {\bibfnamefont {N.~C.}\ \bibnamefont {Brown}}, \bibinfo {author}
  {\bibfnamefont {T.~M.}\ \bibnamefont {Gatterman}}, \bibinfo {author}
  {\bibfnamefont {S.~K.}\ \bibnamefont {Halit}}, \bibinfo {author}
  {\bibfnamefont {K.}~\bibnamefont {Gilmore}}, \bibinfo {author} {\bibfnamefont
  {J.~A.}\ \bibnamefont {Gerber}}, \bibinfo {author} {\bibfnamefont
  {B.}~\bibnamefont {Neyenhuis}}, \bibinfo {author} {\bibfnamefont
  {D.}~\bibnamefont {Hayes}},\ and\ \bibinfo {author} {\bibfnamefont {R.~P.}\
  \bibnamefont {Stutz}},\ }\bibfield  {title} {\bibinfo {title} {Realization of
  real-time fault-tolerant quantum error correction},\ }\href
  {https://doi.org/10.1103/PhysRevX.11.041058} {\bibfield  {journal} {\bibinfo
  {journal} {Phys. Rev. X}\ }\textbf {\bibinfo {volume} {11}},\ \bibinfo
  {pages} {041058} (\bibinfo {year} {2021})}\BibitemShut {NoStop}%
\bibitem [{\citenamefont {Ryan-Anderson}\ \emph {et~al.}(2022)\citenamefont
  {Ryan-Anderson}, \citenamefont {Brown}, \citenamefont {Allman}, \citenamefont
  {Arkin}, \citenamefont {Asa-Attuah}, \citenamefont {Baldwin}, \citenamefont
  {Berg}, \citenamefont {Bohnet}, \citenamefont {Braxton}, \citenamefont
  {Burdick}, \citenamefont {Campora}, \citenamefont {Chernoguzov},
  \citenamefont {Esposito}, \citenamefont {Evans}, \citenamefont {Francois},
  \citenamefont {Gaebler}, \citenamefont {Gatterman}, \citenamefont {Gerber},
  \citenamefont {Gilmore}, \citenamefont {Gresh}, \citenamefont {Hall},
  \citenamefont {Hankin}, \citenamefont {Hostetter}, \citenamefont {Lucchetti},
  \citenamefont {Mayer}, \citenamefont {Myers}, \citenamefont {Neyenhuis},
  \citenamefont {Santiago}, \citenamefont {Sedlacek}, \citenamefont {Skripka},
  \citenamefont {Slattery}, \citenamefont {Stutz}, \citenamefont {Tait},
  \citenamefont {Tobey}, \citenamefont {Vittorini}, \citenamefont {Walker},\
  and\ \citenamefont {Hayes}}]{RyanAnderson22}%
  \BibitemOpen
  \bibfield  {author} {\bibinfo {author} {\bibfnamefont {C.}~\bibnamefont
  {Ryan-Anderson}}, \bibinfo {author} {\bibfnamefont {N.~C.}\ \bibnamefont
  {Brown}}, \bibinfo {author} {\bibfnamefont {M.~S.}\ \bibnamefont {Allman}},
  \bibinfo {author} {\bibfnamefont {B.}~\bibnamefont {Arkin}}, \bibinfo
  {author} {\bibfnamefont {G.}~\bibnamefont {Asa-Attuah}}, \bibinfo {author}
  {\bibfnamefont {C.}~\bibnamefont {Baldwin}}, \bibinfo {author} {\bibfnamefont
  {J.}~\bibnamefont {Berg}}, \bibinfo {author} {\bibfnamefont {J.~G.}\
  \bibnamefont {Bohnet}}, \bibinfo {author} {\bibfnamefont {S.}~\bibnamefont
  {Braxton}}, \bibinfo {author} {\bibfnamefont {N.}~\bibnamefont {Burdick}},
  \bibinfo {author} {\bibfnamefont {J.~P.}\ \bibnamefont {Campora}}, \bibinfo
  {author} {\bibfnamefont {A.}~\bibnamefont {Chernoguzov}}, \bibinfo {author}
  {\bibfnamefont {J.}~\bibnamefont {Esposito}}, \bibinfo {author}
  {\bibfnamefont {B.}~\bibnamefont {Evans}}, \bibinfo {author} {\bibfnamefont
  {D.}~\bibnamefont {Francois}}, \bibinfo {author} {\bibfnamefont {J.~P.}\
  \bibnamefont {Gaebler}}, \bibinfo {author} {\bibfnamefont {T.~M.}\
  \bibnamefont {Gatterman}}, \bibinfo {author} {\bibfnamefont {J.}~\bibnamefont
  {Gerber}}, \bibinfo {author} {\bibfnamefont {K.}~\bibnamefont {Gilmore}},
  \bibinfo {author} {\bibfnamefont {D.}~\bibnamefont {Gresh}}, \bibinfo
  {author} {\bibfnamefont {A.}~\bibnamefont {Hall}}, \bibinfo {author}
  {\bibfnamefont {A.}~\bibnamefont {Hankin}}, \bibinfo {author} {\bibfnamefont
  {J.}~\bibnamefont {Hostetter}}, \bibinfo {author} {\bibfnamefont
  {D.}~\bibnamefont {Lucchetti}}, \bibinfo {author} {\bibfnamefont
  {K.}~\bibnamefont {Mayer}}, \bibinfo {author} {\bibfnamefont
  {J.}~\bibnamefont {Myers}}, \bibinfo {author} {\bibfnamefont
  {B.}~\bibnamefont {Neyenhuis}}, \bibinfo {author} {\bibfnamefont
  {J.}~\bibnamefont {Santiago}}, \bibinfo {author} {\bibfnamefont
  {J.}~\bibnamefont {Sedlacek}}, \bibinfo {author} {\bibfnamefont
  {T.}~\bibnamefont {Skripka}}, \bibinfo {author} {\bibfnamefont
  {A.}~\bibnamefont {Slattery}}, \bibinfo {author} {\bibfnamefont {R.~P.}\
  \bibnamefont {Stutz}}, \bibinfo {author} {\bibfnamefont {J.}~\bibnamefont
  {Tait}}, \bibinfo {author} {\bibfnamefont {R.}~\bibnamefont {Tobey}},
  \bibinfo {author} {\bibfnamefont {G.}~\bibnamefont {Vittorini}}, \bibinfo
  {author} {\bibfnamefont {J.}~\bibnamefont {Walker}},\ and\ \bibinfo {author}
  {\bibfnamefont {D.}~\bibnamefont {Hayes}},\ }\bibfield  {title} {\bibinfo
  {title} {Implementing fault-tolerant entangling gates on the five-qubit code
  and the color code},\ }\href@noop {} {\bibfield  {journal} {\bibinfo
  {journal} {arXiv preprint arXiv:2208.01863}\ } (\bibinfo {year}
  {2022})}\BibitemShut {NoStop}%
\bibitem [{\citenamefont {Postler}\ \emph {et~al.}(2022)\citenamefont
  {Postler}, \citenamefont {Heu$\beta$en}, \citenamefont {Pogorelov},
  \citenamefont {Rispler}, \citenamefont {Feldker}, \citenamefont {Meth},
  \citenamefont {Marciniak}, \citenamefont {Stricker}, \citenamefont
  {Ringbauer}, \citenamefont {Blatt} \emph {et~al.}}]{Postler22}%
  \BibitemOpen
  \bibfield  {author} {\bibinfo {author} {\bibfnamefont {L.}~\bibnamefont
  {Postler}}, \bibinfo {author} {\bibfnamefont {S.}~\bibnamefont
  {Heu$\beta$en}}, \bibinfo {author} {\bibfnamefont {I.}~\bibnamefont
  {Pogorelov}}, \bibinfo {author} {\bibfnamefont {M.}~\bibnamefont {Rispler}},
  \bibinfo {author} {\bibfnamefont {T.}~\bibnamefont {Feldker}}, \bibinfo
  {author} {\bibfnamefont {M.}~\bibnamefont {Meth}}, \bibinfo {author}
  {\bibfnamefont {C.~D.}\ \bibnamefont {Marciniak}}, \bibinfo {author}
  {\bibfnamefont {R.}~\bibnamefont {Stricker}}, \bibinfo {author}
  {\bibfnamefont {M.}~\bibnamefont {Ringbauer}}, \bibinfo {author}
  {\bibfnamefont {R.}~\bibnamefont {Blatt}}, \emph {et~al.},\ }\bibfield
  {title} {\bibinfo {title} {Demonstration of fault-tolerant universal quantum
  gate operations},\ }\href@noop {} {\bibfield  {journal} {\bibinfo  {journal}
  {Nature}\ }\textbf {\bibinfo {volume} {605}},\ \bibinfo {pages} {675}
  (\bibinfo {year} {2022})}\BibitemShut {NoStop}%
\bibitem [{\citenamefont {Erhard}\ \emph {et~al.}(2021)\citenamefont {Erhard},
  \citenamefont {Poulsen~Nautrup}, \citenamefont {Meth}, \citenamefont
  {Postler}, \citenamefont {Stricker}, \citenamefont {Stadler}, \citenamefont
  {Negnevitsky}, \citenamefont {Ringbauer}, \citenamefont {Schindler},
  \citenamefont {Briegel} \emph {et~al.}}]{Erhard21}%
  \BibitemOpen
  \bibfield  {author} {\bibinfo {author} {\bibfnamefont {A.}~\bibnamefont
  {Erhard}}, \bibinfo {author} {\bibfnamefont {H.}~\bibnamefont
  {Poulsen~Nautrup}}, \bibinfo {author} {\bibfnamefont {M.}~\bibnamefont
  {Meth}}, \bibinfo {author} {\bibfnamefont {L.}~\bibnamefont {Postler}},
  \bibinfo {author} {\bibfnamefont {R.}~\bibnamefont {Stricker}}, \bibinfo
  {author} {\bibfnamefont {M.}~\bibnamefont {Stadler}}, \bibinfo {author}
  {\bibfnamefont {V.}~\bibnamefont {Negnevitsky}}, \bibinfo {author}
  {\bibfnamefont {M.}~\bibnamefont {Ringbauer}}, \bibinfo {author}
  {\bibfnamefont {P.}~\bibnamefont {Schindler}}, \bibinfo {author}
  {\bibfnamefont {H.~J.}\ \bibnamefont {Briegel}}, \emph {et~al.},\ }\bibfield
  {title} {\bibinfo {title} {Entangling logical qubits with lattice surgery},\
  }\href@noop {} {\bibfield  {journal} {\bibinfo  {journal} {Nature}\ }\textbf
  {\bibinfo {volume} {589}},\ \bibinfo {pages} {220} (\bibinfo {year}
  {2021})}\BibitemShut {NoStop}%
\bibitem [{\citenamefont {Campbell}(2020)}]{Campbell20}%
  \BibitemOpen
  \bibfield  {author} {\bibinfo {author} {\bibfnamefont {W.~C.}\ \bibnamefont
  {Campbell}},\ }\bibfield  {title} {\bibinfo {title} {Certified quantum
  gates},\ }\href {https://doi.org/10.1103/PhysRevA.102.022426} {\bibfield
  {journal} {\bibinfo  {journal} {Phys. Rev. A}\ }\textbf {\bibinfo {volume}
  {102}},\ \bibinfo {pages} {022426} (\bibinfo {year} {2020})}\BibitemShut
  {NoStop}%
\bibitem [{\citenamefont {Allcock}\ \emph {et~al.}(2021)\citenamefont
  {Allcock}, \citenamefont {Campbell}, \citenamefont {Chiaverini},
  \citenamefont {Chuang}, \citenamefont {Hudson}, \citenamefont {Moore},
  \citenamefont {Ransford}, \citenamefont {Roman}, \citenamefont {Sage},\ and\
  \citenamefont {Wineland}}]{Allcock21}%
  \BibitemOpen
  \bibfield  {author} {\bibinfo {author} {\bibfnamefont {D.~T.~C.}\
  \bibnamefont {Allcock}}, \bibinfo {author} {\bibfnamefont {W.~C.}\
  \bibnamefont {Campbell}}, \bibinfo {author} {\bibfnamefont {J.}~\bibnamefont
  {Chiaverini}}, \bibinfo {author} {\bibfnamefont {I.~L.}\ \bibnamefont
  {Chuang}}, \bibinfo {author} {\bibfnamefont {E.~R.}\ \bibnamefont {Hudson}},
  \bibinfo {author} {\bibfnamefont {I.~D.}\ \bibnamefont {Moore}}, \bibinfo
  {author} {\bibfnamefont {A.}~\bibnamefont {Ransford}}, \bibinfo {author}
  {\bibfnamefont {C.}~\bibnamefont {Roman}}, \bibinfo {author} {\bibfnamefont
  {J.~M.}\ \bibnamefont {Sage}},\ and\ \bibinfo {author} {\bibfnamefont
  {D.~J.}\ \bibnamefont {Wineland}},\ }\bibfield  {title} {\bibinfo {title}
  {omg blueprint for trapped ion quantum computing with metastable states},\
  }\href {https://doi.org/10.1063/5.0069544} {\bibfield  {journal} {\bibinfo
  {journal} {Applied Physics Letters}\ }\textbf {\bibinfo {volume} {119}},\
  \bibinfo {pages} {214002} (\bibinfo {year} {2021})}\BibitemShut {NoStop}%
\bibitem [{\citenamefont {Yang}\ \emph {et~al.}(2022)\citenamefont {Yang},
  \citenamefont {Ma}, \citenamefont {Wu}, \citenamefont {Wang}, \citenamefont
  {Cao}, \citenamefont {Guo}, \citenamefont {Huang}, \citenamefont {Feng},
  \citenamefont {Zhou},\ and\ \citenamefont {Duan}}]{Yang22}%
  \BibitemOpen
  \bibfield  {author} {\bibinfo {author} {\bibfnamefont {H.-X.}\ \bibnamefont
  {Yang}}, \bibinfo {author} {\bibfnamefont {J.-Y.}\ \bibnamefont {Ma}},
  \bibinfo {author} {\bibfnamefont {Y.-K.}\ \bibnamefont {Wu}}, \bibinfo
  {author} {\bibfnamefont {Y.}~\bibnamefont {Wang}}, \bibinfo {author}
  {\bibfnamefont {M.-M.}\ \bibnamefont {Cao}}, \bibinfo {author} {\bibfnamefont
  {W.-X.}\ \bibnamefont {Guo}}, \bibinfo {author} {\bibfnamefont {Y.-Y.}\
  \bibnamefont {Huang}}, \bibinfo {author} {\bibfnamefont {L.}~\bibnamefont
  {Feng}}, \bibinfo {author} {\bibfnamefont {Z.-C.}\ \bibnamefont {Zhou}},\
  and\ \bibinfo {author} {\bibfnamefont {L.-M.}\ \bibnamefont {Duan}},\
  }\bibfield  {title} {\bibinfo {title} {Realizing coherently convertible
  dual-type qubits with the same ion species},\ }\href@noop {} {\bibfield
  {journal} {\bibinfo  {journal} {Nature Physics}\ ,\ \bibinfo {pages} {1}}
  (\bibinfo {year} {2022})}\BibitemShut {NoStop}%
\bibitem [{\citenamefont {Ozeri}\ \emph {et~al.}(2007)\citenamefont {Ozeri},
  \citenamefont {Itano}, \citenamefont {Blakestad}, \citenamefont {Britton},
  \citenamefont {Chiaverini}, \citenamefont {Jost}, \citenamefont {Langer},
  \citenamefont {Leibfried}, \citenamefont {Reichle}, \citenamefont {Seidelin},
  \citenamefont {Wesenberg},\ and\ \citenamefont {Wineland}}]{Ozeri07}%
  \BibitemOpen
  \bibfield  {author} {\bibinfo {author} {\bibfnamefont {R.}~\bibnamefont
  {Ozeri}}, \bibinfo {author} {\bibfnamefont {W.~M.}\ \bibnamefont {Itano}},
  \bibinfo {author} {\bibfnamefont {R.~B.}\ \bibnamefont {Blakestad}}, \bibinfo
  {author} {\bibfnamefont {J.}~\bibnamefont {Britton}}, \bibinfo {author}
  {\bibfnamefont {J.}~\bibnamefont {Chiaverini}}, \bibinfo {author}
  {\bibfnamefont {J.~D.}\ \bibnamefont {Jost}}, \bibinfo {author}
  {\bibfnamefont {C.}~\bibnamefont {Langer}}, \bibinfo {author} {\bibfnamefont
  {D.}~\bibnamefont {Leibfried}}, \bibinfo {author} {\bibfnamefont
  {R.}~\bibnamefont {Reichle}}, \bibinfo {author} {\bibfnamefont
  {S.}~\bibnamefont {Seidelin}}, \bibinfo {author} {\bibfnamefont {J.~H.}\
  \bibnamefont {Wesenberg}},\ and\ \bibinfo {author} {\bibfnamefont {D.~J.}\
  \bibnamefont {Wineland}},\ }\bibfield  {title} {\bibinfo {title} {Errors in
  trapped-ion quantum gates due to spontaneous photon scattering},\ }\href
  {https://doi.org/10.1103/PhysRevA.75.042329} {\bibfield  {journal} {\bibinfo
  {journal} {Phys. Rev. A}\ }\textbf {\bibinfo {volume} {75}},\ \bibinfo
  {pages} {042329} (\bibinfo {year} {2007})}\BibitemShut {NoStop}%
\bibitem [{\citenamefont {Moore}\ \emph {et~al.}(2023)\citenamefont {Moore},
  \citenamefont {Campbell}, \citenamefont {Hudson}, \citenamefont
  {Boguslawski}, \citenamefont {Wineland},\ and\ \citenamefont
  {Allcock}}]{Moore23}%
  \BibitemOpen
  \bibfield  {author} {\bibinfo {author} {\bibfnamefont {I.~D.}\ \bibnamefont
  {Moore}}, \bibinfo {author} {\bibfnamefont {W.~C.}\ \bibnamefont {Campbell}},
  \bibinfo {author} {\bibfnamefont {E.~R.}\ \bibnamefont {Hudson}}, \bibinfo
  {author} {\bibfnamefont {M.~J.}\ \bibnamefont {Boguslawski}}, \bibinfo
  {author} {\bibfnamefont {D.~J.}\ \bibnamefont {Wineland}},\ and\ \bibinfo
  {author} {\bibfnamefont {D.~T.~C.}\ \bibnamefont {Allcock}},\ }\bibfield
  {title} {\bibinfo {title} {Photon scattering errors during stimulated raman
  transitions in trapped-ion qubits},\ }\href
  {https://doi.org/10.1103/PhysRevA.107.032413} {\bibfield  {journal} {\bibinfo
   {journal} {Phys. Rev. A}\ }\textbf {\bibinfo {volume} {107}},\ \bibinfo
  {pages} {032413} (\bibinfo {year} {2023})}\BibitemShut {NoStop}%
\bibitem [{\citenamefont {Uys}\ \emph {et~al.}(2010)\citenamefont {Uys},
  \citenamefont {Biercuk}, \citenamefont {VanDevender}, \citenamefont
  {Ospelkaus}, \citenamefont {Meiser}, \citenamefont {Ozeri},\ and\
  \citenamefont {Bollinger}}]{Uys10}%
  \BibitemOpen
  \bibfield  {author} {\bibinfo {author} {\bibfnamefont {H.}~\bibnamefont
  {Uys}}, \bibinfo {author} {\bibfnamefont {M.~J.}\ \bibnamefont {Biercuk}},
  \bibinfo {author} {\bibfnamefont {A.~P.}\ \bibnamefont {VanDevender}},
  \bibinfo {author} {\bibfnamefont {C.}~\bibnamefont {Ospelkaus}}, \bibinfo
  {author} {\bibfnamefont {D.}~\bibnamefont {Meiser}}, \bibinfo {author}
  {\bibfnamefont {R.}~\bibnamefont {Ozeri}},\ and\ \bibinfo {author}
  {\bibfnamefont {J.~J.}\ \bibnamefont {Bollinger}},\ }\bibfield  {title}
  {\bibinfo {title} {Decoherence due to elastic rayleigh scattering},\ }\href
  {https://doi.org/10.1103/PhysRevLett.105.200401} {\bibfield  {journal}
  {\bibinfo  {journal} {Phys. Rev. Lett.}\ }\textbf {\bibinfo {volume} {105}},\
  \bibinfo {pages} {200401} (\bibinfo {year} {2010})}\BibitemShut {NoStop}%
\bibitem [{\citenamefont {Zhang}\ \emph {et~al.}(2020)\citenamefont {Zhang},
  \citenamefont {Arnold}, \citenamefont {Chanu}, \citenamefont {Kaewuam},
  \citenamefont {Safronova},\ and\ \citenamefont {Barrett}}]{Zhang20}%
  \BibitemOpen
  \bibfield  {author} {\bibinfo {author} {\bibfnamefont {Z.}~\bibnamefont
  {Zhang}}, \bibinfo {author} {\bibfnamefont {K.~J.}\ \bibnamefont {Arnold}},
  \bibinfo {author} {\bibfnamefont {S.~R.}\ \bibnamefont {Chanu}}, \bibinfo
  {author} {\bibfnamefont {R.}~\bibnamefont {Kaewuam}}, \bibinfo {author}
  {\bibfnamefont {M.~S.}\ \bibnamefont {Safronova}},\ and\ \bibinfo {author}
  {\bibfnamefont {M.~D.}\ \bibnamefont {Barrett}},\ }\bibfield  {title}
  {\bibinfo {title} {Branching fractions for ${P}_{3/2}$ decays in
  ${\mathrm{ba}}^{+}$},\ }\href {https://doi.org/10.1103/PhysRevA.101.062515}
  {\bibfield  {journal} {\bibinfo  {journal} {Phys. Rev. A}\ }\textbf {\bibinfo
  {volume} {101}},\ \bibinfo {pages} {062515} (\bibinfo {year}
  {2020})}\BibitemShut {NoStop}%
\bibitem [{\citenamefont {Kreuter}\ \emph {et~al.}(2004)\citenamefont
  {Kreuter}, \citenamefont {Becher}, \citenamefont {Lancaster}, \citenamefont
  {Mundt}, \citenamefont {Russo}, \citenamefont {H\"affner}, \citenamefont
  {Roos}, \citenamefont {Eschner}, \citenamefont {Schmidt-Kaler},\ and\
  \citenamefont {Blatt}}]{Kreuter04}%
  \BibitemOpen
  \bibfield  {author} {\bibinfo {author} {\bibfnamefont {A.}~\bibnamefont
  {Kreuter}}, \bibinfo {author} {\bibfnamefont {C.}~\bibnamefont {Becher}},
  \bibinfo {author} {\bibfnamefont {G.~P.~T.}\ \bibnamefont {Lancaster}},
  \bibinfo {author} {\bibfnamefont {A.~B.}\ \bibnamefont {Mundt}}, \bibinfo
  {author} {\bibfnamefont {C.}~\bibnamefont {Russo}}, \bibinfo {author}
  {\bibfnamefont {H.}~\bibnamefont {H\"affner}}, \bibinfo {author}
  {\bibfnamefont {C.}~\bibnamefont {Roos}}, \bibinfo {author} {\bibfnamefont
  {J.}~\bibnamefont {Eschner}}, \bibinfo {author} {\bibfnamefont
  {F.}~\bibnamefont {Schmidt-Kaler}},\ and\ \bibinfo {author} {\bibfnamefont
  {R.}~\bibnamefont {Blatt}},\ }\bibfield  {title} {\bibinfo {title}
  {Spontaneous emission lifetime of a single trapped ${\mathrm{ca}}^{+}$ ion in
  a high finesse cavity},\ }\href
  {https://doi.org/10.1103/PhysRevLett.92.203002} {\bibfield  {journal}
  {\bibinfo  {journal} {Phys. Rev. Lett.}\ }\textbf {\bibinfo {volume} {92}},\
  \bibinfo {pages} {203002} (\bibinfo {year} {2004})}\BibitemShut {NoStop}%
\bibitem [{\citenamefont {Jin}\ and\ \citenamefont {Church}(1993)}]{Jin93}%
  \BibitemOpen
  \bibfield  {author} {\bibinfo {author} {\bibfnamefont {J.}~\bibnamefont
  {Jin}}\ and\ \bibinfo {author} {\bibfnamefont {D.~A.}\ \bibnamefont
  {Church}},\ }\bibfield  {title} {\bibinfo {title} {Precision lifetimes for
  the {Ca}$^{+}$ 4p $^{2}${P} levels: Experiment challenges theory at the 1\%
  level},\ }\href {https://doi.org/10.1103/PhysRevLett.70.3213} {\bibfield
  {journal} {\bibinfo  {journal} {Phys. Rev. Lett.}\ }\textbf {\bibinfo
  {volume} {70}},\ \bibinfo {pages} {3213} (\bibinfo {year}
  {1993})}\BibitemShut {NoStop}%
\bibitem [{\citenamefont {Gerritsma}\ \emph {et~al.}(2008)\citenamefont
  {Gerritsma}, \citenamefont {Kirchmair}, \citenamefont {Z{\"a}hringer},
  \citenamefont {Benhelm}, \citenamefont {Blatt},\ and\ \citenamefont
  {Roos}}]{Gerritsma08}%
  \BibitemOpen
  \bibfield  {author} {\bibinfo {author} {\bibfnamefont {R.}~\bibnamefont
  {Gerritsma}}, \bibinfo {author} {\bibfnamefont {G.}~\bibnamefont
  {Kirchmair}}, \bibinfo {author} {\bibfnamefont {F.}~\bibnamefont
  {Z{\"a}hringer}}, \bibinfo {author} {\bibfnamefont {J.}~\bibnamefont
  {Benhelm}}, \bibinfo {author} {\bibfnamefont {R.}~\bibnamefont {Blatt}},\
  and\ \bibinfo {author} {\bibfnamefont {C.}~\bibnamefont {Roos}},\ }\bibfield
  {title} {\bibinfo {title} {Precision measurement of the branching fractions
  of the 4p $^2${P}$_{3/2}$ decay of {Ca} {II}},\ }\href@noop {} {\bibfield
  {journal} {\bibinfo  {journal} {The European Physical Journal D}\ }\textbf
  {\bibinfo {volume} {50}},\ \bibinfo {pages} {13} (\bibinfo {year}
  {2008})}\BibitemShut {NoStop}%
\bibitem [{\citenamefont {Knight}\ \emph {et~al.}(2003)\citenamefont {Knight},
  \citenamefont {Hinds}, \citenamefont {Plenio}, \citenamefont {Wineland},
  \citenamefont {Barrett}, \citenamefont {Britton}, \citenamefont {Chiaverini},
  \citenamefont {DeMarco}, \citenamefont {Itano}, \citenamefont {Jelenković},
  \citenamefont {Langer}, \citenamefont {Leibfried}, \citenamefont {Meyer},
  \citenamefont {Rosenband},\ and\ \citenamefont {Schätz}}]{Wineland03}%
  \BibitemOpen
  \bibfield  {author} {\bibinfo {author} {\bibfnamefont {P.~L.}\ \bibnamefont
  {Knight}}, \bibinfo {author} {\bibfnamefont {E.~A.}\ \bibnamefont {Hinds}},
  \bibinfo {author} {\bibfnamefont {M.~B.}\ \bibnamefont {Plenio}}, \bibinfo
  {author} {\bibfnamefont {D.~J.}\ \bibnamefont {Wineland}}, \bibinfo {author}
  {\bibfnamefont {M.}~\bibnamefont {Barrett}}, \bibinfo {author} {\bibfnamefont
  {J.}~\bibnamefont {Britton}}, \bibinfo {author} {\bibfnamefont
  {J.}~\bibnamefont {Chiaverini}}, \bibinfo {author} {\bibfnamefont
  {B.}~\bibnamefont {DeMarco}}, \bibinfo {author} {\bibfnamefont {W.~M.}\
  \bibnamefont {Itano}}, \bibinfo {author} {\bibfnamefont {B.}~\bibnamefont
  {Jelenković}}, \bibinfo {author} {\bibfnamefont {C.}~\bibnamefont {Langer}},
  \bibinfo {author} {\bibfnamefont {D.}~\bibnamefont {Leibfried}}, \bibinfo
  {author} {\bibfnamefont {V.}~\bibnamefont {Meyer}}, \bibinfo {author}
  {\bibfnamefont {T.}~\bibnamefont {Rosenband}},\ and\ \bibinfo {author}
  {\bibfnamefont {T.}~\bibnamefont {Schätz}},\ }\bibfield  {title} {\bibinfo
  {title} {Quantum information processing with trapped ions},\ }\href
  {https://doi.org/10.1098/rsta.2003.1205} {\bibfield  {journal} {\bibinfo
  {journal} {Philosophical Transactions of the Royal Society of London. Series
  A: Mathematical, Physical and Engineering Sciences}\ }\textbf {\bibinfo
  {volume} {361}},\ \bibinfo {pages} {1349} (\bibinfo {year}
  {2003})}\BibitemShut {NoStop}%
\bibitem [{\citenamefont {S\o{}rensen}\ and\ \citenamefont
  {M\o{}lmer}(1999)}]{Sorensen99}%
  \BibitemOpen
  \bibfield  {author} {\bibinfo {author} {\bibfnamefont {A.}~\bibnamefont
  {S\o{}rensen}}\ and\ \bibinfo {author} {\bibfnamefont {K.}~\bibnamefont
  {M\o{}lmer}},\ }\bibfield  {title} {\bibinfo {title} {Quantum computation
  with ions in thermal motion},\ }\href
  {https://doi.org/10.1103/PhysRevLett.82.1971} {\bibfield  {journal} {\bibinfo
   {journal} {Phys. Rev. Lett.}\ }\textbf {\bibinfo {volume} {82}},\ \bibinfo
  {pages} {1971} (\bibinfo {year} {1999})}\BibitemShut {NoStop}%
\bibitem [{Note1()}]{Note1}%
  \BibitemOpen
  \bibinfo {note} {While we use a fixed value of $\eta $ for ground and
  metastable qubits in our calculations, in reality the Lamb-Dicke parameter is
  proportional to $2 \omega _L \sin (\theta /2)/c \times \protect \sqrt {\hbar
  /2M\omega }$ (up to a normalization coefficient that depends on the length of
  ion chain), where $\theta $ ($0 \leq \theta \leq \pi )$ is the angle between
  the two Raman beams, $c$ is the speed of light, $M$ is the ion mass, and
  $\omega $ is the motional-mode frequency. As the laser frequency $\omega _L$
  is smaller for metastable qubits than for ground qubits (by a factor of
  roughly 1.3 for Ba$^+$ and 2.2 for Ca$^+$), fixing $\eta $ between metastable
  and ground qubits may require using larger $\theta $ or smaller $\omega $ for
  metastable qubits than ground qubits. Alternatively, metastable qubits may
  require additional laser power to compensate the effects of smaller $\eta
  $~\cite {Moore23}.}\BibitemShut {Stop}%
\bibitem [{\citenamefont {Wang}\ \emph {et~al.}(2020)\citenamefont {Wang},
  \citenamefont {Crain}, \citenamefont {Fang}, \citenamefont {Zhang},
  \citenamefont {Huang}, \citenamefont {Liang}, \citenamefont {Leung},
  \citenamefont {Brown},\ and\ \citenamefont {Kim}}]{Wang20}%
  \BibitemOpen
  \bibfield  {author} {\bibinfo {author} {\bibfnamefont {Y.}~\bibnamefont
  {Wang}}, \bibinfo {author} {\bibfnamefont {S.}~\bibnamefont {Crain}},
  \bibinfo {author} {\bibfnamefont {C.}~\bibnamefont {Fang}}, \bibinfo {author}
  {\bibfnamefont {B.}~\bibnamefont {Zhang}}, \bibinfo {author} {\bibfnamefont
  {S.}~\bibnamefont {Huang}}, \bibinfo {author} {\bibfnamefont
  {Q.}~\bibnamefont {Liang}}, \bibinfo {author} {\bibfnamefont {P.~H.}\
  \bibnamefont {Leung}}, \bibinfo {author} {\bibfnamefont {K.~R.}\ \bibnamefont
  {Brown}},\ and\ \bibinfo {author} {\bibfnamefont {J.}~\bibnamefont {Kim}},\
  }\bibfield  {title} {\bibinfo {title} {High-fidelity two-qubit gates using a
  microelectromechanical-system-based beam steering system for individual qubit
  addressing},\ }\href {https://doi.org/10.1103/PhysRevLett.125.150505}
  {\bibfield  {journal} {\bibinfo  {journal} {Phys. Rev. Lett.}\ }\textbf
  {\bibinfo {volume} {125}},\ \bibinfo {pages} {150505} (\bibinfo {year}
  {2020})}\BibitemShut {NoStop}%
\bibitem [{\citenamefont {Cetina}\ \emph {et~al.}(2022)\citenamefont {Cetina},
  \citenamefont {Egan}, \citenamefont {Noel}, \citenamefont {Goldman},
  \citenamefont {Biswas}, \citenamefont {Risinger}, \citenamefont {Zhu},\ and\
  \citenamefont {Monroe}}]{Cetina22}%
  \BibitemOpen
  \bibfield  {author} {\bibinfo {author} {\bibfnamefont {M.}~\bibnamefont
  {Cetina}}, \bibinfo {author} {\bibfnamefont {L.~N.}\ \bibnamefont {Egan}},
  \bibinfo {author} {\bibfnamefont {C.}~\bibnamefont {Noel}}, \bibinfo {author}
  {\bibfnamefont {M.~L.}\ \bibnamefont {Goldman}}, \bibinfo {author}
  {\bibfnamefont {D.}~\bibnamefont {Biswas}}, \bibinfo {author} {\bibfnamefont
  {A.~R.}\ \bibnamefont {Risinger}}, \bibinfo {author} {\bibfnamefont
  {D.}~\bibnamefont {Zhu}},\ and\ \bibinfo {author} {\bibfnamefont
  {C.}~\bibnamefont {Monroe}},\ }\bibfield  {title} {\bibinfo {title} {Control
  of transverse motion for quantum gates on individually addressed atomic
  qubits},\ }\href {https://doi.org/10.1103/PRXQuantum.3.010334} {\bibfield
  {journal} {\bibinfo  {journal} {PRX Quantum}\ }\textbf {\bibinfo {volume}
  {3}},\ \bibinfo {pages} {010334} (\bibinfo {year} {2022})}\BibitemShut
  {NoStop}%
\bibitem [{\citenamefont {Kang}\ \emph
  {et~al.}(2023{\natexlab{a}})\citenamefont {Kang}, \citenamefont {Wang},
  \citenamefont {Fang}, \citenamefont {Zhang}, \citenamefont {Khosravani},
  \citenamefont {Kim},\ and\ \citenamefont {Brown}}]{Kang23ff}%
  \BibitemOpen
  \bibfield  {author} {\bibinfo {author} {\bibfnamefont {M.}~\bibnamefont
  {Kang}}, \bibinfo {author} {\bibfnamefont {Y.}~\bibnamefont {Wang}}, \bibinfo
  {author} {\bibfnamefont {C.}~\bibnamefont {Fang}}, \bibinfo {author}
  {\bibfnamefont {B.}~\bibnamefont {Zhang}}, \bibinfo {author} {\bibfnamefont
  {O.}~\bibnamefont {Khosravani}}, \bibinfo {author} {\bibfnamefont
  {J.}~\bibnamefont {Kim}},\ and\ \bibinfo {author} {\bibfnamefont {K.~R.}\
  \bibnamefont {Brown}},\ }\bibfield  {title} {\bibinfo {title} {Designing
  filter functions of frequency-modulated pulses for high-fidelity two-qubit
  gates in ion chains},\ }\href
  {https://doi.org/10.1103/PhysRevApplied.19.014014} {\bibfield  {journal}
  {\bibinfo  {journal} {Phys. Rev. Appl.}\ }\textbf {\bibinfo {volume} {19}},\
  \bibinfo {pages} {014014} (\bibinfo {year} {2023}{\natexlab{a}})}\BibitemShut
  {NoStop}%
\bibitem [{\citenamefont {Jandura}\ \emph {et~al.}(2022)\citenamefont
  {Jandura}, \citenamefont {Thompson},\ and\ \citenamefont
  {Pupillo}}]{Jandura22}%
  \BibitemOpen
  \bibfield  {author} {\bibinfo {author} {\bibfnamefont {S.}~\bibnamefont
  {Jandura}}, \bibinfo {author} {\bibfnamefont {J.~D.}\ \bibnamefont
  {Thompson}},\ and\ \bibinfo {author} {\bibfnamefont {G.}~\bibnamefont
  {Pupillo}},\ }\bibfield  {title} {\bibinfo {title} {Optimizing rydberg gates
  for logical qubit performance},\ }\href@noop {} {\bibfield  {journal}
  {\bibinfo  {journal} {arXiv preprint arXiv:2210.06879}\ } (\bibinfo {year}
  {2022})}\BibitemShut {NoStop}%
\bibitem [{\citenamefont {Brown}\ \emph {et~al.}(2021)\citenamefont {Brown},
  \citenamefont {Chiaverini}, \citenamefont {Sage},\ and\ \citenamefont
  {H{\"a}ffner}}]{Brown21}%
  \BibitemOpen
  \bibfield  {author} {\bibinfo {author} {\bibfnamefont {K.~R.}\ \bibnamefont
  {Brown}}, \bibinfo {author} {\bibfnamefont {J.}~\bibnamefont {Chiaverini}},
  \bibinfo {author} {\bibfnamefont {J.~M.}\ \bibnamefont {Sage}},\ and\
  \bibinfo {author} {\bibfnamefont {H.}~\bibnamefont {H{\"a}ffner}},\
  }\bibfield  {title} {\bibinfo {title} {Materials challenges for trapped-ion
  quantum computers},\ }\href@noop {} {\bibfield  {journal} {\bibinfo
  {journal} {Nature Reviews Materials}\ }\textbf {\bibinfo {volume} {6}},\
  \bibinfo {pages} {892} (\bibinfo {year} {2021})}\BibitemShut {NoStop}%
\bibitem [{\citenamefont {Raussendorf}\ and\ \citenamefont
  {Harrington}(2007)}]{Raussendorf07}%
  \BibitemOpen
  \bibfield  {author} {\bibinfo {author} {\bibfnamefont {R.}~\bibnamefont
  {Raussendorf}}\ and\ \bibinfo {author} {\bibfnamefont {J.}~\bibnamefont
  {Harrington}},\ }\bibfield  {title} {\bibinfo {title} {Fault-tolerant quantum
  computation with high threshold in two dimensions},\ }\href
  {https://doi.org/10.1103/PhysRevLett.98.190504} {\bibfield  {journal}
  {\bibinfo  {journal} {Phys. Rev. Lett.}\ }\textbf {\bibinfo {volume} {98}},\
  \bibinfo {pages} {190504} (\bibinfo {year} {2007})}\BibitemShut {NoStop}%
\bibitem [{\citenamefont {Fowler}\ \emph {et~al.}(2009)\citenamefont {Fowler},
  \citenamefont {Stephens},\ and\ \citenamefont {Groszkowski}}]{Fowler09}%
  \BibitemOpen
  \bibfield  {author} {\bibinfo {author} {\bibfnamefont {A.~G.}\ \bibnamefont
  {Fowler}}, \bibinfo {author} {\bibfnamefont {A.~M.}\ \bibnamefont
  {Stephens}},\ and\ \bibinfo {author} {\bibfnamefont {P.}~\bibnamefont
  {Groszkowski}},\ }\bibfield  {title} {\bibinfo {title} {High-threshold
  universal quantum computation on the surface code},\ }\href
  {https://doi.org/10.1103/PhysRevA.80.052312} {\bibfield  {journal} {\bibinfo
  {journal} {Phys. Rev. A}\ }\textbf {\bibinfo {volume} {80}},\ \bibinfo
  {pages} {052312} (\bibinfo {year} {2009})}\BibitemShut {NoStop}%
\bibitem [{\citenamefont {Fowler}\ \emph {et~al.}(2012)\citenamefont {Fowler},
  \citenamefont {Mariantoni}, \citenamefont {Martinis},\ and\ \citenamefont
  {Cleland}}]{Fowler12}%
  \BibitemOpen
  \bibfield  {author} {\bibinfo {author} {\bibfnamefont {A.~G.}\ \bibnamefont
  {Fowler}}, \bibinfo {author} {\bibfnamefont {M.}~\bibnamefont {Mariantoni}},
  \bibinfo {author} {\bibfnamefont {J.~M.}\ \bibnamefont {Martinis}},\ and\
  \bibinfo {author} {\bibfnamefont {A.~N.}\ \bibnamefont {Cleland}},\
  }\bibfield  {title} {\bibinfo {title} {Surface codes: Towards practical
  large-scale quantum computation},\ }\href
  {https://doi.org/10.1103/PhysRevA.86.032324} {\bibfield  {journal} {\bibinfo
  {journal} {Phys. Rev. A}\ }\textbf {\bibinfo {volume} {86}},\ \bibinfo
  {pages} {032324} (\bibinfo {year} {2012})}\BibitemShut {NoStop}%
\bibitem [{\citenamefont {Tomita}\ and\ \citenamefont
  {Svore}(2014)}]{Tomita14}%
  \BibitemOpen
  \bibfield  {author} {\bibinfo {author} {\bibfnamefont {Y.}~\bibnamefont
  {Tomita}}\ and\ \bibinfo {author} {\bibfnamefont {K.~M.}\ \bibnamefont
  {Svore}},\ }\bibfield  {title} {\bibinfo {title} {Low-distance surface codes
  under realistic quantum noise},\ }\href
  {https://doi.org/10.1103/PhysRevA.90.062320} {\bibfield  {journal} {\bibinfo
  {journal} {Phys. Rev. A}\ }\textbf {\bibinfo {volume} {90}},\ \bibinfo
  {pages} {062320} (\bibinfo {year} {2014})}\BibitemShut {NoStop}%
\bibitem [{\citenamefont {Krinner}\ \emph {et~al.}(2022)\citenamefont
  {Krinner}, \citenamefont {Lacroix}, \citenamefont {Remm}, \citenamefont
  {Di~Paolo}, \citenamefont {Genois}, \citenamefont {Leroux}, \citenamefont
  {Hellings}, \citenamefont {Lazar}, \citenamefont {Swiadek}, \citenamefont
  {Herrmann}, \citenamefont {Norris}, \citenamefont {Andersen}, \citenamefont
  {M{\"u}ller}, \citenamefont {Blais}, \citenamefont {Eichler},\ and\
  \citenamefont {Wallraff}}]{Krinner22}%
  \BibitemOpen
  \bibfield  {author} {\bibinfo {author} {\bibfnamefont {S.}~\bibnamefont
  {Krinner}}, \bibinfo {author} {\bibfnamefont {N.}~\bibnamefont {Lacroix}},
  \bibinfo {author} {\bibfnamefont {A.}~\bibnamefont {Remm}}, \bibinfo {author}
  {\bibfnamefont {A.}~\bibnamefont {Di~Paolo}}, \bibinfo {author}
  {\bibfnamefont {E.}~\bibnamefont {Genois}}, \bibinfo {author} {\bibfnamefont
  {C.}~\bibnamefont {Leroux}}, \bibinfo {author} {\bibfnamefont
  {C.}~\bibnamefont {Hellings}}, \bibinfo {author} {\bibfnamefont
  {S.}~\bibnamefont {Lazar}}, \bibinfo {author} {\bibfnamefont
  {F.}~\bibnamefont {Swiadek}}, \bibinfo {author} {\bibfnamefont
  {J.}~\bibnamefont {Herrmann}}, \bibinfo {author} {\bibfnamefont {G.~J.}\
  \bibnamefont {Norris}}, \bibinfo {author} {\bibfnamefont {C.~K.}\
  \bibnamefont {Andersen}}, \bibinfo {author} {\bibfnamefont {M.}~\bibnamefont
  {M{\"u}ller}}, \bibinfo {author} {\bibfnamefont {A.}~\bibnamefont {Blais}},
  \bibinfo {author} {\bibfnamefont {C.}~\bibnamefont {Eichler}},\ and\ \bibinfo
  {author} {\bibfnamefont {A.}~\bibnamefont {Wallraff}},\ }\bibfield  {title}
  {\bibinfo {title} {Realizing repeated quantum error correction in a
  distance-three surface code},\ }\href
  {https://doi.org/10.1038/s41586-022-04566-8} {\bibfield  {journal} {\bibinfo
  {journal} {Nature}\ }\textbf {\bibinfo {volume} {605}},\ \bibinfo {pages}
  {669} (\bibinfo {year} {2022})}\BibitemShut {NoStop}%
\bibitem [{\citenamefont {Zhao}\ \emph {et~al.}(2022)\citenamefont {Zhao},
  \citenamefont {Ye}, \citenamefont {Huang}, \citenamefont {Zhang},
  \citenamefont {Wu}, \citenamefont {Guan}, \citenamefont {Zhu}, \citenamefont
  {Wei}, \citenamefont {He}, \citenamefont {Cao}, \citenamefont {Chen},
  \citenamefont {Chung}, \citenamefont {Deng}, \citenamefont {Fan},
  \citenamefont {Gong}, \citenamefont {Guo}, \citenamefont {Guo}, \citenamefont
  {Han}, \citenamefont {Li}, \citenamefont {Li}, \citenamefont {Li},
  \citenamefont {Liang}, \citenamefont {Lin}, \citenamefont {Qian},
  \citenamefont {Rong}, \citenamefont {Su}, \citenamefont {Sun}, \citenamefont
  {Wang}, \citenamefont {Wu}, \citenamefont {Xu}, \citenamefont {Ying},
  \citenamefont {Yu}, \citenamefont {Zha}, \citenamefont {Zhang}, \citenamefont
  {Huo}, \citenamefont {Lu}, \citenamefont {Peng}, \citenamefont {Zhu},\ and\
  \citenamefont {Pan}}]{Zhao22}%
  \BibitemOpen
  \bibfield  {author} {\bibinfo {author} {\bibfnamefont {Y.}~\bibnamefont
  {Zhao}}, \bibinfo {author} {\bibfnamefont {Y.}~\bibnamefont {Ye}}, \bibinfo
  {author} {\bibfnamefont {H.-L.}\ \bibnamefont {Huang}}, \bibinfo {author}
  {\bibfnamefont {Y.}~\bibnamefont {Zhang}}, \bibinfo {author} {\bibfnamefont
  {D.}~\bibnamefont {Wu}}, \bibinfo {author} {\bibfnamefont {H.}~\bibnamefont
  {Guan}}, \bibinfo {author} {\bibfnamefont {Q.}~\bibnamefont {Zhu}}, \bibinfo
  {author} {\bibfnamefont {Z.}~\bibnamefont {Wei}}, \bibinfo {author}
  {\bibfnamefont {T.}~\bibnamefont {He}}, \bibinfo {author} {\bibfnamefont
  {S.}~\bibnamefont {Cao}}, \bibinfo {author} {\bibfnamefont {F.}~\bibnamefont
  {Chen}}, \bibinfo {author} {\bibfnamefont {T.-H.}\ \bibnamefont {Chung}},
  \bibinfo {author} {\bibfnamefont {H.}~\bibnamefont {Deng}}, \bibinfo {author}
  {\bibfnamefont {D.}~\bibnamefont {Fan}}, \bibinfo {author} {\bibfnamefont
  {M.}~\bibnamefont {Gong}}, \bibinfo {author} {\bibfnamefont {C.}~\bibnamefont
  {Guo}}, \bibinfo {author} {\bibfnamefont {S.}~\bibnamefont {Guo}}, \bibinfo
  {author} {\bibfnamefont {L.}~\bibnamefont {Han}}, \bibinfo {author}
  {\bibfnamefont {N.}~\bibnamefont {Li}}, \bibinfo {author} {\bibfnamefont
  {S.}~\bibnamefont {Li}}, \bibinfo {author} {\bibfnamefont {Y.}~\bibnamefont
  {Li}}, \bibinfo {author} {\bibfnamefont {F.}~\bibnamefont {Liang}}, \bibinfo
  {author} {\bibfnamefont {J.}~\bibnamefont {Lin}}, \bibinfo {author}
  {\bibfnamefont {H.}~\bibnamefont {Qian}}, \bibinfo {author} {\bibfnamefont
  {H.}~\bibnamefont {Rong}}, \bibinfo {author} {\bibfnamefont {H.}~\bibnamefont
  {Su}}, \bibinfo {author} {\bibfnamefont {L.}~\bibnamefont {Sun}}, \bibinfo
  {author} {\bibfnamefont {S.}~\bibnamefont {Wang}}, \bibinfo {author}
  {\bibfnamefont {Y.}~\bibnamefont {Wu}}, \bibinfo {author} {\bibfnamefont
  {Y.}~\bibnamefont {Xu}}, \bibinfo {author} {\bibfnamefont {C.}~\bibnamefont
  {Ying}}, \bibinfo {author} {\bibfnamefont {J.}~\bibnamefont {Yu}}, \bibinfo
  {author} {\bibfnamefont {C.}~\bibnamefont {Zha}}, \bibinfo {author}
  {\bibfnamefont {K.}~\bibnamefont {Zhang}}, \bibinfo {author} {\bibfnamefont
  {Y.-H.}\ \bibnamefont {Huo}}, \bibinfo {author} {\bibfnamefont {C.-Y.}\
  \bibnamefont {Lu}}, \bibinfo {author} {\bibfnamefont {C.-Z.}\ \bibnamefont
  {Peng}}, \bibinfo {author} {\bibfnamefont {X.}~\bibnamefont {Zhu}},\ and\
  \bibinfo {author} {\bibfnamefont {J.-W.}\ \bibnamefont {Pan}},\ }\bibfield
  {title} {\bibinfo {title} {Realization of an error-correcting surface code
  with superconducting qubits},\ }\href
  {https://doi.org/10.1103/PhysRevLett.129.030501} {\bibfield  {journal}
  {\bibinfo  {journal} {Phys. Rev. Lett.}\ }\textbf {\bibinfo {volume} {129}},\
  \bibinfo {pages} {030501} (\bibinfo {year} {2022})}\BibitemShut {NoStop}%
\bibitem [{Goo(2023)}]{Google23}%
  \BibitemOpen
  \bibfield  {title} {\bibinfo {title} {Suppressing quantum errors by scaling a
  surface code logical qubit},\ }\href@noop {} {\bibfield  {journal} {\bibinfo
  {journal} {Nature}\ }\textbf {\bibinfo {volume} {614}},\ \bibinfo {pages}
  {676} (\bibinfo {year} {2023})}\BibitemShut {NoStop}%
\bibitem [{\citenamefont {Debroy}\ \emph {et~al.}(2020)\citenamefont {Debroy},
  \citenamefont {Li}, \citenamefont {Huang},\ and\ \citenamefont
  {Brown}}]{Debroy20}%
  \BibitemOpen
  \bibfield  {author} {\bibinfo {author} {\bibfnamefont {D.~M.}\ \bibnamefont
  {Debroy}}, \bibinfo {author} {\bibfnamefont {M.}~\bibnamefont {Li}}, \bibinfo
  {author} {\bibfnamefont {S.}~\bibnamefont {Huang}},\ and\ \bibinfo {author}
  {\bibfnamefont {K.~R.}\ \bibnamefont {Brown}},\ }\bibfield  {title} {\bibinfo
  {title} {Logical performance of 9 qubit compass codes in ion traps with
  crosstalk errors},\ }\href {https://doi.org/10.1088/2058-9565/ab7e80}
  {\bibfield  {journal} {\bibinfo  {journal} {Quantum Science and Technology}\
  }\textbf {\bibinfo {volume} {5}},\ \bibinfo {pages} {034002} (\bibinfo {year}
  {2020})}\BibitemShut {NoStop}%
\bibitem [{\citenamefont {Steane}(1997)}]{Steane97}%
  \BibitemOpen
  \bibfield  {author} {\bibinfo {author} {\bibfnamefont {A.~M.}\ \bibnamefont
  {Steane}},\ }\bibfield  {title} {\bibinfo {title} {Active stabilization,
  quantum computation, and quantum state synthesis},\ }\href
  {https://doi.org/10.1103/PhysRevLett.78.2252} {\bibfield  {journal} {\bibinfo
   {journal} {Phys. Rev. Lett.}\ }\textbf {\bibinfo {volume} {78}},\ \bibinfo
  {pages} {2252} (\bibinfo {year} {1997})}\BibitemShut {NoStop}%
\bibitem [{\citenamefont {Dennis}\ \emph {et~al.}(2002)\citenamefont {Dennis},
  \citenamefont {Kitaev}, \citenamefont {Landahl},\ and\ \citenamefont
  {Preskill}}]{Dennis02}%
  \BibitemOpen
  \bibfield  {author} {\bibinfo {author} {\bibfnamefont {E.}~\bibnamefont
  {Dennis}}, \bibinfo {author} {\bibfnamefont {A.}~\bibnamefont {Kitaev}},
  \bibinfo {author} {\bibfnamefont {A.}~\bibnamefont {Landahl}},\ and\ \bibinfo
  {author} {\bibfnamefont {J.}~\bibnamefont {Preskill}},\ }\bibfield  {title}
  {\bibinfo {title} {Topological quantum memory},\ }\href
  {https://doi.org/10.1063/1.1499754} {\bibfield  {journal} {\bibinfo
  {journal} {Journal of Mathematical Physics}\ }\textbf {\bibinfo {volume}
  {43}},\ \bibinfo {pages} {4452} (\bibinfo {year} {2002})}\BibitemShut
  {NoStop}%
\bibitem [{\citenamefont {Delfosse}\ and\ \citenamefont
  {Nickerson}(2021)}]{Delfosse21}%
  \BibitemOpen
  \bibfield  {author} {\bibinfo {author} {\bibfnamefont {N.}~\bibnamefont
  {Delfosse}}\ and\ \bibinfo {author} {\bibfnamefont {N.~H.}\ \bibnamefont
  {Nickerson}},\ }\bibfield  {title} {\bibinfo {title} {Almost-linear time
  decoding algorithm for topological codes},\ }\href@noop {} {\bibfield
  {journal} {\bibinfo  {journal} {Quantum}\ }\textbf {\bibinfo {volume} {5}},\
  \bibinfo {pages} {595} (\bibinfo {year} {2021})}\BibitemShut {NoStop}%
\bibitem [{\citenamefont {Huang}\ \emph {et~al.}(2020)\citenamefont {Huang},
  \citenamefont {Newman},\ and\ \citenamefont {Brown}}]{Huang20faulttolerant}%
  \BibitemOpen
  \bibfield  {author} {\bibinfo {author} {\bibfnamefont {S.}~\bibnamefont
  {Huang}}, \bibinfo {author} {\bibfnamefont {M.}~\bibnamefont {Newman}},\ and\
  \bibinfo {author} {\bibfnamefont {K.~R.}\ \bibnamefont {Brown}},\ }\bibfield
  {title} {\bibinfo {title} {Fault-tolerant weighted union-find decoding on the
  toric code},\ }\href {https://doi.org/10.1103/PhysRevA.102.012419} {\bibfield
   {journal} {\bibinfo  {journal} {Phys. Rev. A}\ }\textbf {\bibinfo {volume}
  {102}},\ \bibinfo {pages} {012419} (\bibinfo {year} {2020})}\BibitemShut
  {NoStop}%
\bibitem [{\citenamefont {Gidney}(2021)}]{Gidney21}%
  \BibitemOpen
  \bibfield  {author} {\bibinfo {author} {\bibfnamefont {C.}~\bibnamefont
  {Gidney}},\ }\bibfield  {title} {\bibinfo {title} {Stim: a fast stabilizer
  circuit simulator},\ }\href@noop {} {\bibfield  {journal} {\bibinfo
  {journal} {Quantum}\ }\textbf {\bibinfo {volume} {5}},\ \bibinfo {pages}
  {497} (\bibinfo {year} {2021})}\BibitemShut {NoStop}%
\bibitem [{\citenamefont {Higgott}(2022)}]{Higgott22}%
  \BibitemOpen
  \bibfield  {author} {\bibinfo {author} {\bibfnamefont {O.}~\bibnamefont
  {Higgott}},\ }\bibfield  {title} {\bibinfo {title} {Pymatching: A python
  package for decoding quantum codes with minimum-weight perfect matching},\
  }\href@noop {} {\bibfield  {journal} {\bibinfo  {journal} {ACM Transactions
  on Quantum Computing}\ }\textbf {\bibinfo {volume} {3}},\ \bibinfo {pages}
  {1} (\bibinfo {year} {2022})}\BibitemShut {NoStop}%
\bibitem [{\citenamefont {Gidney}()}]{stim_erasure}%
  \BibitemOpen
  \bibfield  {author} {\bibinfo {author} {\bibfnamefont {C.}~\bibnamefont
  {Gidney}},\ }\href@noop {} {\bibinfo {title} {{How do I perform an erasure
  error in stim?}}},\ \bibinfo {howpublished}
  {\url{https://quantumcomputing.stackexchange.com/questions/26582}},\ \bibinfo
  {note} {accessed: 2022-10-02}\BibitemShut {NoStop}%
\bibitem [{\citenamefont {Schwerdt}\ \emph {et~al.}(2022)\citenamefont
  {Schwerdt}, \citenamefont {Shapira}, \citenamefont {Manovitz},\ and\
  \citenamefont {Ozeri}}]{Schwerdt22}%
  \BibitemOpen
  \bibfield  {author} {\bibinfo {author} {\bibfnamefont {D.}~\bibnamefont
  {Schwerdt}}, \bibinfo {author} {\bibfnamefont {Y.}~\bibnamefont {Shapira}},
  \bibinfo {author} {\bibfnamefont {T.}~\bibnamefont {Manovitz}},\ and\
  \bibinfo {author} {\bibfnamefont {R.}~\bibnamefont {Ozeri}},\ }\bibfield
  {title} {\bibinfo {title} {Comparing two-qubit and multiqubit gates within
  the toric code},\ }\href {https://doi.org/10.1103/PhysRevA.105.022612}
  {\bibfield  {journal} {\bibinfo  {journal} {Phys. Rev. A}\ }\textbf {\bibinfo
  {volume} {105}},\ \bibinfo {pages} {022612} (\bibinfo {year}
  {2022})}\BibitemShut {NoStop}%
\bibitem [{\citenamefont {Rasmusson}\ \emph {et~al.}(2021)\citenamefont
  {Rasmusson}, \citenamefont {D'Onofrio}, \citenamefont {Xie}, \citenamefont
  {Cui},\ and\ \citenamefont {Richerme}}]{Rasmusson21}%
  \BibitemOpen
  \bibfield  {author} {\bibinfo {author} {\bibfnamefont {A.~J.}\ \bibnamefont
  {Rasmusson}}, \bibinfo {author} {\bibfnamefont {M.}~\bibnamefont
  {D'Onofrio}}, \bibinfo {author} {\bibfnamefont {Y.}~\bibnamefont {Xie}},
  \bibinfo {author} {\bibfnamefont {J.}~\bibnamefont {Cui}},\ and\ \bibinfo
  {author} {\bibfnamefont {P.}~\bibnamefont {Richerme}},\ }\bibfield  {title}
  {\bibinfo {title} {Optimized pulsed sideband cooling and enhanced thermometry
  of trapped ions},\ }\href {https://doi.org/10.1103/PhysRevA.104.043108}
  {\bibfield  {journal} {\bibinfo  {journal} {Phys. Rev. A}\ }\textbf {\bibinfo
  {volume} {104}},\ \bibinfo {pages} {043108} (\bibinfo {year}
  {2021})}\BibitemShut {NoStop}%
\bibitem [{\citenamefont {Christensen}\ \emph {et~al.}(2020)\citenamefont
  {Christensen}, \citenamefont {Hucul}, \citenamefont {Campbell},\ and\
  \citenamefont {Hudson}}]{Christensen20}%
  \BibitemOpen
  \bibfield  {author} {\bibinfo {author} {\bibfnamefont {J.~E.}\ \bibnamefont
  {Christensen}}, \bibinfo {author} {\bibfnamefont {D.}~\bibnamefont {Hucul}},
  \bibinfo {author} {\bibfnamefont {W.~C.}\ \bibnamefont {Campbell}},\ and\
  \bibinfo {author} {\bibfnamefont {E.~R.}\ \bibnamefont {Hudson}},\ }\bibfield
   {title} {\bibinfo {title} {High-fidelity manipulation of a qubit enabled by
  a manufactured nucleus},\ }\href@noop {} {\bibfield  {journal} {\bibinfo
  {journal} {npj Quantum Information}\ }\textbf {\bibinfo {volume} {6}},\
  \bibinfo {pages} {1} (\bibinfo {year} {2020})}\BibitemShut {NoStop}%
\bibitem [{\citenamefont {Ransford}\ \emph {et~al.}(2021)\citenamefont
  {Ransford}, \citenamefont {Roman}, \citenamefont {Dellaert}, \citenamefont
  {McMillin},\ and\ \citenamefont {Campbell}}]{Ransford21}%
  \BibitemOpen
  \bibfield  {author} {\bibinfo {author} {\bibfnamefont {A.}~\bibnamefont
  {Ransford}}, \bibinfo {author} {\bibfnamefont {C.}~\bibnamefont {Roman}},
  \bibinfo {author} {\bibfnamefont {T.}~\bibnamefont {Dellaert}}, \bibinfo
  {author} {\bibfnamefont {P.}~\bibnamefont {McMillin}},\ and\ \bibinfo
  {author} {\bibfnamefont {W.~C.}\ \bibnamefont {Campbell}},\ }\bibfield
  {title} {\bibinfo {title} {Weak dissipation for high-fidelity qubit-state
  preparation and measurement},\ }\href
  {https://doi.org/10.1103/PhysRevA.104.L060402} {\bibfield  {journal}
  {\bibinfo  {journal} {Phys. Rev. A}\ }\textbf {\bibinfo {volume} {104}},\
  \bibinfo {pages} {L060402} (\bibinfo {year} {2021})}\BibitemShut {NoStop}%
\bibitem [{\citenamefont {Jaksch}\ \emph {et~al.}(2000)\citenamefont {Jaksch},
  \citenamefont {Cirac}, \citenamefont {Zoller}, \citenamefont {Rolston},
  \citenamefont {C\^ot\'e},\ and\ \citenamefont {Lukin}}]{Jaksch00}%
  \BibitemOpen
  \bibfield  {author} {\bibinfo {author} {\bibfnamefont {D.}~\bibnamefont
  {Jaksch}}, \bibinfo {author} {\bibfnamefont {J.~I.}\ \bibnamefont {Cirac}},
  \bibinfo {author} {\bibfnamefont {P.}~\bibnamefont {Zoller}}, \bibinfo
  {author} {\bibfnamefont {S.~L.}\ \bibnamefont {Rolston}}, \bibinfo {author}
  {\bibfnamefont {R.}~\bibnamefont {C\^ot\'e}},\ and\ \bibinfo {author}
  {\bibfnamefont {M.~D.}\ \bibnamefont {Lukin}},\ }\bibfield  {title} {\bibinfo
  {title} {Fast quantum gates for neutral atoms},\ }\href
  {https://doi.org/10.1103/PhysRevLett.85.2208} {\bibfield  {journal} {\bibinfo
   {journal} {Phys. Rev. Lett.}\ }\textbf {\bibinfo {volume} {85}},\ \bibinfo
  {pages} {2208} (\bibinfo {year} {2000})}\BibitemShut {NoStop}%
\bibitem [{\citenamefont {Lukin}\ \emph {et~al.}(2001)\citenamefont {Lukin},
  \citenamefont {Fleischhauer}, \citenamefont {Cote}, \citenamefont {Duan},
  \citenamefont {Jaksch}, \citenamefont {Cirac},\ and\ \citenamefont
  {Zoller}}]{Lukin01}%
  \BibitemOpen
  \bibfield  {author} {\bibinfo {author} {\bibfnamefont {M.~D.}\ \bibnamefont
  {Lukin}}, \bibinfo {author} {\bibfnamefont {M.}~\bibnamefont {Fleischhauer}},
  \bibinfo {author} {\bibfnamefont {R.}~\bibnamefont {Cote}}, \bibinfo {author}
  {\bibfnamefont {L.~M.}\ \bibnamefont {Duan}}, \bibinfo {author}
  {\bibfnamefont {D.}~\bibnamefont {Jaksch}}, \bibinfo {author} {\bibfnamefont
  {J.~I.}\ \bibnamefont {Cirac}},\ and\ \bibinfo {author} {\bibfnamefont
  {P.}~\bibnamefont {Zoller}},\ }\bibfield  {title} {\bibinfo {title} {Dipole
  blockade and quantum information processing in mesoscopic atomic ensembles},\
  }\href {https://doi.org/10.1103/PhysRevLett.87.037901} {\bibfield  {journal}
  {\bibinfo  {journal} {Phys. Rev. Lett.}\ }\textbf {\bibinfo {volume} {87}},\
  \bibinfo {pages} {037901} (\bibinfo {year} {2001})}\BibitemShut {NoStop}%
\bibitem [{\citenamefont {Zhang}\ \emph {et~al.}(2012)\citenamefont {Zhang},
  \citenamefont {Gill}, \citenamefont {Isenhower}, \citenamefont {Walker},\
  and\ \citenamefont {Saffman}}]{Zhang12}%
  \BibitemOpen
  \bibfield  {author} {\bibinfo {author} {\bibfnamefont {X.~L.}\ \bibnamefont
  {Zhang}}, \bibinfo {author} {\bibfnamefont {A.~T.}\ \bibnamefont {Gill}},
  \bibinfo {author} {\bibfnamefont {L.}~\bibnamefont {Isenhower}}, \bibinfo
  {author} {\bibfnamefont {T.~G.}\ \bibnamefont {Walker}},\ and\ \bibinfo
  {author} {\bibfnamefont {M.}~\bibnamefont {Saffman}},\ }\bibfield  {title}
  {\bibinfo {title} {Fidelity of a rydberg-blockade quantum gate from simulated
  quantum process tomography},\ }\href
  {https://doi.org/10.1103/PhysRevA.85.042310} {\bibfield  {journal} {\bibinfo
  {journal} {Phys. Rev. A}\ }\textbf {\bibinfo {volume} {85}},\ \bibinfo
  {pages} {042310} (\bibinfo {year} {2012})}\BibitemShut {NoStop}%
\bibitem [{\citenamefont {Graham}\ \emph {et~al.}(2019)\citenamefont {Graham},
  \citenamefont {Kwon}, \citenamefont {Grinkemeyer}, \citenamefont {Marra},
  \citenamefont {Jiang}, \citenamefont {Lichtman}, \citenamefont {Sun},
  \citenamefont {Ebert},\ and\ \citenamefont {Saffman}}]{Graham19}%
  \BibitemOpen
  \bibfield  {author} {\bibinfo {author} {\bibfnamefont {T.~M.}\ \bibnamefont
  {Graham}}, \bibinfo {author} {\bibfnamefont {M.}~\bibnamefont {Kwon}},
  \bibinfo {author} {\bibfnamefont {B.}~\bibnamefont {Grinkemeyer}}, \bibinfo
  {author} {\bibfnamefont {Z.}~\bibnamefont {Marra}}, \bibinfo {author}
  {\bibfnamefont {X.}~\bibnamefont {Jiang}}, \bibinfo {author} {\bibfnamefont
  {M.~T.}\ \bibnamefont {Lichtman}}, \bibinfo {author} {\bibfnamefont
  {Y.}~\bibnamefont {Sun}}, \bibinfo {author} {\bibfnamefont {M.}~\bibnamefont
  {Ebert}},\ and\ \bibinfo {author} {\bibfnamefont {M.}~\bibnamefont
  {Saffman}},\ }\bibfield  {title} {\bibinfo {title} {Rydberg-mediated
  entanglement in a two-dimensional neutral atom qubit array},\ }\href
  {https://doi.org/10.1103/PhysRevLett.123.230501} {\bibfield  {journal}
  {\bibinfo  {journal} {Phys. Rev. Lett.}\ }\textbf {\bibinfo {volume} {123}},\
  \bibinfo {pages} {230501} (\bibinfo {year} {2019})}\BibitemShut {NoStop}%
\bibitem [{\citenamefont {Levine}\ \emph {et~al.}(2019)\citenamefont {Levine},
  \citenamefont {Keesling}, \citenamefont {Semeghini}, \citenamefont {Omran},
  \citenamefont {Wang}, \citenamefont {Ebadi}, \citenamefont {Bernien},
  \citenamefont {Greiner}, \citenamefont {Vuleti\ifmmode~\acute{c}\else
  \'{c}\fi{}}, \citenamefont {Pichler},\ and\ \citenamefont
  {Lukin}}]{Levine19}%
  \BibitemOpen
  \bibfield  {author} {\bibinfo {author} {\bibfnamefont {H.}~\bibnamefont
  {Levine}}, \bibinfo {author} {\bibfnamefont {A.}~\bibnamefont {Keesling}},
  \bibinfo {author} {\bibfnamefont {G.}~\bibnamefont {Semeghini}}, \bibinfo
  {author} {\bibfnamefont {A.}~\bibnamefont {Omran}}, \bibinfo {author}
  {\bibfnamefont {T.~T.}\ \bibnamefont {Wang}}, \bibinfo {author}
  {\bibfnamefont {S.}~\bibnamefont {Ebadi}}, \bibinfo {author} {\bibfnamefont
  {H.}~\bibnamefont {Bernien}}, \bibinfo {author} {\bibfnamefont
  {M.}~\bibnamefont {Greiner}}, \bibinfo {author} {\bibfnamefont
  {V.}~\bibnamefont {Vuleti\ifmmode~\acute{c}\else \'{c}\fi{}}}, \bibinfo
  {author} {\bibfnamefont {H.}~\bibnamefont {Pichler}},\ and\ \bibinfo {author}
  {\bibfnamefont {M.~D.}\ \bibnamefont {Lukin}},\ }\bibfield  {title} {\bibinfo
  {title} {Parallel implementation of high-fidelity multiqubit gates with
  neutral atoms},\ }\href {https://doi.org/10.1103/PhysRevLett.123.170503}
  {\bibfield  {journal} {\bibinfo  {journal} {Phys. Rev. Lett.}\ }\textbf
  {\bibinfo {volume} {123}},\ \bibinfo {pages} {170503} (\bibinfo {year}
  {2019})}\BibitemShut {NoStop}%
\bibitem [{\citenamefont {Jandura}\ and\ \citenamefont
  {Pupillo}(2022)}]{Jandura22Time}%
  \BibitemOpen
  \bibfield  {author} {\bibinfo {author} {\bibfnamefont {S.}~\bibnamefont
  {Jandura}}\ and\ \bibinfo {author} {\bibfnamefont {G.}~\bibnamefont
  {Pupillo}},\ }\bibfield  {title} {\bibinfo {title} {Time-optimal two-and
  three-qubit gates for rydberg atoms},\ }\href@noop {} {\bibfield  {journal}
  {\bibinfo  {journal} {Quantum}\ }\textbf {\bibinfo {volume} {6}},\ \bibinfo
  {pages} {712} (\bibinfo {year} {2022})}\BibitemShut {NoStop}%
\bibitem [{\citenamefont {Ma}\ \emph {et~al.}(2022)\citenamefont {Ma},
  \citenamefont {Burgers}, \citenamefont {Liu}, \citenamefont {Wilson},
  \citenamefont {Zhang},\ and\ \citenamefont {Thompson}}]{Ma22}%
  \BibitemOpen
  \bibfield  {author} {\bibinfo {author} {\bibfnamefont {S.}~\bibnamefont
  {Ma}}, \bibinfo {author} {\bibfnamefont {A.~P.}\ \bibnamefont {Burgers}},
  \bibinfo {author} {\bibfnamefont {G.}~\bibnamefont {Liu}}, \bibinfo {author}
  {\bibfnamefont {J.}~\bibnamefont {Wilson}}, \bibinfo {author} {\bibfnamefont
  {B.}~\bibnamefont {Zhang}},\ and\ \bibinfo {author} {\bibfnamefont {J.~D.}\
  \bibnamefont {Thompson}},\ }\bibfield  {title} {\bibinfo {title} {Universal
  gate operations on nuclear spin qubits in an optical tweezer array of
  $^{171}\mathrm{Yb}$ atoms},\ }\href
  {https://doi.org/10.1103/PhysRevX.12.021028} {\bibfield  {journal} {\bibinfo
  {journal} {Phys. Rev. X}\ }\textbf {\bibinfo {volume} {12}},\ \bibinfo
  {pages} {021028} (\bibinfo {year} {2022})}\BibitemShut {NoStop}%
\bibitem [{\citenamefont {Kang}\ \emph
  {et~al.}(2023{\natexlab{b}})\citenamefont {Kang}, \citenamefont {Liang},
  \citenamefont {Li},\ and\ \citenamefont {Nam}}]{Kang23mode}%
  \BibitemOpen
  \bibfield  {author} {\bibinfo {author} {\bibfnamefont {M.}~\bibnamefont
  {Kang}}, \bibinfo {author} {\bibfnamefont {Q.}~\bibnamefont {Liang}},
  \bibinfo {author} {\bibfnamefont {M.}~\bibnamefont {Li}},\ and\ \bibinfo
  {author} {\bibfnamefont {Y.}~\bibnamefont {Nam}},\ }\bibfield  {title}
  {\bibinfo {title} {Efficient motional-mode characterization for high-fidelity
  trapped-ion quantum computing},\ }\href
  {https://doi.org/10.1088/2058-9565/acb3f1} {\bibfield  {journal} {\bibinfo
  {journal} {Quantum Science and Technology}\ }\textbf {\bibinfo {volume}
  {8}},\ \bibinfo {pages} {024002} (\bibinfo {year}
  {2023}{\natexlab{b}})}\BibitemShut {NoStop}%
\bibitem [{\citenamefont {Zare}(1988)}]{Zare88}%
  \BibitemOpen
  \bibfield  {author} {\bibinfo {author} {\bibfnamefont {R.~N.}\ \bibnamefont
  {Zare}},\ }\href@noop {} {\emph {\bibinfo {title} {Angular momentum:
  understanding spatial aspects in chemistry and physics}}}\ (\bibinfo
  {publisher} {John Wiley \& Sons},\ \bibinfo {year} {1988})\BibitemShut
  {NoStop}%
\bibitem [{\citenamefont {Cohen-Tannoudji}\ \emph {et~al.}(1998)\citenamefont
  {Cohen-Tannoudji}, \citenamefont {Dupont-Roc},\ and\ \citenamefont
  {Grynberg}}]{Cohen98}%
  \BibitemOpen
  \bibfield  {author} {\bibinfo {author} {\bibfnamefont {C.}~\bibnamefont
  {Cohen-Tannoudji}}, \bibinfo {author} {\bibfnamefont {J.}~\bibnamefont
  {Dupont-Roc}},\ and\ \bibinfo {author} {\bibfnamefont {G.}~\bibnamefont
  {Grynberg}},\ }\href@noop {} {\emph {\bibinfo {title} {Atom-photon
  interactions: basic processes and applications}}}\ (\bibinfo  {publisher}
  {John Wiley \& Sons},\ \bibinfo {year} {1998})\BibitemShut {NoStop}%
\bibitem [{\citenamefont {Maslov}(2017)}]{Maslov17}%
  \BibitemOpen
  \bibfield  {author} {\bibinfo {author} {\bibfnamefont {D.}~\bibnamefont
  {Maslov}},\ }\bibfield  {title} {\bibinfo {title} {Basic circuit compilation
  techniques for an ion-trap quantum machine},\ }\href@noop {} {\bibfield
  {journal} {\bibinfo  {journal} {New Journal of Physics}\ }\textbf {\bibinfo
  {volume} {19}},\ \bibinfo {pages} {023035} (\bibinfo {year}
  {2017})}\BibitemShut {NoStop}%
\bibitem [{\citenamefont {Wu}\ \emph {et~al.}(2018)\citenamefont {Wu},
  \citenamefont {Wang},\ and\ \citenamefont {Duan}}]{Wu18}%
  \BibitemOpen
  \bibfield  {author} {\bibinfo {author} {\bibfnamefont {Y.}~\bibnamefont
  {Wu}}, \bibinfo {author} {\bibfnamefont {S.-T.}\ \bibnamefont {Wang}},\ and\
  \bibinfo {author} {\bibfnamefont {L.-M.}\ \bibnamefont {Duan}},\ }\bibfield
  {title} {\bibinfo {title} {Noise analysis for high-fidelity quantum
  entangling gates in an anharmonic linear paul trap},\ }\href
  {https://doi.org/10.1103/PhysRevA.97.062325} {\bibfield  {journal} {\bibinfo
  {journal} {Phys. Rev. A}\ }\textbf {\bibinfo {volume} {97}},\ \bibinfo
  {pages} {062325} (\bibinfo {year} {2018})}\BibitemShut {NoStop}%
\end{thebibliography}%

\end{document}